\input epsf
%
%
%
%

\catcode `\@=11 

\def\@version{1.6}
\def\@verdate{18th September 1995}

%
%


\newif\ifprod@font

\ifx\@typeface\undefined
  \def\@typeface{Comp. Modern}\prod@fontfalse
\else
  \prod@fonttrue 
\fi

\def\newfam{\alloc@8\fam\chardef\sixt@@n} 

\ifprod@font
\font\fiverm=mtr10 at 5pt
\font\fivebf=mtbx10 at 5pt
\font\fiveit=mtti10 at 5pt
\font\fivesl=mtsl10 at 5pt
\font\fivett=cmtt8 at 5pt     \hyphenchar\fivett=-1
\font\fivecsc=mtcsc10 at 5pt
\font\fivesf=mtss10 at 5pt
\font\fivei=mtmi10 at 5pt      \skewchar\fivei='177
\font\fivesy=mtsy10 at 5pt     \skewchar\fivesy='60

\font\sixrm=mtr10 at 6pt
\font\sixbf=mtbx10 at 6pt
\font\sixit=mtti10 at 6pt
\font\sixsl=mtsl10 at 6pt
\font\sixtt=cmtt8 at 6pt      \hyphenchar\sixtt=-1
\font\sixcsc=mtcsc10 at 6pt
\font\sixsf=mtss10 at 6pt
\font\sixi=mtmi10 at 6pt       \skewchar\sixi='177
\font\sixsy=mtsy10 at 6pt      \skewchar\sixsy='60

\font\sevenrm=mtr10 at 7pt
\font\sevenbf=mtbx10 at 7pt
\font\sevenit=mtti10 at 7pt
\font\sevensl=mtsl10 at 7pt
\font\seventt=cmtt8 at 7pt     \hyphenchar\seventt=-1
\font\sevencsc=mtcsc10 at 7pt
\font\sevensf=mtss10 at 7pt
\font\seveni=mtmi10 at 7pt      \skewchar\seveni='177
\font\sevensy=mtsy10 at 7pt     \skewchar\sevensy='60

\font\eightrm=mtr10 at 8pt
\font\eightbf=mtbx10 at 8pt
\font\eightit=mtti10 at 8pt
\font\eighti=mtmi10 at 8pt      \skewchar\eighti='177
\font\eightsy=mtsy10 at 8pt     \skewchar\eightsy='60
\font\eightsl=mtsl10 at 8pt
\font\eighttt=cmtt8             \hyphenchar\eighttt=-1
\font\eightcsc=mtcsc10 at 8pt
\font\eightsf=mtss10 at 8pt

\font\ninerm=mtr10 at 9pt
\font\ninebf=mtbx10 at 9pt
\font\nineit=mtti10 at 9pt
\font\ninei=mtmi10 at 9pt      \skewchar\ninei='177
\font\ninesy=mtsy10 at 9pt     \skewchar\ninesy='60
\font\ninesl=mtsl10 at 9pt
\font\ninett=cmtt9             \hyphenchar\ninett=-1
\font\ninecsc=mtcsc10 at 9pt
\font\ninesf=mtss10 at 9pt

\font\tenrm=mtr10
\font\tenbf=mtbx10
\font\tenit=mtti10
\font\teni=mtmi10		\skewchar\teni='177
\font\tensy=mtsy10		\skewchar\tensy='60
\font\tenex=cmex10
\font\tensl=mtsl10
\font\tentt=cmtt10		\hyphenchar\tentt=-1
\font\tencsc=mtcsc10
\font\tensf=mtss10

\font\elevenrm=mtr10 at 11pt
\font\elevenbf=mtbx10 at 11pt
\font\elevenit=mtti10 at 11pt
\font\eleveni=mtmi10 at 11pt      \skewchar\eleveni='177
\font\elevensy=mtsy10 at 11pt     \skewchar\elevensy='60
\font\elevensl=mtsl10 at 11pt
\font\eleventt=cmtt10 at 11pt     \hyphenchar\eleventt=-1
\font\elevencsc=mtcsc10 at 11pt
\font\elevensf=mtss10 at 11pt

\font\twelverm=mtr10 at 12pt
\font\twelvebf=mtbx10 at 12pt
\font\twelveit=mtti10 at 12pt
\font\twelvesl=mtsl10 at 12pt
\font\twelvett=cmtt12             \hyphenchar\twelvett=-1
\font\twelvecsc=mtcsc10 at 12pt
\font\twelvesf=mtss10 at 12pt
\font\twelvei=mtmi10 at 12pt      \skewchar\twelvei='177
\font\twelvesy=mtsy10 at 12pt     \skewchar\twelvesy='60

\font\fourteenrm=mtr10 at 14pt
\font\fourteenbf=mtbx10 at 14pt
\font\fourteenit=mtti10 at 14pt
\font\fourteeni=mtmi10 at 14pt      \skewchar\fourteeni='177
\font\fourteensy=mtsy10 at 14pt     \skewchar\fourteensy='60
\font\fourteensl=mtsl10 at 14pt
\font\fourteentt=cmtt12 at 14pt     \hyphenchar\fourteentt=-1
\font\fourteencsc=mtcsc10 at 14pt
\font\fourteensf=mtss10 at 14pt

\font\seventeenrm=mtr10 at 17pt
\font\seventeenbf=mtbx10 at 17pt
\font\seventeenit=mtti10 at 17pt
\font\seventeeni=mtmi10 at 17pt      \skewchar\seventeeni='177
\font\seventeensy=mtsy10 at 17pt     \skewchar\seventeensy='60
\font\seventeensl=mtsl10 at 17pt
\font\seventeentt=cmtt12 at 17pt     \hyphenchar\seventeentt=-1
\font\seventeencsc=mtcsc10 at 17pt
\font\seventeensf=mtss10 at 17pt
\else
\font\fiverm=cmr5
\font\fivei=cmmi5             \skewchar\fivei='177
\font\fivesy=cmsy5            \skewchar\fivesy='60
\font\fivebf=cmbx5

\font\sixrm=cmr6
\font\sixi=cmmi6             \skewchar\sixi='177
\font\sixsy=cmsy6            \skewchar\sixsy='60
\font\sixbf=cmbx6

\font\sevenrm=cmr7
\font\sevenit=cmti7
\font\seveni=cmmi7             \skewchar\seveni='177
\font\sevensy=cmsy7            \skewchar\sevensy='60
\font\sevenbf=cmbx7

\font\eightrm=cmr8
\font\eightbf=cmbx8
\font\eightit=cmti8
\font\eighti=cmmi8			\skewchar\eighti='177
\font\eightsy=cmsy8			\skewchar\eightsy='60
\font\eightsl=cmsl8
\font\eighttt=cmtt8			\hyphenchar\eighttt=-1
\font\eightcsc=cmcsc10 at 8pt
\font\eightsf=cmss8

\font\ninerm=cmr9
\font\ninebf=cmbx9
\font\nineit=cmti9
\font\ninei=cmmi9			\skewchar\ninei='177
\font\ninesy=cmsy9			\skewchar\ninesy='60
\font\ninesl=cmsl9
\font\ninett=cmtt9			\hyphenchar\ninett=-1
\font\ninecsc=cmcsc10 at 9pt
\font\ninesf=cmss9

\font\tenrm=cmr10
\font\tenbf=cmbx10
\font\tenit=cmti10
\font\teni=cmmi10		\skewchar\teni='177
\font\tensy=cmsy10		\skewchar\tensy='60
\font\tenex=cmex10
\font\tensl=cmsl10
\font\tentt=cmtt10		\hyphenchar\tentt=-1
\font\tencsc=cmcsc10
\font\tensf=cmss10

\font\elevenrm=cmr10 scaled \magstephalf
\font\elevenbf=cmbx10 scaled \magstephalf
\font\elevenit=cmti10 scaled \magstephalf
\font\eleveni=cmmi10 scaled \magstephalf	\skewchar\eleveni='177
\font\elevensy=cmsy10 scaled \magstephalf	\skewchar\elevensy='60
\font\elevensl=cmsl10 scaled \magstephalf
\font\eleventt=cmtt10 scaled \magstephalf	\hyphenchar\eleventt=-1
\font\elevencsc=cmcsc10 scaled \magstephalf
\font\elevensf=cmss10 scaled \magstephalf

\font\twelverm=cmr10 scaled \magstep1
\font\twelvebf=cmbx10 scaled \magstep1
\font\twelvei=cmmi10 scaled \magstep1      \skewchar\twelvei='177
\font\twelvesy=cmsy10 scaled \magstep1     \skewchar\twelvesy='60

\font\fourteenrm=cmr10 scaled \magstep2
\font\fourteenbf=cmbx10 scaled \magstep2
\font\fourteenit=cmti10 scaled \magstep2
\font\fourteeni=cmmi10 scaled \magstep2		\skewchar\fourteeni='177
\font\fourteensy=cmsy10 scaled \magstep2	\skewchar\fourteensy='60
\font\fourteensl=cmsl10 scaled \magstep2
\font\fourteentt=cmtt10 scaled \magstep2	\hyphenchar\fourteentt=-1
\font\fourteencsc=cmcsc10 scaled \magstep2
\font\fourteensf=cmss10 scaled \magstep2

\font\seventeenrm=cmr10 scaled \magstep3
\font\seventeenbf=cmbx10 scaled \magstep3
\font\seventeenit=cmti10 scaled \magstep3
\font\seventeeni=cmmi10 scaled \magstep3	\skewchar\seventeeni='177
\font\seventeensy=cmsy10 scaled \magstep3	\skewchar\seventeensy='60
\font\seventeensl=cmsl10 scaled \magstep3
\font\seventeentt=cmtt10 scaled \magstep3	\hyphenchar\seventeentt=-1
\font\seventeencsc=cmcsc10 scaled \magstep3
\font\seventeensf=cmss10 scaled \magstep3
\fi

\def\hexnumber#1{\ifcase#1 0\or1\or2\or3\or4\or5\or6\or7\or8\or9\or
  A\or B\or C\or D\or E\or F\fi}

\def\makestrut{%
  \setbox\strutbox=\hbox{%
    \vrule height.7\baselineskip depth.3\baselineskip width \z@}%
}

\def\baselinestretch{1}
\newskip\tmp@bls

\def\b@ls#1{
  \tmp@bls=#1\relax
  \baselineskip=#1\relax\makestrut
  \normalbaselineskip=\baselinestretch\tmp@bls
  \normalbaselines
}

\def\nostb@ls#1{
  \normalbaselineskip=#1\relax
  \normalbaselines
  \makestrut
}

%

\newfam\scfam  
\newfam\sffam  

\def\mit{\fam\@ne}
\def\cal{\fam\tw@}
\def\em{\ifdim\fontdimen1\font>\z@ \rm\else\it\fi}

\textfont3=\tenex
\scriptfont3=\tenex
\scriptscriptfont3=\tenex

\setbox0=\hbox{\tenex B} \p@renwd=\wd0 

\def\eightpoint{
  \def\rm{\fam0\eightrm}%
  \textfont0=\eightrm \scriptfont0=\sixrm \scriptscriptfont0=\fiverm%
  \textfont1=\eighti  \scriptfont1=\sixi  \scriptscriptfont1=\fivei%
  \textfont2=\eightsy \scriptfont2=\sixsy \scriptscriptfont2=\fivesy%
  \textfont\itfam=\eightit\def\it{\fam\itfam\eightit}%
  \ifprod@font
    \scriptfont\itfam=\sixit
      \scriptscriptfont\itfam=\fiveit
  \else
    \scriptfont\itfam=\eightit
      \scriptscriptfont\itfam=\eightit
  \fi
  \textfont\bffam=\eightbf%
    \scriptfont\bffam=\sixbf%
      \scriptscriptfont\bffam=\fivebf%
  \def\bf{\fam\bffam\eightbf}%
  \textfont\slfam=\eightsl\def\sl{\fam\slfam\eightsl}%
  \ifprod@font
    \scriptfont\slfam=\sixsl
      \scriptscriptfont\slfam=\fivesl
  \else
    \scriptfont\slfam=\eightsl
      \scriptscriptfont\slfam=\eightsl
  \fi
  \textfont\ttfam=\eighttt\def\tt{\fam\ttfam\eighttt}%
  \ifprod@font
    \scriptfont\ttfam=\sixtt
      \scriptscriptfont\ttfam=\fivett
  \else
    \scriptfont\ttfam=\eighttt
      \scriptscriptfont\ttfam=\eighttt
  \fi
  \textfont\scfam=\eightcsc\def\sc{\fam\scfam\eightcsc}%
  \ifprod@font
    \scriptfont\scfam=\sixcsc
      \scriptscriptfont\scfam=\fivecsc
  \else
    \scriptfont\scfam=\eightcsc
      \scriptscriptfont\scfam=\eightcsc
  \fi
  \textfont\sffam=\eightsf\def\sf{\fam\sffam\eightsf}%
  \ifprod@font
    \scriptfont\sffam=\sixsf
      \scriptscriptfont\sffam=\fivesf
  \else
    \scriptfont\sffam=\eightsf
      \scriptscriptfont\sffam=\eightsf
  \fi
  \def\oldstyle{\fam\@ne\eighti}%
  \b@ls{10pt}\rm\@viiipt%
}
\def\@viiipt{}

\def\ninepoint{
  \def\rm{\fam0\ninerm}%
  \textfont0=\ninerm \scriptfont0=\sixrm \scriptscriptfont0=\fiverm%
  \textfont1=\ninei  \scriptfont1=\sixi  \scriptscriptfont1=\fivei%
  \textfont2=\ninesy \scriptfont2=\sixsy \scriptscriptfont2=\fivesy%
  \textfont\itfam=\nineit\def\it{\fam\itfam\nineit}%
  \ifprod@font
    \scriptfont\itfam=\sixit
      \scriptscriptfont\itfam=\fiveit
  \else
    \scriptfont\itfam=\nineit
      \scriptscriptfont\itfam=\nineit
  \fi
  \textfont\bffam=\ninebf%
    \scriptfont\bffam=\sixbf%
      \scriptscriptfont\bffam=\fivebf%
  \def\bf{\fam\bffam\ninebf}%
  \textfont\slfam=\ninesl\def\sl{\fam\slfam\ninesl}%
  \ifprod@font
    \scriptfont\slfam=\sixsl
      \scriptscriptfont\slfam=\fivesl
  \else
    \scriptfont\slfam=\ninesl
      \scriptscriptfont\slfam=\ninesl
  \fi
  \textfont\ttfam=\ninett\def\tt{\fam\ttfam\ninett}%
  \ifprod@font
    \scriptfont\ttfam=\sixtt
      \scriptscriptfont\ttfam=\fivett
  \else
    \scriptfont\ttfam=\ninett
      \scriptscriptfont\ttfam=\ninett
  \fi
  \textfont\scfam=\ninecsc\def\sc{\fam\scfam\ninecsc}%
  \ifprod@font
    \scriptfont\scfam=\sixcsc
      \scriptscriptfont\scfam=\fivecsc
  \else
    \scriptfont\scfam=\ninecsc
      \scriptscriptfont\scfam=\ninecsc
  \fi
  \textfont\sffam=\ninesf\def\sf{\fam\sffam\ninesf}%
  \ifprod@font
    \scriptfont\sffam=\sixsf
      \scriptscriptfont\sffam=\fivesf
  \else
    \scriptfont\sffam=\ninesf
      \scriptscriptfont\sffam=\ninesf
  \fi
  \def\oldstyle{\fam\@ne\ninei}%
  \b@ls{\TextLeading plus \Feathering}\rm\@ixpt%
}
\def\@ixpt{}

\def\tenpoint{
  \def\rm{\fam0\tenrm}%
  \textfont0=\tenrm \scriptfont0=\sevenrm \scriptscriptfont0=\fiverm%
  \textfont1=\teni  \scriptfont1=\seveni  \scriptscriptfont1=\fivei%
  \textfont2=\tensy \scriptfont2=\sevensy \scriptscriptfont2=\fivesy%
  \textfont\itfam=\tenit\def\it{\fam\itfam\tenit}%
  \ifprod@font
    \scriptfont\itfam=\sevenit
      \scriptscriptfont\itfam=\fiveit
  \else
    \scriptfont\itfam=\tenit
      \scriptscriptfont\itfam=\tenit
  \fi
  \textfont\bffam=\tenbf%
    \scriptfont\bffam=\sevenbf%
      \scriptscriptfont\bffam=\fivebf%
  \def\bf{\fam\bffam\tenbf}%
  \textfont\slfam=\tensl\def\sl{\fam\slfam\tensl}%
  \ifprod@font
    \scriptfont\slfam=\sevensl
      \scriptscriptfont\slfam=\fivesl
  \else
    \scriptfont\slfam=\tensl
      \scriptscriptfont\slfam=\tensl
  \fi
  \textfont\ttfam=\tentt\def\tt{\fam\ttfam\tentt}%
  \ifprod@font
    \scriptfont\ttfam=\seventt
      \scriptscriptfont\ttfam=\fivett
  \else
    \scriptfont\ttfam=\tentt
      \scriptscriptfont\ttfam=\tentt
  \fi
  \textfont\scfam=\tencsc\def\sc{\fam\scfam\tencsc}%
  \ifprod@font
    \scriptfont\scfam=\sevencsc
      \scriptscriptfont\scfam=\fivecsc
  \else
    \scriptfont\scfam=\tencsc
      \scriptscriptfont\scfam=\tencsc
  \fi
  \textfont\sffam=\tensf\def\sf{\fam\sffam\tensf}%
  \ifprod@font
    \scriptfont\sffam=\sevensf
      \scriptscriptfont\sffam=\fivesf
  \else
    \scriptfont\sffam=\tensf
      \scriptscriptfont\sffam=\tensf
  \fi
  \def\oldstyle{\fam\@ne\teni}%
  \b@ls{11pt}\rm\@xpt%
}
\def\@xpt{}

\def\elevenpoint{
  \def\rm{\fam0\elevenrm}%
  \textfont0=\elevenrm \scriptfont0=\eightrm \scriptscriptfont0=\sixrm%
  \textfont1=\eleveni  \scriptfont1=\eighti  \scriptscriptfont1=\sixi%
  \textfont2=\elevensy \scriptfont2=\eightsy \scriptscriptfont2=\sixsy%
  \textfont\itfam=\elevenit\def\it{\fam\itfam\elevenit}%
  \ifprod@font
    \scriptfont\itfam=\eightit
      \scriptscriptfont\itfam=\sixit
  \else
    \scriptfont\itfam=\elevenit
      \scriptscriptfont\itfam=\elevenit
  \fi
  \textfont\bffam=\elevenbf%
    \scriptfont\bffam=\eightbf%
      \scriptscriptfont\bffam=\sixbf%
  \def\bf{\fam\bffam\elevenbf}%
  \textfont\slfam=\elevensl\def\sl{\fam\slfam\elevensl}%
  \ifprod@font
    \scriptfont\slfam=\eightsl
      \scriptscriptfont\slfam=\sixsl
  \else
    \scriptfont\slfam=\elevensl
      \scriptscriptfont\slfam=\elevensl
  \fi
  \textfont\ttfam=\eleventt\def\tt{\fam\ttfam\eleventt}%
  \ifprod@font
    \scriptfont\ttfam=\eighttt
      \scriptscriptfont\ttfam=\sixtt
  \else
    \scriptfont\ttfam=\eleventt
      \scriptscriptfont\ttfam=\eleventt
  \fi
  \textfont\scfam=\elevencsc\def\sc{\fam\scfam\elevencsc}%
  \ifprod@font
    \scriptfont\scfam=\eightcsc
      \scriptscriptfont\scfam=\sixcsc
  \else
    \scriptfont\scfam=\elevencsc
      \scriptscriptfont\scfam=\elevencsc
  \fi
  \textfont\sffam=\elevensf\def\sf{\fam\sffam\elevensf}%
  \ifprod@font
    \scriptfont\sffam=\eightsf
      \scriptscriptfont\sffam=\sixsf
  \else
    \scriptfont\sffam=\elevensf
      \scriptscriptfont\sffam=\elevensf
  \fi
  \def\oldstyle{\fam\@ne\eleveni}%
  \b@ls{13pt}\rm\@xipt%
}
\def\@xipt{}

\def\fourteenpoint{
  \def\rm{\fam0\fourteenrm}%
  \textfont0\fourteenrm  \scriptfont0\tenrm  \scriptscriptfont0\sevenrm%
  \textfont1\fourteeni   \scriptfont1\teni   \scriptscriptfont1\seveni%
  \textfont2\fourteensy  \scriptfont2\tensy  \scriptscriptfont2\sevensy%
  \textfont\itfam=\fourteenit\def\it{\fam\itfam\fourteenit}%
  \ifprod@font
    \scriptfont\itfam=\tenit
      \scriptscriptfont\itfam=\sevenit
  \else
    \scriptfont\itfam=\fourteenit
      \scriptscriptfont\itfam=\fourteenit
  \fi
  \textfont\bffam=\fourteenbf%
    \scriptfont\bffam=\tenbf%
      \scriptscriptfont\bffam=\sevenbf%
  \def\bf{\fam\bffam\fourteenbf}%
  \textfont\slfam=\fourteensl\def\sl{\fam\slfam\fourteensl}%
  \ifprod@font
    \scriptfont\slfam=\tensl
      \scriptscriptfont\slfam=\sevensl
  \else
    \scriptfont\slfam=\fourteensl
      \scriptscriptfont\slfam=\fourteensl
  \fi
  \textfont\ttfam=\fourteentt\def\tt{\fam\ttfam\fourteentt}%
  \ifprod@font
    \scriptfont\ttfam=\tentt
      \scriptscriptfont\ttfam=\seventt
  \else
    \scriptfont\ttfam=\fourteentt
      \scriptscriptfont\ttfam=\fourteentt
  \fi
  \textfont\scfam=\fourteencsc\def\sc{\fam\scfam\fourteencsc}%
  \ifprod@font
    \scriptfont\scfam=\tencsc
      \scriptscriptfont\scfam=\sevencsc
  \else
    \scriptfont\scfam=\fourteencsc
      \scriptscriptfont\scfam=\fourteencsc
  \fi
  \textfont\sffam=\fourteensf\def\sf{\fam\sffam\fourteensf}%
  \ifprod@font
    \scriptfont\sffam=\tensf
      \scriptscriptfont\sffam=\sevensf
  \else
    \scriptfont\sffam=\fourteensf
      \scriptscriptfont\sffam=\fourteensf
  \fi
  \def\oldstyle{\fam\@ne\fourteeni}%
  \b@ls{17pt}\rm\@xivpt%
}
\def\@xivpt{}

\def\seventeenpoint{
  \def\rm{\fam0\seventeenrm}%
  \textfont0\seventeenrm  \scriptfont0\twelverm  \scriptscriptfont0\tenrm%
  \textfont1\seventeeni   \scriptfont1\twelvei   \scriptscriptfont1\teni%
  \textfont2\seventeensy  \scriptfont2\twelvesy  \scriptscriptfont2\tensy%
  \textfont\itfam=\seventeenit\def\it{\fam\itfam\seventeenit}%
  \ifprod@font
    \scriptfont\itfam=\twelveit
      \scriptscriptfont\itfam=\tenit
  \else
    \scriptfont\itfam=\seventeenit
      \scriptscriptfont\itfam=\seventeenit
  \fi
  \textfont\bffam=\seventeenbf%
    \scriptfont\bffam=\twelvebf%
      \scriptscriptfont\bffam=\tenbf%
  \def\bf{\fam\bffam\seventeenbf}%
  \textfont\slfam=\seventeensl\def\sl{\fam\slfam\seventeensl}%
  \ifprod@font
    \scriptfont\slfam=\twelvesl
      \scriptscriptfont\slfam=\tensl
  \else
    \scriptfont\slfam=\seventeensl
      \scriptscriptfont\slfam=\seventeensl
  \fi
  \textfont\ttfam=\seventeentt\def\tt{\fam\ttfam\seventeentt}%
  \ifprod@font
    \scriptfont\ttfam=\twelvett
      \scriptscriptfont\ttfam=\tentt
  \else
    \scriptfont\ttfam=\seventeentt
      \scriptscriptfont\ttfam=\seventeentt
  \fi
  \textfont\scfam=\seventeencsc\def\sc{\fam\scfam\seventeencsc}%
  \ifprod@font
    \scriptfont\scfam=\twelvecsc
      \scriptscriptfont\scfam=\tencsc
  \else
    \scriptfont\scfam=\seventeencsc
      \scriptscriptfont\scfam=\seventeencsc
  \fi
  \textfont\sffam=\seventeensf\def\sf{\fam\sffam\seventeensf}%
  \ifprod@font
    \scriptfont\sffam=\twelvesf
      \scriptscriptfont\sffam=\tensf
  \else
    \scriptfont\sffam=\seventeensf
      \scriptscriptfont\sffam=\seventeensf
  \fi
  \def\oldstyle{\fam\@ne\seventeeni}%
  \b@ls{20pt}\rm\@xviipt%
}
\def\@xviipt{}

\lineskip=1pt      \normallineskip=\lineskip
\lineskiplimit=\z@ \normallineskiplimit=\lineskiplimit


\def\loadboldmathnames{%
  \def\balpha{{\bmath{\alpha}}}%
  \def\bbeta{{\bmath{\beta}}}%
  \def\bgamma{{\bmath{\gamma}}}%
  \def\bdelta{{\bmath{\delta}}}%
  \def\bepsilon{{\bmath{\epsilon}}}%
  \def\bzeta{{\bmath{\zeta}}}%
  \def\boldeta{{\bmath{\eta}}}%
  \def\btheta{{\bmath{\theta}}}%
  \def\biota{{\bmath{\iota}}}%
  \def\bkappa{{\bmath{\kappa}}}%
  \def\blambda{{\bmath{\lambda}}}%
  \def\bmu{{\bmath{\mu}}}%
  \def\bnu{{\bmath{\nu}}}%
  \def\bxi{{\bmath{\xi}}}%
  \def\bpi{{\bmath{\pi}}}%
  \def\brho{{\bmath{\rho}}}%
  \def\bsigma{{\bmath{\sigma}}}%
  \def\btau{{\bmath{\tau}}}%
  \def\bupsilon{{\bmath{\upsilon}}}%
  \def\bphi{{\bmath{\phi}}}%
  \def\bchi{{\bmath{\chi}}}%
  \def\bpsi{{\bmath{\psi}}}%
  \def\bomega{{\bmath{\omega}}}%
  \def\bvarepsilon{{\bmath{\varepsilon}}}%
  \def\bvartheta{{\bmath{\vartheta}}}%
  \def\bvarpi{{\bmath{\varpi}}}%
  \def\bvarrho{{\bmath{\varrho}}}%
  \def\bvarsigma{{\bmath{\varsigma}}}%
  \def\bvarphi{{\bmath{\varphi}}}%
  \def\baleph{{\bmath{\aleph}}}%
  \def\bimath{{\bmath{\imath}}}%
  \def\bjmath{{\bmath{\jmath}}}%
  \def\bell{{\bmath{\ell}}}%
  \def\bwp{{\bmath{\wp}}}%
  \def\bRe{{\bmath{\Re}}}%
  \def\bIm{{\bmath{\Im}}}%
  \def\bpartial{{\bmath{\partial}}}%
  \def\binfty{{\bmath{\infty}}}%
  \def\bprime{{\bmath{\prime}}}%
  \def\bemptyset{{\bmath{\emptyset}}}%
  \def\bnabla{{\bmath{\nabla}}}%
  \def\btop{{\bmath{\top}}}%
  \def\bbot{{\bmath{\bot}}}%
  \def\btriangle{{\bmath{\triangle}}}%
  \def\bforall{{\bmath{\forall}}}%
  \def\bexists{{\bmath{\exists}}}%
  \def\bneg{{\bmath{\neg}}}%
  \def\bflat{{\bmath{\flat}}}%
  \def\bnatural{{\bmath{\natural}}}%
  \def\bsharp{{\bmath{\sharp}}}%
  \def\bclubsuit{{\bmath{\clubsuit}}}%
  \def\bdiamondsuit{{\bmath{\diamondsuit}}}%
  \def\bheartsuit{{\bmath{\heartsuit}}}%
  \def\bspadesuit{{\bmath{\spadesuit}}}%
  \def\bsmallint{{\bmath{\smallint}}}%
  \def\btriangleleft{{\bmath{\triangleleft}}}%
  \def\btriangleright{{\bmath{\triangleright}}}%
  \def\bbigtriangleup{{\bmath{\bigtriangleup}}}%
  \def\bbigtriangledown{{\bmath{\bigtriangledown}}}%
  \def\bwedge{{\bmath{\wedge}}}%
  \def\bvee{{\bmath{\vee}}}%
  \def\bcap{{\bmath{\cap}}}%
  \def\bcup{{\bmath{\cup}}}%
  \def\bddagger{{\bmath{\ddagger}}}%
  \def\bdagger{{\bmath{\dagger}}}%
  \def\bsqcap{{\bmath{\sqcap}}}%
  \def\bsqcup{{\bmath{\sqcup}}}%
  \def\buplus{{\bmath{\uplus}}}%
  \def\bamalg{{\bmath{\amalg}}}%
  \def\bdiamond{{\bmath{\diamond}}}%
  \def\bbullet{{\bmath{\bullet}}}%
  \def\bwr{{\bmath{\wr}}}%
  \def\bdiv{{\bmath{\div}}}%
  \def\bodot{{\bmath{\odot}}}%
  \def\boslash{{\bmath{\oslash}}}%
  \def\botimes{{\bmath{\otimes}}}%
  \def\bominus{{\bmath{\ominus}}}%
  \def\boplus{{\bmath{\oplus}}}%
  \def\bmp{{\bmath{\mp}}}%
  \def\bpm{{\bmath{\pm}}}%
  \def\bcirc{{\bmath{\circ}}}%
  \def\bbigcirc{{\bmath{\bigcirc}}}%
  \def\bsetminus{{\bmath{\setminus}}}%
  \def\bcdot{{\bmath{\cdot}}}%
  \def\bast{{\bmath{\ast}}}%
  \def\btimes{{\bmath{\times}}}%
  \def\bstar{{\bmath{\star}}}%
  \def\bpropto{{\bmath{\propto}}}%
  \def\bsqsubseteq{{\bmath{\sqsubseteq}}}%
  \def\bsqsupseteq{{\bmath{\sqsupseteq}}}%
  \def\bparallel{{\bmath{\parallel}}}%
  \def\bmid{{\bmath{\mid}}}%
  \def\bdashv{{\bmath{\dashv}}}%
  \def\bvdash{{\bmath{\vdash}}}%
  \def\bnearrow{{\bmath{\nearrow}}}%
  \def\bsearrow{{\bmath{\searrow}}}%
  \def\bnwarrow{{\bmath{\nwarrow}}}%
  \def\bswarrow{{\bmath{\swarrow}}}%
  \def\bLeftrightarrow{{\bmath{\Leftrightarrow}}}%
  \def\bLeftarrow{{\bmath{\Leftarrow}}}%
  \def\bRightarrow{{\bmath{\Rightarrow}}}%
  \def\bleq{{\bmath{\leq}}}%
  \def\bgeq{{\bmath{\geq}}}%
  \def\bsucc{{\bmath{\succ}}}%
  \def\bprec{{\bmath{\prec}}}%
  \def\bapprox{{\bmath{\approx}}}%
  \def\bsucceq{{\bmath{\succeq}}}%
  \def\bpreceq{{\bmath{\preceq}}}%
  \def\bsupset{{\bmath{\supset}}}%
  \def\bsubset{{\bmath{\subset}}}%
  \def\bsupseteq{{\bmath{\supseteq}}}%
  \def\bsubseteq{{\bmath{\subseteq}}}%
  \def\bin{{\bmath{\in}}}%
  \def\bni{{\bmath{\ni}}}%
  \def\bgg{{\bmath{\gg}}}%
  \def\bll{{\bmath{\ll}}}%
  \def\bnot{{\bmath{\not}}}%
  \def\bleftrightarrow{{\bmath{\leftrightarrow}}}%
  \def\bleftarrow{{\bmath{\leftarrow}}}%
  \def\brightarrow{{\bmath{\rightarrow}}}%
  \def\bmapstochar{{\bmath{\mapstochar}}}%
  \def\bsim{{\bmath{\sim}}}%
  \def\bsimeq{{\bmath{\simeq}}}%
  \def\bperp{{\bmath{\perp}}}%
  \def\bequiv{{\bmath{\equiv}}}%
  \def\basymp{{\bmath{\asymp}}}%
  \def\bsmile{{\bmath{\smile}}}%
  \def\bfrown{{\bmath{\frown}}}%
  \def\bleftharpoonup{{\bmath{\leftharpoonup}}}%
  \def\bleftharpoondown{{\bmath{\leftharpoondown}}}%
  \def\brightharpoonup{{\bmath{\rightharpoonup}}}%
  \def\brightharpoondown{{\bmath{\rightharpoondown}}}%
  \def\blhook{{\bmath{\lhook}}}%
  \def\brhook{{\bmath{\rhook}}}%
  \def\bldotp{{\bmath{\ldotp}}}%
  \def\bcdotp{{\bmath{\cdotp}}}%
}

\def\,{\relax\ifmmode \mskip\thinmuskip\else \thinspace\fi}
\let\protect=\relax

\long\def\@ifundefined#1#2#3{\expandafter\ifx\csname
  #1\endcsname\relax#2\else#3\fi}




\newtoks\math@groups \math@groups={}
\def\addtom@thgroup#1#2{#1\expandafter{\the#1#2}} 



\def\addtosizeh@ok#1#2#3#4{%
  \expandafter\def\csname @#1pt\endcsname{%
    \def\s@ze{#2}\def\ss@ze{#3}\def\sss@ze{#4}\the\math@groups%
  }%
}



\let\resetsizehook=\addtosizeh@ok


\ifprod@font
  \addtosizeh@ok{viii} {8} {6}  {5}
  \addtosizeh@ok{ix}   {9} {6}  {5}
  \addtosizeh@ok{x}    {10}{7}  {5}
  \addtosizeh@ok{xi}   {11}{8}  {6}
  \addtosizeh@ok{xiv}  {14}{10} {7}
  \addtosizeh@ok{xvii} {17}{12}{10}
\else
  \addtosizeh@ok{viii} {8}     {6}     {5}
  \addtosizeh@ok{ix}   {9}     {6}     {5}
  \addtosizeh@ok{x}    {10}    {7}     {5}
  \addtosizeh@ok{xi}   {10.95} {8}     {6}
  \addtosizeh@ok{xiv}  {14.4}  {10}    {7}
  \addtosizeh@ok{xvii} {17.28} {12}    {10}
\fi

\def\get@font#1#2#3{%
  \edef\fonts@ze{\romannumeral#3}
  \edef\fontn@me{\fonts@ze#1}
  \@ifundefined{\fontn@me}%
    {
     \global\expandafter\font\csname \fontn@me\endcsname=#2 at #3pt}%
    {}%
}

\def\ass@tfont#1#2{%
  \xdef\fam@name{\csname #1\endcsname}%
  \xdef\font@name{\csname #2\endcsname}%
  \let\textfont@name\font@name
  \textfont\fam@name\textfont@name
}

\def\ass@sfont#1#2{%
  \xdef\fam@name{\csname #1\endcsname}%
  \xdef\font@name{\csname #2\endcsname}%
  \let\textfont@name\font@name
  \scriptfont\fam@name\textfont@name
}

\def\ass@ssfont#1#2{%
  \xdef\fam@name{\csname #1\endcsname}%
  \xdef\font@name{\csname #2\endcsname}%
  \let\textfont@name\font@name
  \scriptscriptfont\fam@name\textfont@name
}


\def\NewSymbolFont#1#2{%
  \expandafter\ifx\csname sym#1fam\endcsname\relax 
    \expandafter\newfam\csname sym#1fam\endcsname
    \expandafter\edef\csname sym#1fam\endcsname{\the\allocationnumber}%
    \addtom@thgroup\math@groups{%
      \get@font{#1}{#2}{\s@ze}%
      \ass@tfont{sym#1fam}{\fontn@me}%
      \get@font{#1}{#2}{\ss@ze}%
      \ass@sfont{sym#1fam}{\fontn@me}%
      \get@font{#1}{#2}{\sss@ze}%
      \ass@ssfont{sym#1fam}{\fontn@me}%
    }%
  \else
    \errmessage{Family `#1' already defined}%
  \fi
}


\def\NewMathSymbol#1#2#3#4{%
  \edef\f@mly{\expandafter\hexnumber{\csname sym#3fam\endcsname}}%
  \mathchardef#1="#2\f@mly#4\relax
}


\newif\ifd@f

\def\NewMathDelimiter#1#2#3#4#5#6{%
  \d@ftrue
  \expandafter\ifx\csname sym#3fam\endcsname\relax
    \d@ffalse \errmessage{Family `#3' is not defined}%
  \fi
  \expandafter\ifx\csname sym#5fam\endcsname\relax
    \d@ffalse \errmessage{Family `#5' is not defined}%
  \fi
  \ifd@f
    \edef\f@mly{\expandafter\hexnumber{\csname sym#3fam\endcsname}}%
    \edef\f@mlytw@{\expandafter\hexnumber{\csname sym#5fam\endcsname}}%
    \xdef#1{\delimiter"#2\f@mly #4\f@mlytw@ #6\relax}%
  \fi
}


\def\setboxz@h{\setbox\z@\hbox}
\def\wdz@{\wd\z@}
\def\boxz@{\box\z@}
\def\setbox@ne{\setbox\@ne}
\def\wd@ne{\wd\@ne}

\def\math@atom#1#2{%
   \binrel@{#1}\binrel@@{#2}}
\def\binrel@#1{\setboxz@h{\thinmuskip0mu
  \medmuskip\m@ne mu\thickmuskip\@ne mu$#1\m@th$}%
 \setbox@ne\hbox{\thinmuskip0mu\medmuskip\m@ne mu\thickmuskip
  \@ne mu${}#1{}\m@th$}%
 \setbox\tw@\hbox{\hskip\wd@ne\hskip-\wdz@}}
\def\binrel@@#1{\ifdim\wd2<\z@\mathbin{#1}\else\ifdim\wd\tw@>\z@
 \mathrel{#1}\else{#1}\fi\fi}

\def\m@thit{1}

\def\set@skchar#1{\global\expandafter\skewchar
  \csname\fontn@me\endcsname=#1\relax}

\def\NewMathAlphabet#1#2#3{%
  \def\tst{#3}%
  \ifx\tst\empty\else 
    \expandafter\gdef\csname #1@sc\endcsname{}
  \fi
  \expandafter\def\csname #1\endcsname{
    \protect\csname @#1\endcsname}%
  \expandafter\def\csname @#1\endcsname##1{
    {%
    \begingroup
      \get@font{#1}{#2}{\s@ze}%
      \@ifundefined{#1@sc}{}{\set@skchar{#3}}%
      \ass@tfont{m@thit}{\fontn@me}%
      \get@font{#1}{#2}{\ss@ze}%
      \@ifundefined{#1@sc}{}{\set@skchar{#3}}%
      \ass@sfont{m@thit}{\fontn@me}%
      \get@font{#1}{#2}{\sss@ze}%
      \@ifundefined{#1@sc}{}{\set@skchar{#3}}%
      \ass@ssfont{m@thit}{\fontn@me}%
      \math@atom{##1}{%
      \mathchoice%
        {\hbox{$\m@th\displaystyle##1$}}%
        {\hbox{$\m@th\textstyle##1$}}%
        {\hbox{$\m@th\scriptstyle##1$}}%
        {\hbox{$\m@th\scriptscriptstyle##1$}}}%
    \endgroup
    }%
  }%
}


\newif\iffirstta  \firsttatrue

\def\set@hchar#1{\global\expandafter\hyphenchar
  \csname\fontn@me\endcsname=#1\relax}

\def\NewTextAlphabet#1#2#3{%
  \iffirstta
    \global\firsttafalse
    \newfam\scratchfam
    \edef\scrt@fam{\the\allocationnumber}%
  \fi
  \def\tst{#3}%
  \ifx\tst\empty\else 
    \expandafter\gdef\csname #1@hc\endcsname{}
  \fi
  \expandafter\def\csname #1\endcsname{
    \protect\csname t@#1\endcsname}%
  \long\expandafter\def\csname t@#1\endcsname##1{
    \ifmmode
      \typeout{Warning: do not use \expandafter\string\csname #1\endcsname
        \space in math mode}\fi%
    {%
      \get@font{#1}{#2}{\s@ze}\let\t@xtfnt=\fontn@me\relax
      \@ifundefined{#1@hc}{}{\set@hchar{#3}}%
      \ass@tfont{scrt@fam}{\fontn@me}%
      \get@font{#1}{#2}{\ss@ze}%
      \@ifundefined{#1@hc}{}{\set@hchar{#3}}%
      \ass@sfont{scrt@fam}{\fontn@me}%
      \get@font{#1}{#2}{\sss@ze}%
      \@ifundefined{#1@hc}{}{\set@hchar{#3}}%
      \ass@ssfont{scrt@fam}{\fontn@me}%
      \fam\scratchfam\csname\t@xtfnt\endcsname
    ##1%
    }%
  }%
  \expandafter\def\csname #1shape
    \endcsname{\protect\csname @#1shape\endcsname}%
  \expandafter\def\csname @#1shape\endcsname{
    \ifmmode
      \typeout{Warning: do not use \expandafter\string\csname
        #1shape\endcsname \space in math mode}\fi
      \get@font{#1}{#2}{\s@ze}\let\t@xtfnt=\fontn@me\relax
      \@ifundefined{#1@hc}{}{\set@hchar{#3}}%
      \ass@tfont{scrt@fam}{\fontn@me}%
      \get@font{#1}{#2}{\ss@ze}%
      \@ifundefined{#1@hc}{}{\set@hchar{#3}}%
      \ass@sfont{scrt@fam}{\fontn@me}%
      \get@font{#1}{#2}{\sss@ze}%
      \@ifundefined{#1@hc}{}{\set@hchar{#3}}%
      \ass@ssfont{scrt@fam}{\fontn@me}%
      \fam\scratchfam\csname\t@xtfnt\endcsname
  }%
}


\ifprod@font
  \def\math@itfnt{mtmib10}
  \def\math@syfnt{mtbsy10}
\else
  \def\math@itfnt{cmmib10}
  \def\math@syfnt{cmbsy10}
\fi

\def\m@thsy{2}

\def\bmath{\protect\@bmath}
\def\@bmath#1{%
  {%
  \begingroup
    \get@font{mthit}{\math@itfnt}{\s@ze}\set@skchar{'177}%
    \ass@tfont{m@thit}{\fontn@me}%
    \get@font{mthit}{\math@itfnt}{\ss@ze}\set@skchar{'177}%
    \ass@sfont{m@thit}{\fontn@me}%
    \get@font{mthit}{\math@itfnt}{\sss@ze}\set@skchar{'177}%
    \ass@ssfont{m@thit}{\fontn@me}%
    \get@font{mthsy}{\math@syfnt}{\s@ze}\set@skchar{'60}%
    \ass@tfont{m@thsy}{\fontn@me}%
    \get@font{mthsy}{\math@syfnt}{\ss@ze}\set@skchar{'60}%
    \ass@sfont{m@thsy}{\fontn@me}%
    \get@font{mthsy}{\math@syfnt}{\sss@ze}\set@skchar{'60}%
    \ass@ssfont{m@thsy}{\fontn@me}%
    \math@atom{#1}{%
    \mathchoice%
      {\hbox{$\m@th\displaystyle#1$}}%
      {\hbox{$\m@th\textstyle#1$}}%
      {\hbox{$\m@th\scriptstyle#1$}}%
      {\hbox{$\m@th\scriptscriptstyle#1$}}}%
  \endgroup
  }%
}



\def\diameter{{\ifmmode\mathchoice
{\ooalign{\hfil\hbox{$\displaystyle/$}\hfil\crcr
{\hbox{$\displaystyle\mathchar"20D$}}}}
{\ooalign{\hfil\hbox{$\textstyle/$}\hfil\crcr
{\hbox{$\textstyle\mathchar"20D$}}}}
{\ooalign{\hfil\hbox{$\scriptstyle/$}\hfil\crcr
{\hbox{$\scriptstyle\mathchar"20D$}}}}
{\ooalign{\hfil\hbox{$\scriptscriptstyle/$}\hfil\crcr
{\hbox{$\scriptscriptstyle\mathchar"20D$}}}}
\else{\ooalign{\hfil/\hfil\crcr\mathhexbox20D}}%
\fi}}

\def\sq{\ifmmode\squareforqed\else{\unskip\nobreak\hfil
\penalty50\hskip1em\null\nobreak\hfil\squareforqed
\parfillskip=0pt\finalhyphendemerits=0\endgraf}\fi}
\def\squareforqed{\hbox{\rlap{$\sqcap$}$\sqcup$}}


\def\bbbr{{\rm I\!R}}

\def\bbbn{{\rm I\!N}}

\def\bbbc{{\mathchoice {\setbox0=\hbox{$\displaystyle\rm C$}\hbox{\hbox
to0pt{\kern0.4\wd0\vrule height0.9\ht0\hss}\box0}}
{\setbox0=\hbox{$\textstyle\rm C$}\hbox{\hbox
to0pt{\kern0.4\wd0\vrule height0.9\ht0\hss}\box0}}
{\setbox0=\hbox{$\scriptstyle\rm C$}\hbox{\hbox
to0pt{\kern0.4\wd0\vrule height0.9\ht0\hss}\box0}}
{\setbox0=\hbox{$\scriptscriptstyle\rm C$}\hbox{\hbox
to0pt{\kern0.4\wd0\vrule height0.9\ht0\hss}\box0}}}}
\def\bbbq{{\mathchoice {\setbox0=\hbox{$\displaystyle\rm
Q$}\hbox{\raise
0.15\ht0\hbox to0pt{\kern0.4\wd0\vrule height0.8\ht0\hss}\box0}}
{\setbox0=\hbox{$\textstyle\rm Q$}\hbox{\raise
0.15\ht0\hbox to0pt{\kern0.4\wd0\vrule height0.8\ht0\hss}\box0}}
{\setbox0=\hbox{$\scriptstyle\rm Q$}\hbox{\raise
0.15\ht0\hbox to0pt{\kern0.4\wd0\vrule height0.7\ht0\hss}\box0}}
{\setbox0=\hbox{$\scriptscriptstyle\rm Q$}\hbox{\raise
0.15\ht0\hbox to0pt{\kern0.4\wd0\vrule height0.7\ht0\hss}\box0}}}}
\def\bbbt{{\mathchoice {\setbox0=\hbox{$\displaystyle\rm
T$}\hbox{\hbox to0pt{\kern0.3\wd0\vrule height0.9\ht0\hss}\box0}}
{\setbox0=\hbox{$\textstyle\rm T$}\hbox{\hbox
to0pt{\kern0.3\wd0\vrule height0.9\ht0\hss}\box0}}
{\setbox0=\hbox{$\scriptstyle\rm T$}\hbox{\hbox
to0pt{\kern0.3\wd0\vrule height0.9\ht0\hss}\box0}}
{\setbox0=\hbox{$\scriptscriptstyle\rm T$}\hbox{\hbox
to0pt{\kern0.3\wd0\vrule height0.9\ht0\hss}\box0}}}}
\def\bbbs{{\mathchoice
{\setbox0=\hbox{$\displaystyle     \rm S$}\hbox{\raise0.5\ht0\hbox
to0pt{\kern0.35\wd0\vrule height0.45\ht0\hss}\hbox
to0pt{\kern0.55\wd0\vrule height0.5\ht0\hss}\box0}}
{\setbox0=\hbox{$\textstyle        \rm S$}\hbox{\raise0.5\ht0\hbox
to0pt{\kern0.35\wd0\vrule height0.45\ht0\hss}\hbox
to0pt{\kern0.55\wd0\vrule height0.5\ht0\hss}\box0}}
{\setbox0=\hbox{$\scriptstyle      \rm S$}\hbox{\raise0.5\ht0\hbox
to0pt{\kern0.35\wd0\vrule height0.45\ht0\hss}\raise0.05\ht0\hbox
to0pt{\kern0.5\wd0\vrule height0.45\ht0\hss}\box0}}
{\setbox0=\hbox{$\scriptscriptstyle\rm S$}\hbox{\raise0.5\ht0\hbox
to0pt{\kern0.4\wd0\vrule height0.45\ht0\hss}\raise0.05\ht0\hbox
to0pt{\kern0.55\wd0\vrule height0.45\ht0\hss}\box0}}}}
\def\bbbz{{\mathchoice {\hbox{$\sf\textstyle Z\kern-0.4em Z$}}
{\hbox{$\sf\textstyle Z\kern-0.4em Z$}}
{\hbox{$\sf\scriptstyle Z\kern-0.3em Z$}}
{\hbox{$\sf\scriptscriptstyle Z\kern-0.2em Z$}}}}


\def\Nulle{0} 
\def\Afe{1}   
\def\Hae{2}   
\def\Hbe{3}   
\def\Hce{4}   
\def\Hde{5}   


\newcount\LastMac       \LastMac=\Nulle

\newskip\half      \half=5.5pt plus 1.5pt minus 2.25pt
\newskip\one       \one=11pt plus 3pt minus 5.5pt
\newskip\onehalf   \onehalf=16.5pt plus 5.5pt minus 8.25pt
\newskip\two       \two=22pt plus 5.5pt minus 11pt

\def\Half{\addvspace{\half}}
\def\One{\addvspace{\one}}
\def\OneHalf{\addvspace{\onehalf}}
\def\Two{\addvspace{\two}}

\def\Raggedright{
  \rightskip=\z@ plus \hsize\relax
}

\def\Fullout{
  \rightskip=\z@\relax
}

\def\Hang#1#2{
  \hangindent=#1%
  \hangafter=#2\relax
}


\newif\ifsp@page
\def\pagestyle#1{\csname ps@#1\endcsname}
\def\thispagestyle#1{\global\sp@pagetrue\gdef\sp@type{#1}}

\def\ps@titlepage{%
  \def\@oddhead{\eightpoint\noindent \the\CatchLine
    \ifprod@font\else\qquad Printed\ \today\qquad
      (MN plain \TeX\ macros\ v\@version)\fi \hfil}%
  \let\@evenhead=\@oddhead
  \def\@oddfoot{\eightpoint\copyright\ \@pubyear\ RAS\hfil}%
  \def\@evenfoot{\hfil\eightpoint\noindent\copyright\ \@pubyear\ RAS}%
}

\def\ps@headings{%
  \def\@oddhead{\elevenpoint\it\noindent
    \hfill\the\RightHeader\hskip1.5em\rm\folio}%
  \def\@evenhead{\elevenpoint\noindent
    \folio\hskip1.5em\it\the\LeftHeader\hfill}%
  \def\@oddfoot{\eightpoint\noindent\copyright\ \@pubyear\ RAS,
    MNRAS {\bf \@volume}, \@pagerange\hfil}%
  \def\@evenfoot{\hfil\eightpoint\copyright\ \@pubyear\ RAS,
    MNRAS {\bf \@volume}, \@pagerange}%
}

\def\ps@plate{%
  \def\@oddhead{\eightpoint\noindent\plt@cap\hfil}%
  \def\@evenhead{\eightpoint\noindent\plt@cap\hfil}%
  \def\@oddfoot{\eightpoint\noindent\copyright\ \@pubyear\ RAS,
    MNRAS {\bf \@volume}, \@pagerange\hfil}%
  \def\@evenfoot{\hfil\eightpoint\copyright\ \@pubyear\ RAS,
    MNRAS {\bf \@volume}, \@pagerange}%
}



\def\title#1{
  \bgroup
    \vbox to 8pt{\vss}%
    \seventeenpoint
    \Raggedright
    \noindent \strut{\bf #1}\par
  \egroup
}

\def\author#1{
  \bgroup
    \ifnum\LastMac=\Afe \OneHalf\else \vskip 21pt\fi
    \fourteenpoint
    \Raggedright
    \noindent \strut #1\par
    \vskip 3pt%
  \egroup
}

\def\affiliation#1{
  \bgroup
    \vskip -4pt%
    \eightpoint
    \Raggedright
    \noindent \strut {\it #1}\par
  \egroup
  \LastMac=\Afe\relax
}

\def\acceptedline#1{
  \bgroup
    \Two
    \eightpoint
    \Raggedright
    \noindent \strut #1\par
  \egroup
}

\long\def\abstract#1{%
  \bgroup
    \vskip 20pt%
    \leftskip 11pc\rightskip\z@
    \noindent{\ninebf ABSTRACT}\par
    \tenpoint
    \Fullout
    \noindent #1\par
  \egroup
}

\long\def\keywords#1{
  \bgroup
    \Half
    \leftskip 11pc\rightskip\z@
    \tenpoint
    \Fullout
    \noindent\hbox{\bf Key words:}\ #1\par
  \egroup
}


\def\maketitle{%
  \EndOpening
  \ifsinglecol \else \MakePage\fi
}



\def\@nameuse#1{\csname #1\endcsname}
\def\arabic#1{\@arabic{\@nameuse{#1}}}
\def\alph#1{\@alph{\@nameuse{#1}}}
\def\Alph#1{\@Alph{\@nameuse{#1}}}
\def\@arabic#1{\number #1}
\def\@Alph#1{\ifcase#1\or A\or B\or C\or D\else\@Ialph{#1}\fi}
\def\@Ialph#1{\ifcase#1\or \or \or \or \or E\or F\or G\or H\or I\or J\or
   K\or L\or M\or N\or O\or P\or Q\or R\or S\or T\or U\or V\or W\or X\or
   Y\or Z\else\errmessage{Counter out of range}\fi}
\def\@alph#1{\ifcase#1\or a\or b\or c\or d\else\@ialph{#1}\fi}
\def\@ialph#1{\ifcase#1\or \or \or \or \or e\or f\or g\or h\or i\or j\or
   k\or l\or m\or n\or o\or p\or q\or r\or s\or t\or u\or v\or w\or x\or y\or
   z\else\errmessage{Counter out of range}\fi}


\newcount\Eqnno
\newcount\SubEqnno

\def\theeq{\arabic{Eqnno}}
\def\thesubeq{\alph{SubEqnno}}

\def\stepeq{\relax
  \global\SubEqnno \z@
  \global\advance\Eqnno \@ne\relax
  {\rm (\theeq)}%
}

\def\startsubeq{\relax
  \global\SubEqnno \z@
  \global\advance\Eqnno \@ne\relax
  \stepsubeq
}

\def\stepsubeq{\relax
  \global\advance\SubEqnno \@ne\relax
  {\rm (\theeq\thesubeq)}%
}


\newcount\Sec        
\newcount\SecSec
\newcount\SecSecSec

\def\thesection{\arabic{Sec}}
\def\thesubsection{\thesection.\arabic{SecSec}}
\def\thesubsubsection{\thesubsection.\arabic{SecSecSec}}

\Sec=\z@

\def\:{\let\@sptoken= } \:  
\def\:{\@xifnch} \expandafter\def\: {\futurelet\@tempc\@ifnch}

\def\@ifnextchar#1#2#3{%
  \let\@tempMACe #1%
  \def\@tempMACa{#2}%
  \def\@tempMACb{#3}%
  \futurelet \@tempMACc\@ifnch%
}

\def\@ifnch{%
\ifx \@tempMACc \@sptoken%
  \let\@tempMACd\@xifnch%
\else%
  \ifx \@tempMACc \@tempMACe%
    \let\@tempMACd\@tempMACa%
  \else%
    \let\@tempMACd\@tempMACb%
  \fi%
\fi%
\@tempMACd%
}

\def\@ifstar#1#2{\@ifnextchar *{\def\@tempMACa*{#1}\@tempMACa}{#2}}

\newskip\@tempskipb

\def\addvspace#1{%
  \ifvmode\else \endgraf\fi%
  \ifdim\lastskip=\z@%
    \vskip #1\relax%
  \else%
    \@tempskipb#1\relax\@xaddvskip%
  \fi%
}

\def\@xaddvskip{%
  \ifdim\lastskip<\@tempskipb%
    \vskip-\lastskip%
    \vskip\@tempskipb\relax%
  \else%
    \ifdim\@tempskipb<\z@%
      \ifdim\lastskip<\z@ \else%
        \advance\@tempskipb\lastskip%
        \vskip-\lastskip\vskip\@tempskipb%
      \fi%
    \fi%
  \fi%
}

\newskip\@tmpSKIP

\def\addpen#1{%
  \ifvmode
    \if@nobreak
    \else
      \ifdim\lastskip=\z@
        \penalty#1\relax
      \else
        \@tmpSKIP=\lastskip
        \vskip -\lastskip
        \penalty#1\vskip\@tmpSKIP
      \fi
    \fi
  \fi
}

\newcount\@clubpen   \@clubpen=\clubpenalty
\newif\if@nobreak    \@nobreakfalse

\def\@noafterindent{%
  \global\@nobreaktrue
  \everypar{\if@nobreak
              \global\@nobreakfalse
              \clubpenalty \@M
              {\setbox\z@\lastbox}%
              \LastMac=\Nulle\relax%
            \else
              \clubpenalty \@clubpen
              \everypar{}%
            \fi}%
}

\newcount\gds@cbrk   \gds@cbrk=-300

\def\@nohdbrk{\interlinepenalty \@M\relax}

\let\@par=\par
\def\@restorepar{\def\par{\@par}}

\newif\if@endpe   \@endpefalse
 
\def\@doendpe{\@endpetrue \@nobreakfalse \LastMac=\Nulle\relax%
     \def\par{\@restorepar\everypar{}\par\@endpefalse}%
              \everypar{\setbox\z@\lastbox\everypar{}\@endpefalse}%
}

\def\section{\@ifstar{\@ssection}{\@section}}

\def\@section#1{
  \if@nobreak
    \everypar{}%
    \ifnum\LastMac=\Hae \addvspace{\half}\fi
  \else
    \addpen{\gds@cbrk}%
    \addvspace{\two}%
  \fi
  \bgroup
    \ninepoint\bf
    \Raggedright
    \global\advance\Sec \@ne
    \ifappendix
      \global\Eqnno=\z@ \global\SubEqnno=\z@\relax
      \def\ch@ck{#1}%
      \ifx\ch@ck\empty \def\c@lon{}\else\def\c@lon{:}\fi
      \noindent\@nohdbrk APPENDIX\ \thesection\c@lon\hskip 0.5em%
        \uppercase{#1}\par
    \else
      \noindent\@nohdbrk\thesection\hskip 1pc \uppercase{#1}\par
    \fi
    \global\SecSec=\z@
  \egroup
  \nobreak
  \vskip\half
  \nobreak
  \@noafterindent
  \LastMac=\Hae\relax
}

\def\@ssection#1{
  \if@nobreak
    \everypar{}%
    \ifnum\LastMac=\Hae \addvspace{\half}\fi
  \else
    \addpen{\gds@cbrk}%
    \addvspace{\two}%
  \fi
  \bgroup
    \ninepoint\bf
    \Raggedright
    \noindent\@nohdbrk\uppercase{#1}\par
  \egroup
  \nobreak
  \vskip\half
  \nobreak
  \@noafterindent
  \LastMac=\Hae\relax
}

\def\subsection{\@ifstar{\@ssubsection}{\@subsection}}

\def\@subsection#1{
  \if@nobreak
    \everypar{}%
    \ifnum\LastMac=\Hae \addvspace{1pt plus 1pt minus .5pt}\fi
  \else
    \addpen{\gds@cbrk}%
    \addvspace{\onehalf}%
  \fi
  \bgroup
    \ninepoint\bf
    \Raggedright
    \global\advance\SecSec \@ne
    \noindent\@nohdbrk\thesubsection \hskip 1pc\relax #1\par
    \global\SecSecSec=\z@
  \egroup
  \nobreak
  \vskip\half
  \nobreak
  \@noafterindent
  \LastMac=\Hbe\relax
}

\def\@ssubsection#1{
  \if@nobreak
    \everypar{}%
    \ifnum\LastMac=\Hae \addvspace{1pt plus 1pt minus .5pt}\fi
  \else
    \addpen{\gds@cbrk}%
    \addvspace{\onehalf}%
  \fi
  \bgroup
    \ninepoint\bf
    \Raggedright
    \noindent\@nohdbrk #1\par
  \egroup
  \nobreak
  \vskip\half
  \nobreak
  \@noafterindent
  \LastMac=\Hbe\relax
}

\def\subsubsection{\@ifstar{\@ssubsubsection}{\@subsubsection}}

\def\@subsubsection#1{
  \if@nobreak
    \everypar{}%
    \ifnum\LastMac=\Hbe \addvspace{1pt plus 1pt minus .5pt}\fi
  \else
    \addpen{\gds@cbrk}%
    \addvspace{\onehalf}%
  \fi
  \bgroup
    \ninepoint\it
    \Raggedright
    \global\advance\SecSecSec \@ne
    \noindent\@nohdbrk\thesubsubsection \hskip 1pc\relax #1\par
  \egroup
  \nobreak
  \vskip\half
  \nobreak
  \@noafterindent
  \LastMac=\Hce\relax
}

\def\@ssubsubsection#1{
  \if@nobreak
    \everypar{}%
    \ifnum\LastMac=\Hbe \addvspace{1pt plus 1pt minus .5pt}\fi
  \else
    \addpen{\gds@cbrk}%
    \addvspace{\onehalf}%
  \fi
  \bgroup
    \ninepoint\it
    \Raggedright
    \noindent\@nohdbrk #1\par
  \egroup
  \nobreak
  \vskip\half
  \nobreak
  \@noafterindent
  \LastMac=\Hce\relax
}

\def\paragraph#1{
  \if@nobreak
    \everypar{}%
  \else
    \addpen{\gds@cbrk}%
    \addvspace{\one}%
  \fi%
  \bgroup%
    \ninepoint\it
    \noindent #1\ \nobreak%
  \egroup
  \LastMac=\Hde\relax
  \ignorespaces
}


\newif\ifappendix

\def\appendix{%
  \global\appendixtrue
  \def\thesection{\Alph{Sec}}%
  \def\thesubsection{\thesection\arabic{SecSec}}%
  \def\theeq{\thesection\arabic{Eqnno}}%
  \Sec=\z@ \SecSec=\z@ \SecSecSec=\z@ \Eqnno=\z@ \SubEqnno=\z@\relax
}




\def\beginlist{%
  \par\if@nobreak \else\addvspace{\half}\fi%
  \bgroup%
    \ninepoint
    \let\item=\list@item%
}

\def\list@item{%
  \par\noindent\hskip 1em\relax%
  \ignorespaces%
}

\def\endlist{\par\egroup\addvspace{\half}\@doendpe}


\def\beginrefs{%
  \par
  \bgroup
    \eightpoint
    \Fullout
    \let\bibitem=\bib@item
}

\def\bib@item{%
  \par\parindent=1.5em\Hang{1.5em}{1}%
  \everypar={\Hang{1.5em}{1}\ignorespaces}%
  \noindent\ignorespaces
}

\def\endrefs{\par\egroup\@doendpe}


\newtoks\CatchLine

\def\@journal{Mon.\ Not.\ R.\ Astron.\ Soc.\ }  
\def\@pubyear{1994}        
\def\@pagerange{000--000}  
\def\@volume{000}          
\def\@microfiche{}         %

\def\pubyear#1{\gdef\@pubyear{#1}\@makecatchline}
\def\pagerange#1{\gdef\@pagerange{#1}\@makecatchline}
\def\volume#1{\gdef\@volume{#1}\@makecatchline}
\def\microfiche#1{\gdef\@microfiche{and Microfiche\ #1}\@makecatchline}

\def\@makecatchline{%
  \global\CatchLine{%
    {\rm \@journal {\bf \@volume},\ \@pagerange\ (\@pubyear)\ \@microfiche}}%
}

\@makecatchline 

\newtoks\LeftHeader
\def\shortauthor#1{
  \global\LeftHeader{#1}%
}

\newtoks\RightHeader
\def\shorttitle#1{
  \global\RightHeader{#1}%
}

\def\PageHead{
  \begingroup
    \ifsp@page
      \csname ps@\sp@type\endcsname
    \fi
    \ifodd\pageno
      \let\the@head=\@oddhead
    \else
      \let\the@head=\@evenhead
    \fi
    \vbox to \z@{\vskip-22.5\p@%
      \hbox to \PageWidth{\vbox to8.5\p@{}%
        \the@head
      }%
    \vss}%
  \endgroup
  \nointerlineskip
}

\gdef\PageFoot{%
  \nointerlineskip%
  \begingroup
  \ifsp@page
    \csname ps@\sp@type\endcsname
    \global\sp@pagefalse
  \fi
  \vbox to 22pt{\vfil%
    \hbox to \PageWidth{%
      \eightpoint\strut\noindent
      \ifodd\pageno
        \@oddfoot
      \else
        \@evenfoot
      \fi
    }%
  }%
  \endgroup
}

\def\today{%
  \number\day\space
  \ifcase\month\or January\or February\or March\or April\or May\or June\or
    July\or August\or September\or October\or November\or December\fi
  \space\number\year%
}

\def\authorcomment#1{%
  \gdef\PageFoot{%
    \nointerlineskip%
    \vbox to 20pt{\vfil%
      \hbox to \PageWidth{\elevenpoint\noindent \hfil #1 \hfil}}%
  }%
}


\newif\ifplate@page
\newbox\plt@box

\def\beginplatepage{%
  \let\plate=\plate@head
  \let\caption=\fig@caption
  \global\setbox\plt@box=\vbox\bgroup
  \TEMPDIMEN=\PageWidth 
  \hsize=\PageWidth\relax
}

\def\endplatepage{\par\egroup\global\plate@pagetrue}
\def\plate@head#1{\gdef\plt@cap{#1}}


\def\letters{%
  \gdef\folio{\ifnum\pageno<\z@ L\romannumeral-\pageno
    \else L\number\pageno \fi}%
}


\newdimen\mathindent

\global\mathindent=\z@
\global\everydisplay{\global\@dspwd=\displaywidth\displaysetup}


\def\@displaylines#1{
  {}$\displ@y\hbox{\vbox{\halign{$\@lign\hfil\displaystyle##\hfil$\crcr
  #1\crcr}}}${}%
}

\def\@eqalign#1{\null\vcenter{\openup\jot\m@th
  \ialign{\strut\hfil$\displaystyle{##}$&$\displaystyle{{}##}$\hfil
      \crcr#1\crcr}}%
}

\def\@eqalignno#1{
  \global\advance\@dspwd by -\mathindent%
  {}$\displ@y\hbox{\vbox{\halign to\@dspwd%
  {\hfil$\@lign\displaystyle{##}$\tabskip\z@skip
  &$\@lign\displaystyle{{}##}$\hfil\tabskip\centering
  &\llap{$\@lign##$}\tabskip\z@skip\crcr
  #1\crcr}}}${}%
}


\global\let\displaylines=\@displaylines
\global\let\eqalign=\@eqalign
\global\let\eqalignno=\@eqalignno
\global\let\leqalignno=\@eqalignno

\newdimen\@dspwd   \@dspwd=\z@
\newif\if@eqno
\newif\if@leqno
\newtoks\@eqn
\newtoks\@eq

\def\displaysetup#1$${\displaytest#1\eqno\eqno\displaytest}

\def\displaytest#1\eqno#2\eqno#3\displaytest{%
 \if!#3!\ldisplaytest#1\leqno\leqno\ldisplaytest
 \else\@eqnotrue\@leqnofalse\@eqn={#2}\@eq={#1}\fi
 \generaldisplay$$}

\def\ldisplaytest#1\leqno#2\leqno#3\ldisplaytest{%
\@eq={#1}%
 \if!#3!\@eqnofalse\else\@eqnotrue\@leqnotrue
  \@eqn={#2}\fi}

\def\generaldisplay{%
  \if@eqno
    \if@leqno
      \hbox to \displaywidth{\noindent
        \rlap{$\displaystyle\the\@eqn$}%
        \hskip\mathindent$\displaystyle\the\@eq$\hfil}%
    \else
      \hbox to \displaywidth{\noindent
        \hskip\mathindent
        $\displaystyle\the\@eq$\hfil$\displaystyle\the\@eqn$}%
    \fi
  \else
    \hbox to \displaywidth{\noindent
      \hskip\mathindent$\displaystyle\the\@eq$\hfil}%
  \fi
}


\def\@notice{%
  \par\Two%
  \noindent{\b@ls{11pt}\ninerm This paper has been produced using the
    Royal Astronomical Society/Blackwell Science \TeX\ macros.\par}%
}

\outer\def\bye{\@notice\par\vfill\supereject\end}


\def\start@mess{%
  Monthly notices of the RAS journal style (\@typeface)\space
    v\@version,\space \@verdate.%
}

\everyjob{\Warn{\start@mess}}



\newif\if@debug \@debugfalse  

\def\Print#1{\if@debug\immediate\write16{#1}\else \fi}
\def\Warn#1{\immediate\write16{#1}}
\def\wlog#1{}

\newcount\Iteration 

\def\Single{0} \def\Double{1}                 
\def\Figure{0} \def\Table{1}                  

\def\InStack{0}  
\def\InZoneA{1}
\def\InZoneB{2}
\def\InZoneC{3}

\newcount\TEMPCOUNT 
\newdimen\TEMPDIMEN 
\newbox\TEMPBOX     
\newbox\VOIDBOX     

\newcount\LengthOfStack 
\newcount\MaxItems      
\newcount\StackPointer
\newcount\Point         
\newcount\NextFigure    
\newcount\NextTable     
\newcount\NextItem      

\newcount\StatusStack   
\newcount\NumStack      
\newcount\TypeStack     
\newcount\SpanStack     
\newcount\BoxStack      

\newcount\ItemSTATUS    
\newcount\ItemNUMBER    
\newcount\ItemTYPE      
\newcount\ItemSPAN      
\newbox\ItemBOX         
\newdimen\ItemSIZE      

\newdimen\PageHeight    
\newdimen\TextLeading   
\newdimen\Feathering    
\newcount\LinesPerPage  
\newdimen\ColumnWidth   
\newdimen\ColumnGap     
\newdimen\PageWidth     
\newdimen\BodgeHeight   
\newcount\Leading       

\newdimen\ZoneBSize  
\newdimen\TextSize   
\newbox\ZoneABOX     
\newbox\ZoneBBOX     
\newbox\ZoneCBOX     

\newif\ifFirstSingleItem
\newif\ifFirstZoneA
\newif\ifMakePageInComplete
\newif\ifMoreFigures \MoreFiguresfalse 
\newif\ifMoreTables  \MoreTablesfalse  

\newif\ifFigInZoneB 
\newif\ifFigInZoneC 
\newif\ifTabInZoneB 
\newif\ifTabInZoneC

\newif\ifZoneAFullPage

\newbox\MidBOX    
\newbox\LeftBOX
\newbox\RightBOX
\newbox\PageBOX   

\newif\ifLeftCOL  
\LeftCOLtrue

\newdimen\ZoneBAdjust

\newcount\ItemFits
\def\Yes{1}
\def\No{2}


\MaxItems=15
\NextFigure=\z@        
\NextTable=\@ne

\BodgeHeight=6pt
\TextLeading=11pt    
\Leading=11
\Feathering=\z@      
\LinesPerPage=61     
\topskip=\TextLeading
\ColumnWidth=20pc    
\ColumnGap=2pc       

\newskip\ItemSepamount  
\ItemSepamount=\TextLeading plus \TextLeading minus 4pt

\parskip=\z@ plus .1pt
\parindent=18pt
\widowpenalty=\z@
\clubpenalty=10000
\tolerance=1500
\hbadness=1500
\abovedisplayskip=6pt plus 2pt minus 1pt
\belowdisplayskip=6pt plus 2pt minus 1pt
\abovedisplayshortskip=6pt plus 2pt minus 1pt
\belowdisplayshortskip=6pt plus 2pt minus 1pt

\frenchspacing

\ninepoint 

\PageHeight=682pt
\PageWidth=2\ColumnWidth
\advance\PageWidth by \ColumnGap

\pagestyle{headings}




\newcount\DUMMY \StatusStack=\allocationnumber
\newcount\DUMMY \newcount\DUMMY \newcount\DUMMY 
\newcount\DUMMY \newcount\DUMMY \newcount\DUMMY 
\newcount\DUMMY \newcount\DUMMY \newcount\DUMMY
\newcount\DUMMY \newcount\DUMMY \newcount\DUMMY 
\newcount\DUMMY \newcount\DUMMY \newcount\DUMMY

\newcount\DUMMY \NumStack=\allocationnumber
\newcount\DUMMY \newcount\DUMMY \newcount\DUMMY 
\newcount\DUMMY \newcount\DUMMY \newcount\DUMMY 
\newcount\DUMMY \newcount\DUMMY \newcount\DUMMY 
\newcount\DUMMY \newcount\DUMMY \newcount\DUMMY 
\newcount\DUMMY \newcount\DUMMY \newcount\DUMMY

\newcount\DUMMY \TypeStack=\allocationnumber
\newcount\DUMMY \newcount\DUMMY \newcount\DUMMY 
\newcount\DUMMY \newcount\DUMMY \newcount\DUMMY 
\newcount\DUMMY \newcount\DUMMY \newcount\DUMMY 
\newcount\DUMMY \newcount\DUMMY \newcount\DUMMY 
\newcount\DUMMY \newcount\DUMMY \newcount\DUMMY

\newcount\DUMMY \SpanStack=\allocationnumber
\newcount\DUMMY \newcount\DUMMY \newcount\DUMMY 
\newcount\DUMMY \newcount\DUMMY \newcount\DUMMY 
\newcount\DUMMY \newcount\DUMMY \newcount\DUMMY 
\newcount\DUMMY \newcount\DUMMY \newcount\DUMMY 
\newcount\DUMMY \newcount\DUMMY \newcount\DUMMY

\newbox\DUMMY   \BoxStack=\allocationnumber
\newbox\DUMMY   \newbox\DUMMY \newbox\DUMMY 
\newbox\DUMMY   \newbox\DUMMY \newbox\DUMMY 
\newbox\DUMMY   \newbox\DUMMY \newbox\DUMMY 
\newbox\DUMMY   \newbox\DUMMY \newbox\DUMMY 
\newbox\DUMMY   \newbox\DUMMY \newbox\DUMMY

\def\wlog{\immediate\write\m@ne}


\def\GetItemAll#1{%
 \GetItemSTATUS{#1}
 \GetItemNUMBER{#1}
 \GetItemTYPE{#1}
 \GetItemSPAN{#1}
 \GetItemBOX{#1}
}

\def\GetItemSTATUS#1{%
 \Point=\StatusStack
 \advance\Point by #1
 \global\ItemSTATUS=\count\Point
}

\def\GetItemNUMBER#1{%
 \Point=\NumStack
 \advance\Point by #1
 \global\ItemNUMBER=\count\Point
}

\def\GetItemTYPE#1{%
 \Point=\TypeStack
 \advance\Point by #1
 \global\ItemTYPE=\count\Point
}

\def\GetItemSPAN#1{%
 \Point\SpanStack
 \advance\Point by #1
 \global\ItemSPAN=\count\Point
}

\def\GetItemBOX#1{%
 \Point=\BoxStack
 \advance\Point by #1
 \global\setbox\ItemBOX=\vbox{\copy\Point}
 \global\ItemSIZE=\ht\ItemBOX
 \global\advance\ItemSIZE by \dp\ItemBOX
 \TEMPCOUNT=\ItemSIZE
 \divide\TEMPCOUNT by \Leading
 \divide\TEMPCOUNT by 65536
 \advance\TEMPCOUNT \@ne
 \ItemSIZE=\TEMPCOUNT pt
 \global\multiply\ItemSIZE by \Leading
}


\def\JoinStack{%
 \ifnum\LengthOfStack=\MaxItems 
  \Warn{WARNING: Stack is full...some items will be lost!}
 \else
  \Point=\StatusStack
  \advance\Point by \LengthOfStack
  \global\count\Point=\ItemSTATUS
  \Point=\NumStack
  \advance\Point by \LengthOfStack
  \global\count\Point=\ItemNUMBER
  \Point=\TypeStack
  \advance\Point by \LengthOfStack
  \global\count\Point=\ItemTYPE
  \Point\SpanStack
  \advance\Point by \LengthOfStack
  \global\count\Point=\ItemSPAN
  \Point=\BoxStack
  \advance\Point by \LengthOfStack
  \global\setbox\Point=\vbox{\copy\ItemBOX}
  \global\advance\LengthOfStack \@ne
  \ifnum\ItemTYPE=\Figure 
   \global\MoreFigurestrue
  \else
   \global\MoreTablestrue
  \fi
 \fi
}


\def\LeaveStack#1{%
 {\Iteration=#1
 \loop
 \ifnum\Iteration<\LengthOfStack
  \advance\Iteration \@ne
  \GetItemSTATUS{\Iteration}
   \advance\Point by \m@ne
   \global\count\Point=\ItemSTATUS
  \GetItemNUMBER{\Iteration}
   \advance\Point by \m@ne
   \global\count\Point=\ItemNUMBER
  \GetItemTYPE{\Iteration}
   \advance\Point by \m@ne
   \global\count\Point=\ItemTYPE
  \GetItemSPAN{\Iteration}
   \advance\Point by \m@ne
   \global\count\Point=\ItemSPAN
  \GetItemBOX{\Iteration}
   \advance\Point by \m@ne
   \global\setbox\Point=\vbox{\copy\ItemBOX}
 \repeat}
 \global\advance\LengthOfStack by \m@ne
}


\newif\ifStackNotClean

\def\CleanStack{%
 \StackNotCleantrue
 {\Iteration=\z@
  \loop
   \ifStackNotClean
    \GetItemSTATUS{\Iteration}
    \ifnum\ItemSTATUS=\InStack
     \advance\Iteration \@ne
     \else
      \LeaveStack{\Iteration}
    \fi
   \ifnum\LengthOfStack<\Iteration
    \StackNotCleanfalse
   \fi
 \repeat}
}


\def\FindItem#1#2{%
 \global\StackPointer=\m@ne 
 {\Iteration=\z@
  \loop
  \ifnum\Iteration<\LengthOfStack
   \GetItemSTATUS{\Iteration}
   \ifnum\ItemSTATUS=\InStack
    \GetItemTYPE{\Iteration}
    \ifnum\ItemTYPE=#1
     \GetItemNUMBER{\Iteration}
     \ifnum\ItemNUMBER=#2
      \global\StackPointer=\Iteration
      \Iteration=\LengthOfStack 
     \fi
    \fi
   \fi
  \advance\Iteration \@ne
 \repeat}
}


\def\FindNext{%
 \global\StackPointer=\m@ne 
 {\Iteration=\z@
  \loop
  \ifnum\Iteration<\LengthOfStack
   \GetItemSTATUS{\Iteration}
   \ifnum\ItemSTATUS=\InStack
    \GetItemTYPE{\Iteration}
   \ifnum\ItemTYPE=\Figure
    \ifMoreFigures
      \global\NextItem=\Figure
      \global\StackPointer=\Iteration
      \Iteration=\LengthOfStack 
    \fi
   \fi
   \ifnum\ItemTYPE=\Table
    \ifMoreTables
      \global\NextItem=\Table
      \global\StackPointer=\Iteration
      \Iteration=\LengthOfStack 
    \fi
   \fi
  \fi
  \advance\Iteration \@ne
 \repeat}
}


\def\ChangeStatus#1#2{%
 \Point=\StatusStack
 \advance\Point by #1
 \global\count\Point=#2
}



\def\Zone{\InZoneA}

\ZoneBAdjust=\z@

\def\MakePage{
 \global\ZoneBSize=\PageHeight
 \global\TextSize=\ZoneBSize
 \global\ZoneAFullPagefalse
 \global\topskip=\TextLeading
 \MakePageInCompletetrue
 \MoreFigurestrue
 \MoreTablestrue
 \FigInZoneBfalse
 \FigInZoneCfalse
 \TabInZoneBfalse
 \TabInZoneCfalse
 \global\FirstSingleItemtrue
 \global\FirstZoneAtrue
 \global\setbox\ZoneABOX=\box\VOIDBOX
 \global\setbox\ZoneBBOX=\box\VOIDBOX
 \global\setbox\ZoneCBOX=\box\VOIDBOX
 \loop
  \ifMakePageInComplete
 \FindNext
 \ifnum\StackPointer=\m@ne
  \NextItem=\m@ne
  \MoreFiguresfalse
  \MoreTablesfalse
 \fi
 \ifnum\NextItem=\Figure
   \FindItem{\Figure}{\NextFigure}
   \ifnum\StackPointer=\m@ne \global\MoreFiguresfalse
   \else
    \GetItemSPAN{\StackPointer}
    \ifnum\ItemSPAN=\Single \def\Zone{\InZoneB}\relax
     \ifFigInZoneC \global\MoreFiguresfalse\fi
    \else
     \def\Zone{\InZoneA}
     \ifFigInZoneB \def\Zone{\InZoneC}\fi
    \fi
   \fi
   \ifMoreFigures\Print{}\FigureItems\fi
 \fi
\ifnum\NextItem=\Table
   \FindItem{\Table}{\NextTable}
   \ifnum\StackPointer=\m@ne \global\MoreTablesfalse
   \else
    \GetItemSPAN{\StackPointer}
    \ifnum\ItemSPAN=\Single\relax
     \ifTabInZoneC \global\MoreTablesfalse\fi
    \else
     \def\Zone{\InZoneA}
     \ifTabInZoneB \def\Zone{\InZoneC}\fi
    \fi
   \fi
   \ifMoreTables\Print{}\TableItems\fi
 \fi
   \MakePageInCompletefalse 
   \ifMoreFigures\MakePageInCompletetrue\fi
   \ifMoreTables\MakePageInCompletetrue\fi
 \repeat
 \ifZoneAFullPage
  \global\TextSize=\z@
  \global\ZoneBSize=\z@
  \global\vsize=\z@\relax
  \global\topskip=\z@\relax
  \vbox to \z@{\vss}
  \eject
 \else
 \global\advance\ZoneBSize by -\ZoneBAdjust
 \global\vsize=\ZoneBSize
 \global\hsize=\ColumnWidth
 \global\ZoneBAdjust=\z@
 \ifdim\TextSize<23pt
 \Warn{}
 \Warn{* Making column fall short: TextSize=\the\TextSize *}
 \vskip-\lastskip\eject\fi
 \fi
}

\def\MakeRightCol{
 \global\TextSize=\ZoneBSize
 \MakePageInCompletetrue
 \MoreFigurestrue
 \MoreTablestrue
 \global\FirstSingleItemtrue
 \global\setbox\ZoneBBOX=\box\VOIDBOX
 \def\Zone{\InZoneB}
 \loop
  \ifMakePageInComplete
 \FindNext
 \ifnum\StackPointer=\m@ne
  \NextItem=\m@ne
  \MoreFiguresfalse
  \MoreTablesfalse
 \fi
 \ifnum\NextItem=\Figure
   \FindItem{\Figure}{\NextFigure}
   \ifnum\StackPointer=\m@ne \MoreFiguresfalse
   \else
    \GetItemSPAN{\StackPointer}
    \ifnum\ItemSPAN=\Double\relax
     \MoreFiguresfalse\fi
   \fi
   \ifMoreFigures\Print{}\FigureItems\fi
 \fi
 \ifnum\NextItem=\Table
   \FindItem{\Table}{\NextTable}
   \ifnum\StackPointer=\m@ne \MoreTablesfalse
   \else
    \GetItemSPAN{\StackPointer}
    \ifnum\ItemSPAN=\Double\relax
     \MoreTablesfalse\fi
   \fi
   \ifMoreTables\Print{}\TableItems\fi
 \fi
   \MakePageInCompletefalse 
   \ifMoreFigures\MakePageInCompletetrue\fi
   \ifMoreTables\MakePageInCompletetrue\fi
 \repeat
 \ifZoneAFullPage
  \global\TextSize=\z@
  \global\ZoneBSize=\z@
  \global\vsize=\z@\relax
  \global\topskip=\z@\relax
  \vbox to \z@{\vss}
  \eject
 \else
 \global\vsize=\ZoneBSize
 \global\hsize=\ColumnWidth
 \ifdim\TextSize<23pt
 \Warn{}
 \Warn{* Making column fall short: TextSize=\the\TextSize *}
 \vskip-\lastskip\eject\fi
\fi
}

\def\FigureItems{
 \Print{Considering...}
 \ShowItem{\StackPointer}
 \GetItemBOX{\StackPointer} 
 \GetItemSPAN{\StackPointer}
  \CheckFitInZone 
  \ifnum\ItemFits=\Yes
   \ifnum\ItemSPAN=\Single
     \ChangeStatus{\StackPointer}{\InZoneB} 
     \global\FigInZoneBtrue
     \ifFirstSingleItem
      \hbox{}\vskip-\BodgeHeight
     \global\advance\ItemSIZE by \TextLeading
     \fi
     \unvbox\ItemBOX\ItemSep
     \global\FirstSingleItemfalse
     \global\advance\TextSize by -\ItemSIZE
     \global\advance\TextSize by -\TextLeading
   \else
    \ifFirstZoneA
     \global\advance\ItemSIZE by \TextLeading
     \global\FirstZoneAfalse\fi
    \global\advance\TextSize by -\ItemSIZE
    \global\advance\TextSize by -\TextLeading
    \global\advance\ZoneBSize by -\ItemSIZE
    \global\advance\ZoneBSize by -\TextLeading
    \ifFigInZoneB\relax
     \else
     \ifdim\TextSize<3\TextLeading
     \global\ZoneAFullPagetrue
     \fi
    \fi
    \ChangeStatus{\StackPointer}{\Zone}
    \ifnum\Zone=\InZoneC \global\FigInZoneCtrue\fi
  \fi
   \Print{TextSize=\the\TextSize}
   \Print{ZoneBSize=\the\ZoneBSize}
  \global\advance\NextFigure \@ne
   \Print{This figure has been placed.}
  \else
   \Print{No space available for this figure...holding over.}
   \Print{}
   \global\MoreFiguresfalse
  \fi
}

\def\TableItems{
 \Print{Considering...}
 \ShowItem{\StackPointer}
 \GetItemBOX{\StackPointer} 
 \GetItemSPAN{\StackPointer}
  \CheckFitInZone 
  \ifnum\ItemFits=\Yes
   \ifnum\ItemSPAN=\Single
    \ChangeStatus{\StackPointer}{\InZoneB}
     \global\TabInZoneBtrue
     \ifFirstSingleItem
      \hbox{}\vskip-\BodgeHeight
     \global\advance\ItemSIZE by \TextLeading
     \fi
     \unvbox\ItemBOX\ItemSep
     \global\FirstSingleItemfalse
     \global\advance\TextSize by -\ItemSIZE
     \global\advance\TextSize by -\TextLeading
   \else
    \ifFirstZoneA
    \global\advance\ItemSIZE by \TextLeading
    \global\FirstZoneAfalse\fi
    \global\advance\TextSize by -\ItemSIZE
    \global\advance\TextSize by -\TextLeading
    \global\advance\ZoneBSize by -\ItemSIZE
    \global\advance\ZoneBSize by -\TextLeading
    \ifFigInZoneB\relax
     \else
     \ifdim\TextSize<3\TextLeading
     \global\ZoneAFullPagetrue
     \fi
    \fi
    \ChangeStatus{\StackPointer}{\Zone}
    \ifnum\Zone=\InZoneC \global\TabInZoneCtrue\fi
   \fi
  \global\advance\NextTable \@ne
   \Print{This table has been placed.}
  \else
  \Print{No space available for this table...holding over.}
   \Print{}
   \global\MoreTablesfalse
  \fi
}


\def\CheckFitInZone{%
{\advance\TextSize by -\ItemSIZE
 \advance\TextSize by -\TextLeading
 \ifFirstSingleItem
  \advance\TextSize by \TextLeading
 \fi
 \ifnum\Zone=\InZoneA\relax
  \else \advance\TextSize by -\ZoneBAdjust
 \fi
 \ifdim\TextSize<3\TextLeading \global\ItemFits=\No
 \else \global\ItemFits=\Yes\fi}
}

\def\BeginOpening{%
  \ninepoint
  \thispagestyle{titlepage}%
  \global\setbox\ItemBOX=\vbox\bgroup%
    \hsize=\PageWidth%
    \hrule height \z@
    \ifsinglecol\vskip 6pt\fi 
}

\let\begintopmatter=\BeginOpening  

\def\EndOpening{%
  \One
  \egroup
  \ifsinglecol
    \box\ItemBOX%
    \vskip\TextLeading plus 2\TextLeading
    \@noafterindent
  \else
    \ItemNUMBER=\z@%
    \ItemTYPE=\Figure
    \ItemSPAN=\Double
    \ItemSTATUS=\InStack
    \JoinStack
  \fi
}


\newif\if@here  \@herefalse

\def\no@float{\global\@heretrue}
\let\nofloat=\relax 

\def\beginfigure{%
  \@ifstar{\global\@dfloattrue \@bfigure}{\global\@dfloatfalse \@bfigure}%
}

\def\@bfigure#1{%
  \par
  \if@dfloat
    \ItemSPAN=\Double
    \TEMPDIMEN=\PageWidth
  \else
    \ItemSPAN=\Single
    \TEMPDIMEN=\ColumnWidth
  \fi
  \ifsinglecol
    \TEMPDIMEN=\PageWidth
  \else
    \ItemSTATUS=\InStack
    \ItemNUMBER=#1%
    \ItemTYPE=\Figure
  \fi
  \bgroup
    \hsize=\TEMPDIMEN
    \global\setbox\ItemBOX=\vbox\bgroup
      \eightpoint\nostb@ls{10pt}%
      \let\caption=\fig@caption
      \ifsinglecol \let\nofloat=\no@float\fi
}

\def\fig@caption#1{%
  \vskip 5.5pt plus 6pt%
  \bgroup 
    \eightpoint\nostb@ls{10pt}%
    \setbox\TEMPBOX=\hbox{#1}%
    \ifdim\wd\TEMPBOX>\TEMPDIMEN
      \noindent \unhbox\TEMPBOX\par
    \else
      \hbox to \hsize{\hfil\unhbox\TEMPBOX\hfil}%
    \fi
  \egroup
}

\def\endfigure{%
  \par\egroup 
  \egroup
  \ifsinglecol
    \if@here \midinsert\global\@herefalse\else \topinsert\fi
      \unvbox\ItemBOX
    \endinsert
  \else
    \JoinStack
    \Print{Processing source for figure \the\ItemNUMBER}%
  \fi
}


\newbox\tab@cap@box
\def\tab@caption#1{\global\setbox\tab@cap@box=\hbox{#1\par}}

\newtoks\tab@txt@toks
\long\def\tab@txt#1{\global\tab@txt@toks={#1}\global\table@txttrue}

\newif\iftable@txt  \table@txtfalse
\newif\if@dfloat    \@dfloatfalse

\def\begintable{%
  \@ifstar{\global\@dfloattrue \@btable}{\global\@dfloatfalse \@btable}%
}

\def\@btable#1{%
  \par
  \if@dfloat
    \ItemSPAN=\Double
    \TEMPDIMEN=\PageWidth
  \else
    \ItemSPAN=\Single
    \TEMPDIMEN=\ColumnWidth
  \fi
  \ifsinglecol
    \TEMPDIMEN=\PageWidth
  \else
    \ItemSTATUS=\InStack
    \ItemNUMBER=#1%
    \ItemTYPE=\Table
  \fi
  \bgroup
    \eightpoint\nostb@ls{10pt}%
    \global\setbox\ItemBOX=\vbox\bgroup
      \let\caption=\tab@caption
      \let\tabletext=\tab@txt
      \ifsinglecol \let\nofloat=\no@float\fi
}

\def\endtable{%
  \par\egroup 
  \egroup
  \setbox\TEMPBOX=\hbox to \TEMPDIMEN{%
    \eightpoint\nostb@ls{10pt}%
    \hss
    \vbox{%
      \hsize=\wd\ItemBOX
      \ifvoid\tab@cap@box
      \else
        \noindent\unhbox\tab@cap@box
        \vskip 5.5pt plus 6pt%
      \fi
      \box\ItemBOX
      \iftable@txt
        \vskip 10pt%
        \noindent\the\tab@txt@toks
        \global\table@txtfalse
      \fi
    }%
    \hss
  }%
  \ifsinglecol
    \if@here \midinsert\global\@herefalse\else \topinsert\fi
      \box\TEMPBOX
    \endinsert
  \else
    \global\setbox\ItemBOX=\box\TEMPBOX
    \JoinStack
    \Print{Processing source for table \the\ItemNUMBER}%
  \fi
}

\def\UnloadZoneA{%
\FirstZoneAtrue
 \Iteration=\z@
  \loop
   \ifnum\Iteration<\LengthOfStack
    \GetItemSTATUS{\Iteration}
    \ifnum\ItemSTATUS=\InZoneA
     \GetItemBOX{\Iteration}
     \ifFirstZoneA \vbox to \BodgeHeight{\vfil}%
     \FirstZoneAfalse\fi
     \unvbox\ItemBOX\ItemSep
     \LeaveStack{\Iteration}
     \else
     \advance\Iteration \@ne
   \fi
 \repeat
}

\def\UnloadZoneC{%
\Iteration=\z@
  \loop
   \ifnum\Iteration<\LengthOfStack
    \GetItemSTATUS{\Iteration}
    \ifnum\ItemSTATUS=\InZoneC
     \GetItemBOX{\Iteration}
     \ItemSep\unvbox\ItemBOX
     \LeaveStack{\Iteration}
     \else
     \advance\Iteration \@ne
   \fi
 \repeat
}


\def\ShowItem#1{
  {\GetItemAll{#1}
  \Print{\the#1:
  {TYPE=\ifnum\ItemTYPE=\Figure Figure\else Table\fi}
  {NUMBER=\the\ItemNUMBER}
  {SPAN=\ifnum\ItemSPAN=\Single Single\else Double\fi}
  {SIZE=\the\ItemSIZE}}}
}

\def\ShowStack{%
 \Print{}
 \Print{LengthOfStack = \the\LengthOfStack}
 \ifnum\LengthOfStack=\z@ \Print{Stack is empty}\fi
 \Iteration=\z@
 \loop
 \ifnum\Iteration<\LengthOfStack
  \ShowItem{\Iteration}
  \advance\Iteration \@ne
 \repeat
}

\def\B#1#2{%
\hbox{\vrule\kern-0.4pt\vbox to #2{%
\hrule width #1\vfill\hrule}\kern-0.4pt\vrule}
}


\newif\ifsinglecol   \singlecolfalse

\def\onecolumn{%
  \global\output={\singlecoloutput}%
  \global\hsize=\PageWidth
  \global\vsize=\PageHeight
  \global\ColumnWidth=\hsize
  \global\TextLeading=12pt
  \global\Leading=12
  \global\singlecoltrue
  \global\let\onecolumn=\relax
  \global\let\footnote=\sing@footnote
  \global\let\vfootnote=\sing@vfootnote
  \ninepoint 
  \message{(Single column)}%
}

\def\singlecoloutput{%
  \shipout\vbox{\PageHead\vbox to \PageHeight{\pagebody\vss}\PageFoot}%
  \advancepageno
  \ifplate@page
    \shipout\vbox{%
      \sp@pagetrue
      \def\sp@type{plate}%
      \global\plate@pagefalse
      \PageHead\vbox to \PageHeight{\unvbox\plt@box\vfil}\PageFoot%
    }%
    \message{[plate]}%
    \advancepageno
  \fi
  \ifnum\outputpenalty>-\@MM \else\dosupereject\fi%
}

\def\ItemSep{\vskip\ItemSepamount\relax}

\def\ItemSepbreak{\par\ifdim\lastskip<\ItemSepamount
  \removelastskip\penalty-200\ItemSep\fi%
}


\let\@@endinsert=\endinsert 

\def\endinsert{\egroup 
  \if@mid \dimen@\ht\z@ \advance\dimen@\dp\z@ \advance\dimen@12\p@
    \advance\dimen@\pagetotal \advance\dimen@-\pageshrink
    \ifdim\dimen@>\pagegoal\@midfalse\p@gefalse\fi\fi
  \if@mid \ItemSep\box\z@\ItemSepbreak
  \else\insert\topins{\penalty100 
    \splittopskip\z@skip
    \splitmaxdepth\maxdimen \floatingpenalty\z@
    \ifp@ge \dimen@\dp\z@
    \vbox to\vsize{\unvbox\z@\kern-\dimen@}
    \else \box\z@\nobreak\ItemSep\fi}\fi\endgroup%
}


\def\gobbleone#1{}
\def\gobbletwo#1#2{}
\let\footnote=\gobbletwo 
\let\vfootnote=\gobbleone

\def\sing@footnote#1{\let\@sf\empty 
  \ifhmode\edef\@sf{\spacefactor\the\spacefactor}\/\fi
  \hbox{$^{\hbox{\eightpoint #1}}$}\@sf\sing@vfootnote{#1}%
}

\def\sing@vfootnote#1{\insert\footins\bgroup\eightpoint\b@ls{9pt}%
  \interlinepenalty\interfootnotelinepenalty
  \splittopskip\ht\strutbox 
  \splitmaxdepth\dp\strutbox \floatingpenalty\@MM
  \leftskip\z@skip \rightskip\z@skip \spaceskip\z@skip \xspaceskip\z@skip
  \noindent $^{\scriptstyle\hbox{#1}}$\hskip 4pt%
    \footstrut\futurelet\next\fo@t%
}

\def\footnoterule{\kern-3\p@ \hrule height \z@ \kern 3\p@}

\skip\footins=19.5pt plus 12pt minus 1pt
\count\footins=1000
\dimen\footins=\maxdimen

\def\note#1#2{%
  \let\@sf=\empty \ifhmode\edef\@sf{\spacefactor\the\spacefactor}\/\fi
  #1\insert\footins\bgroup
    \eightpoint\b@ls{10pt}\rm
    \interlinepenalty\interfootnotelinepenalty
    \splitmaxdepth\dp\strutbox \floatingpenalty\@MM
    \leftskip\z@skip \rightskip\z@skip \spaceskip\z@skip \xspaceskip\z@skip
    \noindent\footstrut #1$\,$#2\strut\par
  \egroup
  \@sf\relax}


\def\landscape{%
  \global\TEMPDIMEN=\PageWidth
  \global\PageWidth=\PageHeight
  \global\PageHeight=\TEMPDIMEN
  \global\let\landscape=\relax
  \onecolumn
  \message{(landscape)}%
  \raggedbottom
}


\output{%
  \ifLeftCOL
    \global\setbox\LeftBOX=\vbox to \ZoneBSize{\box255\unvbox\ZoneBBOX
      \ifvoid\footins\else
        \vskip\skip\footins\unvbox\footins\fi
    }%
    \global\LeftCOLfalse
    \MakeRightCol
  \else
    \setbox\RightBOX=\vbox to \ZoneBSize{\box255\unvbox\ZoneBBOX
      \ifvoid\footins\else
        \vskip\skip\footins\unvbox\footins\fi
    }%
    \setbox\MidBOX=\hbox{\box\LeftBOX\hskip\ColumnGap\box\RightBOX}%
    \setbox\PageBOX=\vbox to \PageHeight{%
      \UnloadZoneA\box\MidBOX\UnloadZoneC}%
    \shipout\vbox{\PageHead\vbox to \PageHeight{\box\PageBOX\vss}\PageFoot}%
    \advancepageno
    \ifplate@page
      \shipout\vbox{%
        \sp@pagetrue
        \def\sp@type{plate}%
        \global\plate@pagefalse
        \PageHead\vbox to \PageHeight{\unvbox\plt@box\vfil}\PageFoot%
      }%
      \message{[plate]}%
      \advancepageno
    \fi
    \global\LeftCOLtrue
    \CleanStack
    \MakePage
  \fi
}


\Warn{\start@mess}

\newif\ifCUPmtplainloaded 
\ifprod@font
  \global\CUPmtplainloadedtrue
\fi


\catcode `\@=12 



\loadboldmathnames

\def\GIOscriptD{{\cal D}}
\def\GIOscriptS{{\cal S}}

\onecolumn
\begintopmatter
\title{Waves and instabilities in a differentially rotating disc containing a poloidal magnetic field}
\author{G. I. Ogilvie$^{1,2}$}
\affiliation{$^1$Institute of Astronomy, University of Cambridge, Madingley Road, Cambridge CB3 0HA}
\vskip1mm
\affiliation{$^2$Department of Applied Mathematics and Theoretical Physics, University of Cambridge, Silver Street, Cambridge CB3 9EW}
\shortauthor{G. I. Ogilvie}
\shorttitle{Waves and instabilities in a differentially rotating disc}
\abstract{The theory of waves and instabilities in a differentially rotating disc containing a poloidal magnetic field is developed within the framework of ideal magnetohydrodynamics.  A continuous spectrum, for which the eigenfunctions are localized on individual magnetic surfaces, is identified but is found not to contain any instabilities associated with differential rotation.  The normal modes of a weakly magnetized thin disc are studied by extending the asymptotic methods used previously to describe the equilibria.  Waves propagate radially in the disc according to a dispersion relation which is determined by solving an eigenvalue problem at each radius.  The dispersion relation for a hydrodynamic disc is re-examined and the modes are classified according to their behaviour in the limit of large wavenumber.  The addition of a magnetic field introduces new, potentially unstable, modes and also breaks up the dispersion diagram by causing avoided crossings.  The stability boundary to the magnetorotational instability in the parameter space of polytropic equilibria is located by solving directly for marginally stable equilibria.  For a given vertical magnetic field in the disc, bending of the field lines has a stabilizing effect and it is shown that stable equilibria exist which are capable of launching a predominantly centrifugally driven wind.}
\keywords {accretion, accretion discs -- hydrodynamics -- instabilities -- MHD -- waves.}
\maketitle

\section{Introduction}

There are good reasons for believing that magnetic fields are important to the physics of accretion discs.  Magnetohydrodynamic (MHD) mechanisms provide the most convincing explanations for the anomalous transport of angular momentum that is required for accretion to proceed.  One possibility is that angular momentum is removed from the disc by a rotating MHD wind (Blandford \& Payne 1982).  These flows have the property of collimating into jets perpendicular to the disc (Heyvaerts \& Norman 1989), which is attractive in view of the observed association of discs and jets.  A rather more convincing mechanism for angular momentum transport is provided by the magnetorotational instability (Velikhov 1959; Chandrasekhar 1960; Balbus \& Hawley 1991).  A Keplerian shear flow, while hydrodynamically stable, is destabilized by the presence of a magnetic field, provided that the magnetic field is sufficiently weak and the disc is sufficiently ionized.  The instability is linear, dynamical and operates for toroidal as well as poloidal magnetic fields (Foglizzo \& Tagger 1995; Ogilvie \& Pringle 1996; Terquem \& Papaloizou 1996).  In local simulations, the non-linear development leads to turbulence, and the associated Reynolds and Maxwell stresses transport angular momentum radially outwards, as is required for accretion (Brandenburg et al. 1995, 1996; Hawley, Gammie \& Balbus 1995, 1996; Stone et al. 1996).

In an earlier work (Ogilvie 1997; hereafter, Paper~I) some idealized equilibrium models of accretion discs containing magnetic fields were presented, with the aim of studying the magnetorotational instability in more realistic geometry.  The model system consists of a perfectly conducting, non-self-gravitating fluid in differential rotation about a massive central object.  The fluid contains a purely poloidal magnetic field of dipolar symmetry, which bends as it passes through the disc, enforcing isorotation on magnetic surfaces.  This rather general model is governed by a non-linear, elliptic partial differential equation in two dimensions, a version of the Grad--Shafranov equation.  Two classes of special solutions of this general problem were described.

First, asymptotic solutions were obtained in the limit of a thin disc, using as the small parameter $\epsilon$ a characteristic value of $H(r)/r$, where $H(r)$ is the height of the upper surface of the disc above the equatorial plane at radius $r$.  It was shown that two families of solutions resembling accretion discs exist in this limit, with different asymptotic scalings representing different balances of forces in the radial and vertical directions.  The {\it weakly magnetized discs\/} are the natural generalization of the standard, hydrodynamic thin discs (Pringle 1981).  The sound speed and Alfv\'en velocity are both $O(\epsilon)$ [given that the Keplerian velocity is $O(1)$], while the fractional deviation from Keplerian rotation, caused by the radial Lorentz force, is $O(\epsilon)$.  The {\it strongly magnetized discs\/} are not directly related to hydrodynamic thin discs but resemble models of solar prominences in which a sheet of matter is supported against gravity by a bending magnetic field (Kippenhahn \& Schl\"uter 1957).  The sound speed and Alfv\'en velocity are both $O(\epsilon^{1/2})$, while the fractional deviation from Keplerian rotation is $O(1)$.  Previous approaches based on a vertical integration of the equations (e.g. Heyvaerts \& Priest 1989) described only the strongly magnetized discs.  It is, however, the weakly magnetized discs that are capable of launching a predominantly centrifugally driven wind if the magnetic field lines at the surface of the disc are inclined to the vertical by an angle greater than $\pi/6$.

Secondly, solutions were obtained by assuming self-similarity in the spherical radial coordinate, as is often done in the analysis of winds and jets from accretion discs.  This is a convenient method of studying thick equilibria and also of verifying the asymptotic results for thin discs.

Although the accretion flow, the resistivity of the fluid, and any toroidal magnetic field are neglected in this model, it is expected that these additional terms constitute only small perturbations to the internal equilibrium of the disc.  It was shown in Paper~I that a model of a wind-driven accretion disc can be built up by superimposing these additional features on the solution for a weakly magnetized thin disc without disturbing the equilibrium at leading order in $\epsilon$.

The primary purpose of this paper is to determine the spectrum of waves and instabilities in a weakly magnetized thin disc, within the framework of ideal MHD, by extending the asymptotic methods used to describe the equilibria.  In particular, the stability of thin discs to the magnetorotational instability is to be studied.  There may, of course, be other types of instability in magnetized accretion discs.  A non-ax\-isym\-met\-ric interchange instability (Spruit, Stehle \& Papaloizou 1995) or bending instability (Agapitou, Papaloizou \& Terquem 1997) may be present if the magnetic field provides a significant amount of support against gravity in the radial direction.  A poloidal magnetic field is also expected to alter the criterion for convective instability (Moss \& Tayler 1969).  Finally, a global, non-ax\-isym\-met\-ric instability, related to the Papaloizou--Pringle instability (Papaloizou \& Pringle 1984) may be anticipated (Curry \& Pudritz 1996).  However, not all of these instabilities will feature in the analysis that follows, either because they are not present in weakly magnetized discs, or because their growth rates are too small to be detected.

The energy principle of Bernstein et~al. (1958) has played an important role in the analysis of the stability of magnetostatic equilibria relevant to astrophysics.  Moss \& Tayler (1969) examined the case of an axisymmetric, non-rotating star containing a poloidal magnetic field, and demonstrated that, if self-grav\-i\-ta\-tion is neglected, the most unstable (or least stable) modes are those for which the azimuthal wavenumber $m$ tends to infinity, and that the problem reduces to the consideration of a system of ordinary differential equations on each magnetic field line separately.  Tayler (1973) considered the stability of a non-rotating star with a purely toroidal magnetic field, and found that the problem reduces to a system of algebraic equations at each separate point in the meridional plane.  In each case, the most important displacements are those localized on a single magnetic field line or magnetic surface, and the equations obtained are closely related to those governing the continuous spectrum, which could be derived directly using the methods outlined in Section~2 below.

In an accretion disc the differential rotation is the most important feature of the dynamics and cannot be ignored or approximated by uniform rotation.  The energy principle of Bernstein et~al. (1958), which applies only to static equilibria, cannot be used.  Nevertheless, Papaloizou \& Szuszkiewicz (1992) showed that, for a differentially rotating, non-self-grav\-i\-ta\-ting fluid containing a poloidal magnetic field, a generalization of the energy principle exists in the form of a variational principle for the frequency eigenvalues of axisymmetric normal modes.  Using the variational principle, they deduced some sufficient conditions for stability to axisymmetric perturbations.  In this case, however, the problem does not reduce to a consideration of each magnetic surface separately.  This suggests that the most unstable (or least stable) part of the spectrum of axisymmetric modes is discrete rather than continuous.  This is consistent with the fact that the axisymmetric magnetorotational instability, although often described as a local instability, requires a relatively long wavelength in the direction perpendicular to the magnetic field, rather than being localized on a single magnetic surface.  Nevertheless, it is possible to analyse the continuous spectrum directly without reference to an energy principle.

This paper is concerned principally with the weakly magnetized thin discs and will involve an extension of the asymptotic methods used in Paper~I.  Primary consideration is given to axisymmetric modes, but non-ax\-isym\-met\-ric perturbations are also discussed briefly.  The continuous spectrum requires a separate analysis which can be made for a much more general equilibrium, and which is not restricted to axisymmetric modes.  This is presented in Section~2.  In Section~3 the equations and boundary conditions for axisymmetric modes in a weakly magnetized thin disc are derived.  Some general properties of these equations are discussed in Section~4 and the numerical method of solution is described.  In Section~5 some analytical results are obtained, principally for non-magnetized discs, which allow the modes to be enumerated and classified meaningfully in certain circumstances.  For weakly magnetized discs, the most important result of this paper is the stability boundary in the parameter space of polytropic equilibria, which is obtained in Section~6.  Non-ax\-isym\-met\-ric modes are discussed briefly in Section~7, and a concluding discussion is given in Section~8.

Throughout this paper, physical quantities are written in SI units with the permeability of free space, $\mu_0$, omitted for convenience.  The coordinates used are cylindrical polar coordinates $(r,\phi,z)$ and the magnetic flux coordinates $(\psi,\phi,\chi)$ defined in Paper~I.  The magnetic flux coordinates form a right-handed orthogonal coordinate system such that the magnetic field is ${\bmath B}=\nabla\psi\times\nabla\phi=B(\psi,\chi)\,{\bmath e}_\chi$, where $\phi$ is the usual azimuthal angular coordinate.  The Jacobian of the coordinate system is $J=1/|\nabla\psi||\nabla\phi||\nabla\chi|$.

\section{The continuous spectrum}

The linearized equation governing the Lagrangian displacement $\bxi({\bmath r},t)$ corresponding to a small departure from any state of ideal MHD may be written
$$\rho{{\rm D}^2\bxi\over{\rm D}t^2}=-\nabla\delta\Pi-(\nabla\!\cdot\!\bxi)\nabla\Pi-\bxi\!\cdot\!\nabla\nabla\Pi-\rho\bxi\!\cdot\!\nabla\nabla\Phi+{\bmath B}\!\cdot\!\nabla\left[{\bmath B}\!\cdot\!\nabla\bxi-(\nabla\!\cdot\!\bxi){\bmath B}\right],\eqno(2.1)$$
where $\Pi$ is the total pressure and
$$\delta\Pi=-(\gamma p+{\textstyle{1\over2}}B^2)\nabla\!\cdot\!\bxi-\bxi\!\cdot\!\nabla\Pi+{\bmath B}\cdot({\bmath B}\!\cdot\!\nabla\bxi)\eqno(2.2)$$
is its linearized Eulerian perturbation.  Here $\rho$, $p$, ${\bmath B}$ and $\gamma$ are the density, pressure, magnetic field and adiabatic exponent, respectively, and ${\rm D}/{\rm D}t$ is the Lagrangian time derivative.  This equation is equivalent to that given by Frieman \& Rotenberg (1960), but is more general in that the gravitational potential $\Phi$ is included, and the equation holds for an arbitrary basic flow.  The self-grav\-i\-ta\-tion of the fluid is neglected here, as it was neglected throughout the construction of equilibria in Paper~I.  Although self-grav\-i\-ta\-tion has, in general, a destabilizing influence on long-wavelength perturbations, its effect on continuum modes, which are localized on a single magnetic surface, is nil.  The equilibrium state under consideration is of the general class described in Section~2 of Paper~I.

The continuum modes may be described using a technique which was introduced by Papaloizou \& Pringle (1982) and developed by Lin, Papaloizou \& Kley (1993) and Terquem \& Papaloizou (1996).  The modes take the form of wave packets which are infinitely localized in the $\psi$-direction so that they are effectively confined to a single magnetic surface.  Localization in the $\chi$-direction cannot be achieved because of the infinite restoring force that would result from bending the magnetic field lines.  Using the magnetic flux coordinates $(\psi,\phi,\chi)$ of Paper~I, one considers a sequence of functions of the form
$$\bxi({\bmath r},t)={\rm Re}\left[{\bxi(\chi)f\left({\psi-\psi_0\over w(k)}\right)\exp({\rm i}k\psi+{\rm i}m\phi-{\rm i}\omega t)}\right],\eqno(2.3)$$
where $\omega\in\bbbc$ is the frequency eigenvalue, $m\in\bbbz$ the azimuthal wavenumber, and $k\in\bbbr$ a parameter which is allowed to tend to infinity.  The function $f$ is a unimodal `wavelet'\note{$^1$}{For example, $f(x)=\exp(-x^2)$.} of width $w(k)$ centered on $\psi=\psi_0$, and forms the envelope of the wave packet.  In the limit $k\to\infty$, the function $f$ is to become infinitely localized at $\psi=\psi_0$, but the width $w(k)$ of the function should tend to zero more slowly than $k^{-1}$, perhaps $w\propto k^{-1/2}$.  Then the number of oscillations under the envelope tends to infinity, and a derivative of $\bxi$ with respect to $\psi$ corresponds at leading order to multiplication by ${\rm i}k$.  The generalized function defined in the limit $k\to\infty$ represents an infinitely localized wave packet.  It is now possible to obtain the equations satisfied by the mode in this limit.

It is convenient to adopt a normalization such that $|\bxi|=O(1)$ at $\psi=\psi_0$.  If one is to obtain a solution of equation (2.1) with a frequency eigenvalue that has a finite limit (and a finite value of $m$), then it is clear that the ordering must be such that $\nabla\!\cdot\!\bxi=O(1)$ [rather than $O(k)$] and $\delta\Pi=O(k^{-1})$ [rather than $O(k)$ or $O(1)$].  Therefore the fast magnetoacoustic wave is filtered out in this limit, as in the magneto-Boussinesq approximation (Spiegel \& Weiss 1982).  It also follows that the component $\xi_\psi$ is $O(k^{-1})$, so the displacement is confined within the magnetic surface (at leading order).  One may write
$$\nabla\!\cdot\!\bxi=\Delta\qquad\hbox{and}\qquad\delta\Pi=k^{-1}\varpi,\eqno(2.4)$$
where $\Delta$ and $\varpi$ are both $O(1)$.

In order to write equation (2.1) in magnetic flux coordinates, it is convenient to introduce the angle of inclination $i(\psi,\chi)$ of the magnetic field to the vertical, such that
$$\cos i={\bmath e}_r\cdot{\bmath e}_\psi=rB{\partial r\over\partial\psi}\qquad\hbox{and}\qquad\sin i={\bmath e}_r\cdot{\bmath e}_\chi={1\over JB}{\partial r\over\partial\chi}.\eqno(2.5)$$
Then the derivatives of the unit vectors are given by
$$\eqalignno{{\rm d}{\bmath e}_\psi&=-\left({\partial i\over\partial\psi}\right){\bmath e}_\chi\,{\rm d}\psi+\cos i\,{\bmath e}_\phi\,{\rm d}\phi-\left({\partial i\over\partial\chi}\right){\bmath e}_\chi\,{\rm d}\chi,&(2.6)\cr
{\rm d}{\bmath e}_\phi&=-\cos i\,{\bmath e}_\psi\,{\rm d}\phi-\sin i\,{\bmath e}_\chi\,{\rm d}\phi,&(2.7)\cr
{\rm d}{\bmath e}_\chi&=\left({\partial i\over\partial\psi}\right){\bmath e}_\psi\,{\rm d}\psi+\sin i\,{\bmath e}_\phi\,{\rm d}\phi+\left({\partial i\over\partial\chi}\right){\bmath e}_\psi\,{\rm d}\chi,&(2.8)\cr}$$
with
$${\partial i\over\partial\psi}={1\over JB}{\partial\over\partial\chi}\left({1\over rB}\right)\qquad\hbox{and}\qquad{\partial i\over\partial\chi}=-rB{\partial(JB)\over\partial\psi}.\eqno(2.9)$$

The quantity $\varpi$ appears at leading order only in the $\psi$-component of equation (2.1), which therefore serves to define $\varpi$ in terms of $\xi_\phi$, $\xi_\chi$ and $\Delta$, but is otherwise unimportant.  The remaining equations at leading order are
$$-\rho\hat\omega^2\xi_\phi-2{\rm i}\rho\hat\omega\Omega\sin i\,\xi_\chi-\rho\Omega^2\xi_\phi=-\cos i\,rB\left({\partial\Pi\over\partial\psi}+\rho{\partial\Phi\over\partial\psi}\right){\xi_\phi\over r}-\sin i\,{1\over JB}\left({\partial\Pi\over\partial\chi}+\rho{\partial\Phi\over\partial\chi}\right){\xi_\phi\over r}+{1\over J}{\partial\over\partial\chi}\left({1\over J}{\partial\xi_\phi\over\partial\chi}\right),\eqno(2.10)$$
$$\eqalignno{&-\rho\hat\omega^2\xi_\chi+2{\rm i}\rho\hat\omega\Omega\sin i\,\xi_\phi-\rho\Omega^2\sin^2i\,\xi_\chi=-{1\over JB}{\partial\Pi\over\partial\chi}\Delta-{\xi_\chi\over JB}{\partial\over\partial\chi}\left({1\over JB}{\partial\Pi\over\partial\chi}\right)-\rho{\xi_\chi\over JB}{\partial\over\partial\chi}\left({1\over JB}{\partial\Phi\over\partial\chi}\right)&\cr
&\qquad\qquad-rB\left({\partial\Pi\over\partial\psi}-\rho{\partial\Phi\over\partial\psi}\right){\xi_\chi\over JB}\left({\partial i\over\partial\chi}\right)+{1\over J}{\partial\over\partial\chi}\left({1\over J}{\partial\xi_\chi\over\partial\chi}-B\Delta\right)-\left({1\over J}{\partial i\over\partial\chi}\right)^{\!2}\xi_\chi&(2.11)\cr}$$
and
$$0=-(\gamma p+B^2)\Delta-{\xi_\chi\over JB}{\partial\Pi\over\partial\chi}+{B\over J}{\partial\xi_\chi\over\partial\chi},\eqno(2.12)$$
where $\hat\omega=\omega-m\Omega$ is the intrinsic frequency.  These are the exact equations defining the continuous spectrum.  It should be understood that all quantities are to be evaluated on the magnetic surface $\psi=\psi_0$; the equations are essentially ordinary differential equations with $\psi$ treated as a parameter.  Once $\Delta$ has been eliminated, the remaining equations may be written
$$\hat\omega^2\xi_\phi+2{\rm i}\hat\omega\Omega\sin i\,\xi_\chi=-{1\over\rho rJ}{\partial\over\partial\chi}\left[{r^2\over J}{\partial\over\partial\chi}\left({\xi_\phi\over r}\right)\right]\eqno(2.13)$$
and
$$\hat\omega^2\xi_\chi-2{\rm i}\hat\omega\Omega\sin i\,\xi_\phi=-{1\over\rho JB}{\partial\over\partial\chi}\left[\left({v_{\rm s}^2\over v_{\rm s}^2+v_{\rm A}^2}\right){B^2\over J}{\partial\over\partial\chi}\left({\xi_\chi\over B}\right)\right]+\left\{\left({v_{\rm s}^2\over v_{\rm s}^2+v_{\rm A}^2}\right)N^2_\chi-{1\over JB^2}{\partial\over\partial\chi}\left[\left({v_{\rm A}^2\over v_{\rm s}^2+v_{\rm A}^2}\right)Bg_\chi\right]\right\}\xi_\chi,\eqno(2.14)$$
where $v_{\rm s}=(\gamma p/\rho)^{1/2}$ and $v_{\rm A}=(B^2/\rho)^{1/2}$ are the sound speed and the Alfv\'en velocity, respectively,
$$g_\chi=-{1\over JB}{\partial\Phi\over\partial\chi}+r\Omega^2\sin i\eqno(2.15)$$
is the effective gravitational acceleration parallel to the magnetic field, and
$$N^2_\chi=g_\chi\left[{1\over JB}{\partial\ln\rho\over\partial\chi}-{g_\chi\over v_{\rm s}^2}\right]\eqno(2.16)$$
is a quantity analogous to the square of the Brunt--V\"ais\"al\"a frequency for displacements within the magnetic surface.

Boundary conditions must be supplied for these equations.  The poloidal magnetic field line may either form a closed loop or extend to infinity, and in general will have segments both inside and outside the disc.  The exterior region is to be treated as a force-free medium of zero density and pressure, in which equation (2.10) reduces to
$${\partial\over\partial\chi}(r\delta B_\phi)={\partial\over\partial\chi}\left[{r^2\over J}{\partial\over\partial\chi}\left({\xi_\phi\over r}\right)\right]=0,\eqno(2.17)$$
while equation (2.11) becomes vacuous.  If the field line crosses the surface $\GIOscriptS$ of the disc at points $\chi=\chi_1$ and $\chi=\chi_2$, say, and extends to infinity, then the relevant solution of equation (2.17) is $\delta B_\phi=0$, precisely as if the exterior were a vacuum.  Since $\delta{\bmath B}$ must be continuous on $\GIOscriptS$, the boundary condition on $\xi_\phi$ is
$${\partial\over\partial\chi}\left({\xi_\phi\over r}\right)=0\qquad\hbox{at $\chi=\chi_1$ and $\chi=\chi_2$}.\eqno(2.18)$$
If, instead, the field line is closed, then periodic boundary conditions should be applied, with $\delta{\bmath B}$ being continuous on $\GIOscriptS$.  It can be shown that $\delta B_\chi$ vanishes automatically on $\GIOscriptS$ (at leading order), and no boundary condition on $\xi_\chi$ is obtained.  The explanation is that equation (2.14) is singular at the surface, where $v_{\rm s}$ vanishes, and a regularity condition applies to $\xi_\chi$ there.

In the absence of rotation, equations (2.13) and (2.14) are two uncoupled, second-order differential equations in Sturm--Liouville form, describing the Alfv\'en continuum and the cusp continuum, respectively, for the polarization for which the displacement is confined to the magnetic surface.  These equations have been given by Poedts, Hermans \& Goossens (1985), who also showed that the addition of a toroidal magnetic field results in a coupling between the equations.  Similarly, in this case, the equations are coupled by rotation, if the magnetic field is not purely vertical.  A second effect of rotation is to modify the effective gravitational acceleration $g_\chi$.

It should be noted, however, that there is a significant difference in the continuous spectrum between the case of no rotation (or uniform rotation) and that of differential rotation.  In the former case, one may consider functions of the form (2.3) and take the simultaneous limits $k\to\infty$ and $m\to\infty$.  A continuous range of polarizations is obtained by choosing different limiting values of the ratio $k/m$, and the corresponding eigenfunctions need not have displacements confined within the magnetic surface.  This leads naturally to the conclusions of Moss \& Tayler (1969).  The equations given by Poedts et~al. (1985) in the case of a purely poloidal magnetic field describe only that part of the continuous spectrum with the polarization for which $k/m\to\infty$, since $m$ is assumed finite.  When the fluid is in differential rotation, however, the limit $m\to\infty$ cannot be taken, because an infinite differential Doppler shift would result.  The presence of a toroidal magnetic field would also prevent the limit $m\to\infty$ from being taken.

It is clear that the azimuthal wavenumber $m$ appears in equations (2.13), (2.14) and (2.18) only in the combination $\hat\omega=\omega-m\Omega$.  Moreover, $\omega$, $\Omega$ and $\hat\omega$ are all constant on the magnetic surface.  It follows that the eigenfunctions are the same for all values of $m$, while the frequency eigenvalues of non-ax\-isym\-met\-ric modes are related to those of axisymmetric modes simply by a Doppler shift.  In the same way that Papaloizou \& Szuszkiewicz (1992) derived a variational principle for axisymmetric modes, so it is possible to derive a variational principle for modes of arbitrary $m$, provided that attention is restricted to the continuous spectrum.  One may proceed by defining $\{\omega_n^2(\psi):n\in\bbbz^+\}$ and $\{u_n(\psi,\chi):n\in\bbbz^+\}$ to be the ordered, real eigenvalues and normalized, real eigenfunctions of the `Alfv\'en operator' that appears in equation (2.13).  This is a Sturm--Liouville problem,
$${\partial\over\partial\chi}\left({r^2\over J}{\partial u_n\over\partial\chi}\right)+\omega_n^2\rho r^2Ju_n=0,\eqno(2.19)$$
subject to the boundary condition $\partial u_n/\partial\chi=0$ at $\chi=\chi_1$ and $\chi=\chi_2$ (or periodic boundary conditions for a closed field line), and the normalization condition
$$\int_{\chi_1}^{\chi_2}\rho r^2Ju_mu_n\,{\rm d}\chi=\delta_{mn},\eqno(2.20)$$
for each value of $\psi$.  Let $\{a_n\}$ and $\{b_n\}$ be the components of $(\sin i\,\xi_\chi/r)$ and $(\xi_\phi/r)$ with respect to this set of eigenfunctions, such that
$$\sin i\,{\xi_\chi\over r}=\sum_{n=0}^\infty a_nu_n(\psi,\chi)\qquad\hbox{and}\qquad{\xi_\phi\over r}=\sum_{n=0}^\infty b_nu_n(\psi,\chi).\eqno(2.21)$$
Then the solution of the inhomogeneous equation (2.13) is given by
$$b_n=-{2{\rm i}\hat\omega\Omega(\psi)\over\hat\omega^2-\omega_n^2(\psi)}a_n,\eqno(2.22)$$
and exists provided that the appropriate coefficient $a_n$ vanishes should $\hat\omega^2$ be equal to any of the eigenvalues of the Sturm--Liouville problem.  The spectrum of the Alfv\'en operator has the following characteristics.  The eigenvalues $\{\omega_n^2(\psi)\}$ for each magnetic surface form an ordered, denumerably infinite sequence of distinct, non-negative real numbers with limit $+\infty$.  The lowest eigenvalue is always $\omega_0^2(\psi)=0$, the corresponding eigenfunction $u_0(\psi,\chi)$ being independent of $\chi$.  This zero-frequency mode is a trivial displacement in which the entire magnetic surface is rotated through an infinitesimal angle about the $z$-axis.\note{$^2$}{In the absence of a magnetic field, the infinitesimal rotation of any ring of fluid about the axis is a trivial displacement, but these trivial displacements are easily eliminated from the analysis by discarding the root $\hat\omega=0$.}  A standard method (Courant \& Hilbert 1953) leads to the asymptotic expression
$$\omega_n^2(\psi)={n^2\pi^2\over\tau_{\rm A}^2(\psi)}+O(1)\eqno(2.23)$$
for the eigenvalues in the limit $n\to\infty$, where
$$\tau_{\rm A}(\psi)=\int_{\chi_1}^{\chi_2}\rho^{1/2}J\,{\rm d}\chi=\int_{\chi_1}^{\chi_2}{h_\chi\,{\rm d}\chi\over v_{\rm A}}\eqno(2.24)$$
is the Alfv\'en time of the magnetic field line.

The eigenfunction expansions (2.21) may be substituted into equation (2.14), the equation multiplied by $\rho J\xi_\chi^*$ and integrated with respect to $\chi$ to yield
$$L[\xi_\chi;\hat\omega^2]=R[\xi_\chi],\eqno(2.25)$$
where the two functionals
$$L[\xi_\chi;\hat\omega^2]=\hat\omega^2\int_{\chi_1}^{\chi_2}\rho|\xi_\chi|^2\,J\,{\rm d}\chi-\sum_{n=1}^\infty{4\hat\omega^2\Omega^2\over\hat\omega^2-\omega_n^2}|a_n|^2\eqno(2.26)$$
and
$$\eqalignno{&R[\xi_\chi]=4\Omega^2|a_0|^2+\int_{\chi_1}^{\chi_2}\left({v_{\rm s}^2\over v_{\rm s}^2+v_{\rm A}^2}\right){B^2\over J}\left|{\partial\over\partial\chi}\left({\xi_\chi\over B}\right)\right|^2\,{\rm d}\chi+\int_{\chi_1}^{\chi_2}\left\{\left({v_{\rm s}^2\over v_{\rm s}^2+v_{\rm A}^2}\right)N^2_\chi-{1\over JB^2}{\partial\over\partial\chi}\left[\left({v_{\rm A}^2\over v_{\rm s}^2+v_{\rm A}^2}\right)Bg_\chi\right]\right\}\rho|\xi_\chi|^2\,J\,{\rm d}\chi&\cr
&&(2.27)\cr}$$
have been introduced.  Note that the infinite sum on the left-hand side starts at $n=1$, the term $n=0$, which is independent of $\hat\omega^2$,  having been transferred to the right-hand side.  If $\hat\omega^2=(\hat\omega^2)_{\rm r}+{\rm i}(\hat\omega^2)_{\rm i}$, then the imaginary part of equation (2.25) is
$$(\hat\omega^2)_{\rm i}\int_{\chi_1}^{\chi_2}\rho|\xi_\chi|^2\,J\,{\rm d}\chi+(\hat\omega^2)_{\rm i}\sum_{n=1}^\infty{4\omega_n^2\Omega^2\over|\hat\omega^2-\omega_n^2|^2}|a_n|^2=0,\eqno(2.28)$$
from which it follows that the eigenvalues $\hat\omega^2$ are real.  The eigenfunctions $\xi_\chi$ may also be taken to be real.  Now consider $L[\xi_\chi;\hat\omega^2]$ as a function of $\hat\omega^2$ for a given trial function $\xi_\chi$.  Its derivative with respect to $\hat\omega^2$ is
$${\partial\over\partial\hat\omega^2}L[\xi_\chi;\hat\omega^2]=\int_{\chi_1}^{\chi_2}\rho|\xi_\chi|^2\,J\,{\rm d}\chi+\sum_{n=1}^\infty{4\omega_n^2\Omega^2\over(\hat\omega^2-\omega_n^2)^2}|a_n|^2>0,\eqno(2.29)$$
and so the function is monotonic wherever the derivative is defined.  The function has a pole, however, at each of the values $\hat\omega^2=\omega_n^2$ (other than $n=0$) for which the coefficient $a_n$ does not vanish.  The function therefore increases once through all real values in each of the intervals $(-\infty,p_1)$, $(p_1,p_2)$, $(p_2,p_3),\dots\,$, where $p_1$, $p_2$, $p_3,\dots\,$ are the abscissae of the poles.  Equation (2.25) may therefore be understood as defining a multiple-valued functional $\hat\omega^2[\xi_\chi]$.  Moreover, the stationary values of this functional may be demonstrated by standard techniques to be identical to the true eigenvalues $\hat\omega^2$.  Corresponding to each eigenfunction $\xi_\chi$ is one eigenvalue in each of the intervals $(-\infty,p_1)$, $(p_1,p_2)$, $(p_2,p_3),\dots\,$.\note{$^3$}{Although these modes share the same displacement $\xi_\chi$, they have different displacements $\xi_\phi$, according to the relation (2.22).}  Now $L[\xi_\chi;0]=0$, which is the reason for transferring the term $n=0$ to $R[\xi_\chi]$.  A necessary condition for the existence of a negative eigenvalue $\hat\omega^2$ is that the functional $R[\xi_\chi]$ admit negative values for trial functions satisfying the boundary conditions.  The variational principle ensures that this condition is also sufficient.  Therefore a necessary and sufficient condition for instability in the continuous spectrum is that the functional $R[\xi_\chi]$ admit negative values for trial functions satisfying the boundary conditions.  In particular, stability is assured if
$$\left({v_{\rm s}^2\over v_{\rm s}^2+v_{\rm A}^2}\right)N^2_\chi-{1\over JB^2}{\partial\over\partial\chi}\left[\left({v_{\rm A}^2\over v_{\rm s}^2+v_{\rm A}^2}\right)Bg_\chi\right]>0\eqno(2.30)$$
throughout $\chi_1<\chi<\chi_2$, for each magnetic surface.

The only type of instability that can be present in the continuous spectrum is a magnetoconvective (Parker) instability modified by rotation.  The instability is due to the release of gravitational potential energy by displacements parallel to the magnetic field.  The first term in $R[\xi_\chi]$, which may be expressed in the form
$$4\Omega^2|a_0|^2=\left(|a_0|^2\bigg/\sum_{n=0}^\infty|a_n|^2\right)\int_{\chi_1}^{\chi_2}4\Omega^2\sin^2i\,\rho|\xi_\chi|^2\,J\,{\rm d}\chi,\eqno(2.31)$$
represents the stabilizing effect of rotation, but only on modes for which $a_0\ne0$.  In particular, if the disc is symmetrical about the equatorial plane, then this stabilizing effect is zero for all modes with odd symmetry.

Significantly, the confinement of the displacements within the magnetic surface implies that no instabilities associated with differential rotation are found.  In particular, the magnetorotational instability does not appear in the continuous spectrum.  This is in contrast with the case of a purely toroidal magnetic field (Terquem \& Papaloizou 1996).  Neither is a magnetoconvective instability associated with the $\psi$-component of gravity found.

\section{Axisymmetric waves and instabilities in thin discs}

The principal aim of this paper is to determine the spectrum of waves and instabilities in a magnetized thin disc of the type described in Paper~I.  Although asymptotic analysis in the parameter $\epsilon$ underlies the mathematics, calculations are made to leading order only, and therefore few explicit references to $\epsilon$ are required.  Of the two families of thin discs, the weakly magnetized discs have the more interesting spectrum, for they are potentially unstable to the magnetorotational instability.  Indeed, a major objective is to determine the stability boundary in the parameter space of weakly magnetized, polytropic discs.  Strongly magnetized discs are stable to the magnetorotational instability and so are not considered in detail.

The analysis presented here may be seen as an extension of the work of Lubow \& Pringle (1993; hereafter, LP) and Korycansky \& Pringle (1995; hereafter, KP) on the equivalent problem for hydrodynamic thin discs.  Previously, Ruden, Papaloizou \& Lin (1988) had examined convectively unstable modes in a thin disc using a similar method.  This analysis also extends the work of Gammie \& Balbus (1994), who examined the magnetorotational instability of a stratified disc containing a uniform magnetic field.  The present analysis is concerned with a polytropic disc containing a poloidal magnetic field that bends, in general, as it passes through the disc.

\subsection{Review of weakly magnetized discs}

The analysis of thin discs in Paper~I is based on a small parameter $\epsilon$ which may be defined as either the maximum value, or a characteristic value, of $H(r)/r$, where $z=H(r)$ is the location of the upper surface of the disc at radius $r$.  The internal structure of the disc is resolved by introducing a stretched vertical coordinate $\zeta=\epsilon^{-1}z$ whose value at the surface of the disc is $\zeta_{\rm s}=\epsilon^{-1}H$.  There are two aspects to obtaining a solution at leading order.  First, there are ordinary differential equations in $\zeta$ that determine the vertical equilibrium of the disc at each radius.  Secondly, there is an integral relation that determines the global magnetic structure; however, this is not directly relevant to the analysis that follows.

It is assumed that the pressure and density satisfy a polytropic relation on each magnetic surface.  Then the leading-order equations for a weakly magnetized disc are
$$\Omega_0=\left({GM\over r^3}\right)^{1/2},\eqno(3.1)$$
$${\partial p_0\over\partial\zeta}=-\rho_0\Omega_0^2\zeta+{2\rho_0r\Omega_0\Omega_1B_{r0}\over B_{z0}},\eqno(3.2)$$
$${\partial B_{r0}\over\partial\zeta}=-{2\rho_0r\Omega_0\Omega_1\over B_{z0}},\eqno(3.3)$$
$${\partial\Omega_1\over\partial\zeta}={3\Omega_0B_{r0}\over2rB_{z0}}\eqno(3.4)$$
and
$$p_0=K_0\rho_0^\Gamma,\eqno(3.5)$$
where the angular velocity, density, pressure and magnetic field are given by
$$\Omega(r,z)=\Omega_0(r)+\epsilon\Omega_1(r,\zeta)+O(\epsilon^2),\eqno(3.6)$$
$$\rho(r,z)=\epsilon^{2s}\left[\rho_0(r,\zeta)+\epsilon\rho_1(r,\zeta)+O(\epsilon^2)\right],\eqno(3.7)$$
$$p(r,z)=\epsilon^{2s+2}\left[p_0(r,\zeta)+\epsilon p_1(r,\zeta)+O(\epsilon^2)\right]\eqno(3.8)$$
and
$${\bmath B}(r,z)=\epsilon^{s+1}\left[B_{r0}(r,\zeta)+O(\epsilon)\right]{\bmath e}_r+\epsilon^{s+1}\left[B_{z0}(r)+O(\epsilon)\right]{\bmath e}_z.\eqno(3.9)$$
Here $s$ is a free parameter which should be positive if self-gravitation is to be unimportant.

These equations are to be solved on $0<\zeta<\zeta_{\rm s}(r)$, with boundary conditions
$$B_{r0}(r,0)=0,\qquad\rho_0(r,\zeta_{\rm s}(r))=0\qquad\hbox{and}\qquad B_{r0}(r,\zeta_{\rm s}(r))=B_{r0{\rm s}}(r),\eqno(3.10)$$
where $B_{r0{\rm s}}(r)$ denotes the value of $B_{r0}$ on the upper surface, and is supposed to be known from the global magnetic structure.  The solution in $-\zeta_{\rm s}<\zeta<0$ is inferred from symmetry, $\rho_0$, $p_0$ and $\Omega_1$ being even functions of $\zeta$, while $B_{r0}$ is odd.

\subsection{Normal modes}

The asymptotic method used to construct the equilibria can be applied to study their normal modes.  For a weakly magnetized disc, the angular velocity, the buoyancy frequency, and the characteristic frequencies of acoustic and Alfv\'en waves with wavelengths comparable to $H$ are all $O(1)$, which implies that this is the appropriate scaling for the frequency eigenvalue of a normal mode.  Only if a mode is found with eigenvalue $0$ or $\infty$ in this scaling need a different scaling be considered.  In deciding the form of the eigenfunctions of axisymmetric normal modes at leading order, it is natural to assume that they take the form
$$\bxi({\bmath r},t)\sim{\rm Re}\left[\bxi_0(r,\zeta)\exp(-{\rm i}\omega t)\right].\eqno(3.11)$$
If this is substituted into equation (2.1), there results at leading order a sixth-order differential system\note{$^4$}{Identical to equations (3.13)--(3.15) below, but with $k=0$.} on $-\zeta_{\rm s}<\zeta<\zeta_{\rm s}$ at each radius separately, which, when supplemented by appropriate boundary conditions, would constitute an eigenvalue problem for $\omega$.  The eigenvalues would be discrete, but would depend on $r$ in general, which means that a global solution, connecting a range of radii, could not be obtained other than in exceptional circumstances.  One is led to conclude that the eigenfunction must depend more strongly on $r$.  The correct form of the eigenfunction is
$$\bxi({\bmath r},t)\sim{\rm Re}\left\{\bxi_0(r,\zeta)\exp\left[-{\rm i}\omega t+{\rm i}\epsilon^{-1}\int k(r)\,{\rm d}r\right]\right\},\eqno(3.12)$$
where $k(r)$ is a radial wavenumber.  This is, of course, a WKB function, but it should be emphasized that no additional approximation or limit is involved here; this form arises simply from the fact that the eigenfunction varies on a radial length scale that is much shorter than that of the equilibrium disc.

Strictly speaking, equation (3.12) represents a wave propagating in the $r$-direction (when $\omega$ is real), whereas a normal mode is formed from the superposition of two such waves propagating in opposite directions.  However, the local dispersion relation derived below does not depend on the sign of $k$, and so it is sufficient to consider travelling waves of this form.  When $\omega$ is imaginary, equation (3.12) represents an exponentially growing or decaying normal mode rather than a travelling wave.  The scaling of $\bxi$ with $\epsilon$ is arbitrary, since this is a linear problem, but is chosen to be $O(1)$ for convenience.  For simplicity of notation, the subscript `0' on all quantities is omitted hereafter.

When the expression (3.12) is substituted into equation (2.1), the following differential system is obtained at leading order:
$$\rho\left[(\omega^2+3\Omega^2)\xi_r-2{\rm i}\omega\Omega\xi_\phi\right]={\rm i}k\,\delta\Pi-\GIOscriptD(\GIOscriptD\xi_r-B_r\Delta),\eqno(3.13)$$
$$\rho(\omega^2\xi_\phi+2{\rm i}\omega\Omega\xi_r)=-\GIOscriptD^2\xi_\phi,\eqno(3.14)$$
$$\rho\left(\omega^2+\Omega^2{\partial\ln\rho\over\partial\ln\zeta}\right)\xi_z={\partial\,\delta\Pi\over\partial\zeta}-\rho\Omega^2\zeta\Delta-\GIOscriptD(\GIOscriptD\xi_z-B_z\Delta),\eqno(3.15)$$
where
$$\Delta={\rm i}k\xi_r+{\partial\xi_z\over\partial\zeta},\eqno(3.16)$$
$$\delta\Pi=-(\gamma p+B^2)\Delta+\rho\Omega^2\zeta\xi_z+B_r\GIOscriptD\xi_r+B_z\GIOscriptD\xi_z\eqno(3.17)$$
and
$$\GIOscriptD={\rm i}kB_r+B_z{\partial\over\partial\zeta}.\eqno(3.18)$$
This system is of sixth order and requires six boundary conditions (apart from the arbitrary normalization condition that applies to all linear eigenvalue problems).  There are infinitely many discrete eigenvalues $\omega$ for each value of $k$, and these form a local dispersion relation $\omega(k;r)$ for the disc which has infinitely many branches.  The result of Papaloizou \& Szuszkiewicz (1992) that $\omega^2$ must be real for any global normal mode is reflected in the fact, demonstrated below, that the eigenvalues $\omega$ of the system (3.13)--(3.15) are either purely real or purely imaginary when $k$ is itself real.  As in a standard WKB problem, a global solution here is trapped in a `wave region' between two points, each of which may be either a turning point (at which $k$ goes to zero) or a boundary point (either the inner or outer radius of the disc), and within which $k$ is real and non-zero.  The variation of $\bxi_0$ with $r$ is determined by higher-order terms which could be calculated in principle.

The equations for an incompressible fluid ($\gamma\to\infty$) are obtained by imposing the constraint $\Delta=0$ on $\bxi$, and making $\delta\Pi$ a distinct variable in its own right, rather than using the indeterminate equation (3.17).

The boundary conditions for equations (3.13)--(3.15) are complicated because there is a free surface with an inclined magnetic field.  In the `corona' above the disc, where $\zeta>\zeta_{\rm s}$ and $\rho=0$, the magnetic field is uniform on the spatial scale $H$ and equations (3.13)--(3.15) reduce to the two equations
$$B_z^2{\partial^2\xi_\psi\over\partial\zeta^2}=k^2B_z^2\xi_\psi\eqno(3.19)$$
and
$$\left({\rm i}kB_{r{\rm s}}+B_z{\partial\over\partial\zeta}\right)^{\!2}\xi_\phi=0,\eqno(3.20)$$
where $\xi_\psi$ is the component of $\bxi$ in the meridional plane perpendicular to the magnetic field.  The appropriate solution of equation (3.19) is
$$\xi_\psi\propto\exp\left(-|k|\zeta\right),\eqno(3.21)$$
when $k$ is real.  The appropriate solution of equation (3.20) is
$$\xi_\phi\propto\exp\left[-{\rm i}(B_{r{\rm s}}/B_z)k\zeta\right],\eqno(3.22)$$
since this corresponds to a vanishing Eulerian perturbation
$$\delta B_\phi=\left({\rm i}kB_{r{\rm s}}+B_z{\partial\over\partial\zeta}\right)\xi_\phi=0.\eqno(3.23)$$
In each case the solution is such that the components of the Eulerian perturbation of the magnetic field go to zero at infinity.  The corresponding boundary conditions for the interior solution at $\zeta=\zeta_{\rm s}$ are easily obtained by applying the continuity of $\bxi$ and its first derivatives.  Thus
$$B_z{\partial\xi_r\over\partial\zeta}-B_{r{\rm s}}{\partial\xi_z\over\partial\zeta}=-|k|(B_z\xi_r-B_{r{\rm s}}\xi_z)\eqno(3.24)$$
and
$$B_z{\partial\xi_\phi\over\partial\zeta}=-{\rm i}kB_{r{\rm s}}\xi_\phi.\eqno(3.25)$$
The third boundary condition at $\zeta=\zeta_{\rm s}$ is a regularity condition, which is necessary because the sound speed goes to zero at the surface.  The equation for $\xi_\chi$, the component of $\bxi$ parallel to the magnetic field, has a singular point there and only the regular solution is acceptable.

The three remaining boundary conditions are the equivalents of the first three at the lower surface $\zeta=-\zeta_{\rm s}$.  Alternatively, these may be replaced by symmetry conditions at $\zeta=0$, since the solutions are either even or odd.  It is convenient to define an `even' mode as one which preserves the symmetry of the equilibrium about the equatorial plane.  For such a mode, the Eulerian perturbation of any quantity must have the same symmetry as that quantity has in the equilibrium state.  This implies that $\xi_r$ and $\xi_\phi$ are even functions of $\zeta$, while $\xi_z$ is odd.  An `odd' mode is one that breaks the symmetry of the equilibrium, implying that $\xi_r$ and $\xi_\phi$ are odd, while $\xi_z$ is even.  The boundary conditions at $\zeta=0$ are therefore
$${\partial\xi_r\over\partial\zeta}={\partial\xi_\phi\over\partial\zeta}=\xi_z=0\eqno(3.26)$$
for an even mode, and
$$\xi_r=\xi_\phi={\partial\xi_z\over\partial\zeta}=0\eqno(3.27)$$
for an odd mode.  Note that this convention is opposite to that used by LP and KP.

The relation between the local dispersion relation and the spectrum of global normal modes is quite straightforward.  As in any WKB problem, a quantization condition of the form
$$\epsilon^{-1}\int_{r_1}^{r_2}k(r)\,{\rm d}r=n\pi+\delta\eqno(3.28)$$
applies, where $r_1$ and $r_2$ are the limits of the wave region, $n$ is an integer and $\delta$ is a phase constant.  This implies that $n=O(\epsilon^{-1})$ for all modes with the scalings considered, and therefore the fact that it is an integer is irrelevant at this level of approximation.  If at any radius the local dispersion relation has a root $\omega$ corresponding to some real value of $k$, then a global normal mode must exist with that frequency eigenvalue at leading order.  The spectrum is dense in the asymptotic limit under consideration.  This also means that a necessary and sufficient condition for the existence of an unstable global normal mode is that the local dispersion relation should have an imaginary root $\omega$ for some real value of $k$.  This is true, at least, for axisymmetric modes with dynamical [i.e. $O(1)$] growth rates.

For completeness, the equations for axisymmetric waves in a strongly magnetized thin disc should be mentioned.  In such a disc, the angular velocity and the buoyancy frequency are both $O(1)$, but the characteristic frequencies of acoustic and Alfv\'en waves with wavelengths comparable to $H$ are both $O(\epsilon^{-1/2})$.  This implies that an appropriate scaling for the frequency eigenvalue of a normal mode is $O(\epsilon^{-1/2})$.  A set of equations is then obtained, equivalent to equations (3.13)--(3.15) but with $\Omega$ set to zero.  The modes are either Alfv\'en waves with $\xi_r=\xi_z=0$ or magnetoacoustic waves with $\xi_\phi=0$; neither rotation nor buoyancy has any effect, and there is no instability.  However, there is a mode with exactly zero frequency in this scaling, having an eigenfunction corresponding to a uniform displacement.  This allows a class of modes to exist that have frequency eigenvalues $\omega=O(1)$ and vary on a long radial length scale, as in equation (3.11).  These are the bending modes, which can be described using a two-dimensional treatment, and which can be unstable if the magnetic field provides a significant amount of support against gravity (Agapitou et~al. 1997).

\section{Properties of the local dispersion relation}

\subsection{Overview}

The sixth-order eigenvalue problem defined by equations (3.13)--(3.15) and the boundary conditions is considerably more complicated than the hydrodynamic equivalent.  An attempt is made in Section~5 below to classify the hydrodynamic modes, but little progress can be made when a magnetic field is included.  It is useful, however, to derive a variational principle for the frequency eigenvalues.  This not only shows that $\omega^2$ must be real when $k$ is real, but also helps to clarify under what circumstances unstable modes are expected.  The method here is an adaptation of the analysis of Papaloizou \& Szuszkiewicz (1992) and closely follows the method of Section~2.

Before deriving the variational principle it may be noted that the dispersion relation $\omega(k)$ has reflectional symmetry in both $\omega$- and $k$-axes.  This can be seen from the fact that equations (3.13)--(3.15) have the symmetries $(\xi_\phi,\omega)\mapsto(-\xi_\phi,-\omega)$ and {\it either\/} $(\bxi,\omega,k)\mapsto(\bxi^*,-\omega,-k)$, if $\omega$ is real, {\it or\/} $(\bxi,\omega,k)\mapsto(\bxi^*,\omega,-k)$, if $\omega$ is imaginary.  When presenting the numerical results it is therefore sufficient to restrict attention to positive values of $\omega$ (or $\omega/{\rm i}$) and of $k$.

\subsection{A variational principle}

To invert equation (3.14), which is subject to the boundary condition (3.25), one may define the ordered, real eigenvalues $\{\omega_n^2(r):n\in\bbbz^+\}$ and the real eigenfunctions $\{u_n(r,\zeta):n\in\bbbz^+\}$ of the Sturm--Liouville equation
$$B_z^2{\partial^2u_n\over\partial\zeta^2}+\omega_n^2\rho u_n=0,\eqno(4.1)$$
subject to the boundary conditions
$${\partial u_n\over\partial\zeta}=0\qquad\hbox{at}\quad\zeta=\pm\zeta_{\rm s}\eqno(4.2)$$
and the normalization condition
$$\int_{-\zeta_{\rm s}}^{\zeta_{\rm s}}\rho u_m u_n\,{\rm d}\zeta=\delta_{mn}\eqno(4.3)$$
for each $r$.  This is entirely analogous to the situation in Section~2.  Physically, the eigenfunctions represent Alfv\'en waves in a fictitious, non-rotating disc with the same density and vertical magnetic field, but without a radial magnetic field.  The eigenvalues $\{\omega_n^2\}$ are the corresponding squared frequency eigenvalues.

The equation
$$\GIOscriptD^2\tilde u_n+\omega_n^2\rho\tilde u_n=0,\eqno(4.4)$$
subject to the boundary conditions
$$\GIOscriptD\tilde u_n=0\qquad\hbox{at}\quad\zeta=\pm\zeta_{\rm s}\eqno(4.5)$$
is related to the Sturm--Liouville equation by a unitary transformation.  It has the same eigenvalues and its eigenfunctions are simply
$$\tilde u_n(r,\zeta)=u_n(r,\zeta)\exp\left[-{\rm i}k\int\left({B_r\over B_z}\right)\,{\rm d}\zeta\right]=u_n(r,\zeta)\exp\left[-{2\Omega_1(r,\zeta)\over3\Omega(r)}\,{\rm i}kr\right].\eqno(4.6)$$
Equation (3.14) may be written in the form
$$\GIOscriptD^2\xi_\phi+\omega^2\rho\xi_\phi=-2{\rm i}\omega\Omega\rho\xi_r,\eqno(4.7)$$
and is conveniently solved using the eigenfunction expansions
$$\xi_r(r,\zeta)=\sum_{n=0}^\infty a_n(r)u_n(r,\zeta)\exp\left[-{2\Omega_1(r,\zeta)\over3\Omega(r)}\,{\rm i}kr\right]\eqno(4.8)$$
and
$$\xi_\phi(r,\zeta)=\sum_{n=0}^\infty b_n(r)u_n(r,\zeta)\exp\left[-{2\Omega_1(r,\zeta)\over3\Omega(r)}\,{\rm i}kr\right].\eqno(4.9)$$
The solution is
$$b_n(r)=-{2{\rm i}\omega\Omega(r)\over\omega^2-\omega_n^2(r)}a_n(r),\eqno(4.10)$$
and exists at any particular value of $r$ provided that the appropriate coefficient $a_n(r)$ vanishes should $\omega^2$ be equal to any of the eigenvalues $\{\omega_n^2(r)\}$.  One can now substitute for $\xi_\phi$ in equation (3.13) and form the integral relation [cf. equation (2.25)]
$$L[\xi_r,\xi_z;\omega^2,k]=R[\xi_r,\xi_z;k]\eqno(4.11)$$
involving the two functionals
$$L[\xi_r,\xi_z;\omega^2,k]=\omega^2\int_{-\zeta_{\rm s}}^{\zeta_{\rm s}}\rho\left(|\xi_r|^2+|\xi_z|^2\right)\,{\rm d}\zeta-\sum_{n=1}^\infty{4\omega^2\Omega^2\over\omega^2-\omega_n^2}|a_n|^2\eqno(4.12)$$
and
$$\eqalignno{&R[\xi_r,\xi_z;k]=4\Omega^2|a_0|^2+\left[|k||B_z\xi_r-B_r\xi_z|^2\right]_{-\zeta_{\rm s}}^{\zeta_{\rm s}}&\cr
&\qquad+\int_{-\zeta_{\rm s}}^{\zeta_{\rm s}}\left[{|\delta\Pi|^2\over\gamma p+B^2}+|\GIOscriptD\xi_r|^2+|\GIOscriptD\xi_z|^2-\left({1\over\gamma p+B^2}\right)|B_r\GIOscriptD\xi_r+B_z\GIOscriptD\xi_z+\rho\Omega^2\zeta\xi_z|^2\right.&\cr
&\left.\qquad\qquad\qquad\qquad-3\rho\Omega^2|\xi_r|^2-\Omega^2\zeta{\partial\rho\over\partial\zeta}|\xi_z|^2\right]\,{\rm d}\zeta,&(4.13)\cr}$$
where it has been assumed that $k$ is real.  The imaginary part of equation (4.11) is
$$(\omega^2)_{\rm i}\int_{-\zeta_{\rm s}}^{\zeta_{\rm s}}\rho\left(|\xi_r|^2+|\xi_z|^2\right)\,{\rm d}\zeta+(\omega^2)_{\rm i}\sum_{n=1}^\infty{4\omega_n^2\Omega^2\over|\omega^2-\omega_n^2|^2}|a_n|^2=0,\eqno(4.14)$$
where $\omega^2=(\omega^2)_{\rm r}+{\rm i}(\omega^2)_{\rm i}$, and it follows that $\omega^2$ is real.  Moreover, it can be shown that equation (4.11) has a variational property which implies that the necessary and sufficient condition for the existence of an unstable mode at any radius is that there exists a trial displacement $\bxi$, satisfying the boundary conditions, which makes the functional $R[\xi_r,\xi_z;k]$ negative for some real value of $k$.

The functional $R[\xi_r,\xi_z;k]$ can be expressed a number of different ways which are useful for some purposes.  One alternative is to write
$$\eqalignno{R[\xi_r,\xi_z;k]=&4\Omega^2|a_0|^2+\left[|k||B_z\xi_r-B_r\xi_z|^2\right]_{-\zeta_{\rm s}}^{\zeta_{\rm s}}&\cr
&+\int_{-\zeta_{\rm s}}^{\zeta_{\rm s}}\left\{{|\delta\Pi|^2\over\gamma p+B^2}+{1\over B^2}|B_z\GIOscriptD\xi_r-B_r\GIOscriptD\xi_z|^2\right.&\cr
&\left.\qquad\qquad\qquad+\left({\gamma p\over\gamma p+B^2}\right){1\over B^2}\left|B_r\GIOscriptD\xi_r+B_z\GIOscriptD\xi_z-\left({B^2\over\gamma p}\right)\rho\Omega^2\zeta\xi_z\right|^2\right.&\cr
&\left.\qquad\qquad\qquad-3\rho\Omega^2|\xi_r|^2-\left[\Omega^2\zeta{\partial\rho\over\partial\zeta}+{(\rho\Omega^2\zeta)^2\over\gamma p}\right]|\xi_z|^2\right\}\,{\rm d}\zeta.&(4.15)\cr}$$
The coefficient of $|\xi_z|^2$ in the integrand is
$$-\left[\Omega^2\zeta{\partial\rho\over\partial\zeta}+{(\rho\Omega^2\zeta)^2\over\gamma p}\right]=\rho\Omega^2\zeta\left({1\over\gamma}{\partial\ln p\over\partial\zeta}-{\partial\ln\rho\over\partial\zeta}+{B_r\over\gamma p}{\partial B_r\over\partial\zeta}\right).\eqno(4.16)$$
If both the specific entropy and the (radial) magnetic field strength are non-decreasing functions of $\zeta$ for $0<\zeta<\zeta_{\rm s}$, then this coefficient is non-negative and it follows that only the term $-3\rho\Omega^2|\xi_r|^2$ in the integrand of equation (4.15) can lead to instability.  This implies that, in a disc with a convectively stable stratification, any instability is due to the differential rotation.

A second useful variation is to extend the integral in $R[\xi_r,\xi_z;k]$ over the interval $-\infty<\zeta<\infty$ and write
$$\eqalignno{&R[\xi_r,\xi_z;k]=4\Omega^2|a_0|^2&\cr
&\qquad+\int_{-\infty}^\infty\left[{|\delta\Pi|^2\over\gamma p+B^2}+|\GIOscriptD\xi_r|^2+|\GIOscriptD\xi_z|^2-\left({1\over\gamma p+B^2}\right)|B_r\GIOscriptD\xi_r+B_z\GIOscriptD\xi_z+\rho\Omega^2\zeta\xi_z|^2\right.&\cr
&\left.\qquad\qquad\qquad\qquad-3\rho\Omega^2|\xi_r|^2-\Omega^2\zeta{\partial\rho\over\partial\zeta}|\xi_z|^2\right]\,{\rm d}\zeta,&(4.17)\cr}$$
etc., in which case the trial displacement must be defined on $-\infty<\zeta<\infty$ and go to zero as $|\zeta|\to\infty$, but need not satisfy any boundary conditions at $\zeta=\pm\zeta_{\rm s}$.

Finally, the version for an incompressible fluid is
$$R[\xi_r,\xi_z;k]=4\Omega^2|a_0|^2+\left[|k||B_z\xi_r-B_r\xi_z|^2\right]_{-\zeta_{\rm s}}^{\zeta_{\rm s}}+\int_{-\zeta_{\rm s}}^{\zeta_{\rm s}}\left(|\GIOscriptD\xi_r|^2+|\GIOscriptD\xi_z|^2-3\rho\Omega^2|\xi_r|^2-\Omega^2\zeta{\partial\rho\over\partial\zeta}|\xi_z|^2\right)\,{\rm d}\zeta,\eqno(4.18)$$
where $\bxi$ is subject to the constraint $\Delta=0$.

Despite the unconventional location of the frequency eigenvalue in the variational principle, the orthogonality principle is straightforward.  If $\bxi_1$ and $\bxi_2$ are two eigenfunctions at the same radius, and for the same real value of $k$, which correspond to frequency eigenvalues $\omega_1$ and $\omega_2$, then
$$(\omega_1^2-\omega_2^2)\int_{-\zeta_{\rm s}}^{\zeta_{\rm s}}\rho\bxi_1^*\cdot\bxi_2\,{\rm d}\zeta=0,\eqno(4.19)$$
which implies that the eigenfunctions are orthogonal provided that the squared eigenvalues are distinct.  Note that this orthogonality principle involves all three components of $\bxi$, not just the meridional part.  It is conjectured that the eigenfunctions at each radius, and for each real value of $k$, form a complete set in the space of continuous, vector-valued functions of $\zeta$ on $-\zeta_{\rm s}<\zeta<\zeta_{\rm s}$.

\subsection{Numerical method}

The sixth-order eigenvalue problem (3.13)--(3.15) is solved numerically by the following method.  First, the equilibrium itself must be computed, as in Paper~I.  When $B_{r{\rm s}}$ and $B_z$ are specified, equations (3.1)--(3.5) constitute a third-order, non-linear eigenvalue problem for $\Omega_{1{\rm s}}$ (the value of $\Omega_1$ at $\zeta=\zeta_{\rm s}$).  This is solved using the shooting method: a value of $\Omega_{1{\rm s}}$ is guessed and the equations are integrated from $\zeta=\zeta_{\rm s}$ to $\zeta=0$.  The amount by which the boundary condition $B_{r0}(r,0)=0$ fails to be satisfied defines a mismatch function whose derivative with respect to $\Omega_{1{\rm s}}$ is obtained by simultaneously integrating the equations differentiated with respect to this parameter.  Newton--Raphson convergence is then applied to obtain the eigenvalue.  A similar technique is applied to equations (3.13)--(3.15).  However, the coefficients in these equations are known only numerically, and so to avoid the use of interpolation one must integrate equations (3.1)--(3.5) simultaneously (with the previously determined eigenvalue).  When $k$ is specified, equations (3.13)--(3.15) contain only one parameter, $\omega$.  However, since only three boundary conditions apply at $\zeta=\zeta_{\rm s}$, one must also guess two boundary values (the sixth being given by a normalization condition).  The full problem requires shooting in a three-dimensional complex space.  However, Newton--Raphson iteration can still be applied successfully.

The most difficult part of the numerical method is the treatment of the singular point at $\zeta=\zeta_{\rm s}$.  One can re-write equations (3.13) and (3.15) in the form
$${\partial^2\xi_r\over\partial\zeta^2}={N_r\over\gamma p},\eqno(4.20)$$
$${\partial^2\xi_z\over\partial\zeta^2}={N_z\over\gamma p},\eqno(4.21)$$
where the numerators $N_r$ and $N_z$ are regular and are linear functions of $\xi_r$, $\xi_\phi$, $\xi_z$ and their first derivatives.  The condition that $N_r$ and $N_z$ both go to zero as fast as $\gamma p$ as $\zeta\to\zeta_{\rm s}$ defines a regularity condition.  However, to start the numerical integration at $\zeta=\zeta_{\rm s}$, these quotients must be evaluated.  This is done by expanding all quantities in power series (not Taylor series) about the singular point, a procedure that involves much tedious algebra.

Once a mode has been computed for some value of $k$, it can be followed quasi-continuously as $k$ is varied, so tracing out a single branch of the dispersion relation.  This is then repeated for all branches within some limited region of the dispersion diagram.  Fortunately there are certain limits in which the modes can be enumerated using analytical methods.

All numerical calculations are carried out in units such that $GM=K_0=r=1$, and $\epsilon$ is defined such that $\zeta_{\rm s}=1$ at the value of $r$ under consideration.  In particular, this means that ${\bmath B}_0$ is expressed in units of
$$(GM)^{\Gamma/2(\Gamma-1)}K_0^{-1/2(\Gamma-1)}r^{-3\Gamma/2(\Gamma-1)}\zeta_{\rm s}^{\Gamma/(\Gamma-1)}.\eqno(4.22)$$
Taking into account the factors of $\epsilon$, this means that the true unit of magnetic field strength is
$$(GM)^{\Gamma/2(\Gamma-1)}K^{-1/2(\Gamma-1)}r^{-3\Gamma/2(\Gamma-1)}H^{\Gamma/(\Gamma-1)}.\eqno(4.23)$$
Similarly, frequencies and growth rates are expressed in units of the local Keplerian angular velocity, and the radial wavenumber in units of $H^{-1}$.

\section{Classification of modes}

\subsection{Hydrodynamic discs revisited}

The spectrum of axisymmetric waves in a polytropic, hydrodynamic thin disc has been described by KP.  When the magnetic field is absent, the solution for the equilibrium is
$$\rho_0=\left[\left({\Gamma-1\over\Gamma}\right){h_0\over K_0}\right]^{1/(\Gamma-1)},\eqno(5.1)$$
$$p_0=K_0\left[\left({\Gamma-1\over\Gamma}\right){h_0\over K_0}\right]^{\Gamma/(\Gamma-1)},\eqno(5.2)$$
where the enthalpy is
$$h_0={\textstyle{1\over2}}\Omega_0^2(\zeta_{\rm s}^2-\zeta^2).\eqno(5.3)$$
Equations (3.13)--(3.15) then become equivalent to those solved by KP in terms of Eulerian perturbations.  In their analysis, which was for a convectively stable disc, they identified three distinct classes of modes: p modes (acoustic modes, due to compressibility and driven mainly by pressure forces) and g modes (gravity modes, due to buoyancy and driven mainly by gravitational forces), both with $\omega^2\ge\Omega^2$, and also r modes\note{$^5$}{Called g modes by LP.} (inertial modes, due to rotation and driven mainly by inertial forces), with $\omega^2\le\Omega^2$.  Examples of the eigenfunctions for these modes can be found in KP.  They found no modes analogous to the f mode (the fundamental, surface gravity mode) of stellar oscillations, but reported certain peculiarities concerning the ordering of the eigenvalues.  The aim of this subsection is to clarify the situation and demonstrate that there are in fact two f modes.

An example of the local dispersion relation for a hydrodynamic disc is shown in Fig.~1.  The disc is convectively stable, with $\Gamma=4/3$ and $\gamma=5/3$, and has a spectrum that is qualitatively similar to that calculated by KP using slightly different parameters.  The different classes of modes may be distinguished (in a convectively stable disc) by the following characteristics:

\item{(i)} f, p and g modes have $\omega^2\ge\Omega^2$, while r modes have $\omega^2\le\Omega^2$;

\item{(ii)} in the limit $\gamma\to\Gamma$, in which the disc becomes adiabatically stratified, the g modes all collapse to $\omega^2=\Omega^2$, while all other modes have non-trivial limits;

\item{(iii)} in the limit $\gamma\to\infty$, in which the fluid becomes incompressible, the p modes have $\omega^2\to\infty$, while all other modes have non-trivial limits.

\noindent The f modes are the only modes with $\omega^2\ge\Omega^2$ to survive both limits (ii) and (iii).  There are two of them, one of each parity.

\beginfigure{1}
\centerline{\epsfysize=8cm\epsfbox{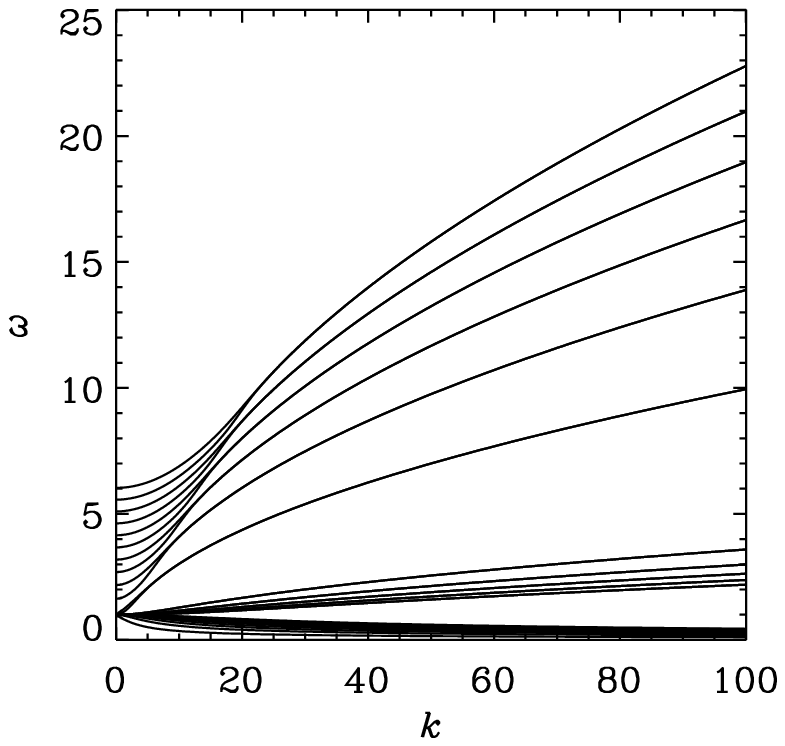}\qquad\qquad\epsfysize=8cm\epsfbox{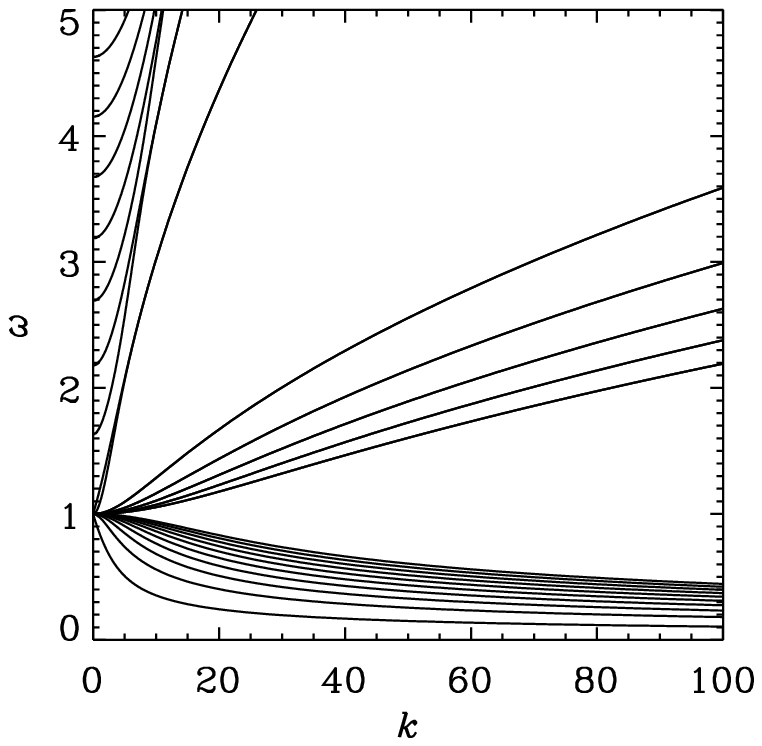}}
\caption{{\bf Figure~1.}  Left: local dispersion diagram for a polytropic, hydrodynamic thin disc.  Right: an expanded view of the lower-frequency branches.  The disc is convectively stable, with $\Gamma=4/3$ and $\gamma=5/3$.  The frequency eigenvalues of various branches of modes, in units of the local angular velocity, are plotted against the radial wavenumber, in units of $H^{-1}$.  The two f modes, and the first five p, g and r modes of each parity are shown.  In order of increasing frequency these are ${\rm r}_1^{\rm o}$, ${\rm r}_1^{\rm e}$, ${\rm r}_2^{\rm o}$, ${\rm r}_2^{\rm e}$, ${\rm r}_3^{\rm o}$, ${\rm r}_3^{\rm e}$, ${\rm r}_4^{\rm o}$, ${\rm r}_4^{\rm e}$, ${\rm r}_5^{\rm o}$, ${\rm r}_5^{\rm e}$, ${\rm g}_5^{\rm e}$, ${\rm g}_5^{\rm o}$, ${\rm g}_4^{\rm e}$, ${\rm g}_4^{\rm o}$, ${\rm g}_3^{\rm e}$, ${\rm g}_3^{\rm o}$, ${\rm g}_2^{\rm e}$, ${\rm g}_2^{\rm o}$, ${\rm g}_1^{\rm e}$, ${\rm g}_1^{\rm o}$, ${\rm f}^{\rm e}$, ${\rm f}^{\rm o}$, ${\rm p}_1^{\rm e}$, ${\rm p}_1^{\rm o}$, ${\rm p}_2^{\rm e}$, ${\rm p}_2^{\rm o}$, ${\rm p}_3^{\rm e}$, ${\rm p}_3^{\rm o}$, ${\rm p}_4^{\rm e}$, ${\rm p}_4^{\rm o}$, ${\rm p}_5^{\rm e}$, ${\rm p}_5^{\rm o}$, although even and odd f, p and g modes coalesce for sufficiently large $k$.}
\endfigure

The classification of the modes is particularly clear in the limit $k\zeta_{\rm s}\to\infty$.  All modes with $\omega^2>\Omega^2$ are then trapped near the surfaces of the disc.  As explained by KP, the loss of contact between the two surfaces implies that the frequency eigenvalues of adjacent even and odd modes must coalesce so that a single surface wave of neither symmetry can exist.  The WKB method used by KP gives the asymptotic form of the dispersion relation for large vertical mode number, but is not appropriate for studying modes of small vertical mode number such as the f modes.  Instead, one can use an asymptotic method based on the large parameter $k\zeta_{\rm s}$.  It is found numerically that all modes with $\omega^2>\Omega^2$ are trapped in a layer of extent $O\left((k\zeta_{\rm s})^{-1}\right)$ near each surface of the disc, and have $\omega^2=O(k\zeta_{\rm s})$ in this limit.  This is consistent with the interpretation of

\item{(i)} f modes as surface gravity modes (`$\omega^2\sim gk$'), since $g=O(1)$ in the layer;

\item{(ii)} p modes as acoustic modes (`$\omega^2\sim v_{\rm s}^2k^2$'), since $v_{\rm s}^2=O\left((k\zeta_{\rm s})^{-1}\right)$ in the layer;

\item{(iii)} g modes as buoyancy modes (`$\omega^2\sim N^2$'), since $N^2=O(k\zeta_{\rm s})$ in the layer.

\noindent The scalings imply that neither rotation nor the non-uniformity of gravity is significant to these modes in the limit $k\zeta_{\rm s}\to\infty$, and the modes therefore obey the same equations at leading order as in a static, polytropic atmosphere with uniform gravity.  This problem has been solved by Lamb (1932) in terms of confluent hypergeometric functions.  It was noted by Christensen-Dalsgaard (1980) that Lamb's analysis is valid in an asymptotic sense for modes of large degree $\ell$ in an arbitrary stellar model.  The same is true of modes of large $k\zeta_{\rm s}$ in an accretion disc.  To prove this, one should introduce a stretched coordinate
$$x_1=k(\zeta_{\rm s}-\zeta)\eqno(5.4)$$
which is $O(1)$ in the layer in which the modes are trapped.  Equations (3.13)--(3.15) are then solved asymptotically with
$$\xi_r(r,\zeta)=\xi_{r0}(r,x_1)+O\left((k\zeta_{\rm s})^{-1}\right),\eqno(5.5)$$
$$\xi_z(r,\zeta)=\xi_{z0}(r,x_1)+O\left((k\zeta_{\rm s})^{-1}\right)\eqno(5.6)$$
and
$$\omega^2=\lambda\Omega^2k\zeta_{\rm s}+O(1),\eqno(5.7)$$
while $\xi_\phi=O\left((k\zeta_{\rm s})^{-1/2}\right)$.  The quantity $\lambda$ is the dimensionless eigenvalue of the leading-order equations, which, following Lamb (1932), can be combined to obtain
$$x_1{\partial^2\Delta_0\over\partial x_1^2}+c{\partial\Delta_0\over\partial x_1}+\left[2(n-1)+c-x_1\right]\Delta_0=0,\eqno(5.8)$$
where
$$\Delta_0={\rm i}\xi_{r0}-{\partial\xi_{z0}\over\partial x_1}\eqno(5.9)$$
is the leading-order part of $\nabla\!\cdot\!\bxi$, while $c$ and $n$ are given by
$$c={2\Gamma-1\over\Gamma-1}\eqno(5.10)$$
and
$$\Gamma\lambda^2-\gamma\left[2(\Gamma-1)n+1\right]\lambda+(\gamma-\Gamma)=0.\eqno(5.11)$$
The solution regular at $x_1=0$ is (e.g. Erd\'elyi et~al. 1953a)
$$\Delta_0\propto{\rm e}^{-x_1}\,{}_1\!F_1\left(1-n;c;2x_1\right).\eqno(5.12)$$
It is exponentially small as $x_1\to+\infty$, thereby justifying the layer analysis, if and only if $n$ is a positive integer, in which case the confluent hypergeometric function is proportional to a generalized Laguerre polynomial of degree $n-1$.  Equation (5.11) gives the corresponding eigenvalues $\lambda_n^\pm$ as the roots of a quadratic equation.  Provided that $\gamma>\Gamma>1$, these roots are real and ordered such that
$$\dots<\lambda_2^-<\lambda_1^-<1<\lambda_1^+<\lambda_2^+<\dots,\eqno(5.13)$$
with
$$\lambda_n^+\sim{2\gamma(\Gamma-1)\over\Gamma}n\qquad\hbox{and}\qquad\lambda_n^-\sim{\gamma-\Gamma\over2\gamma(\Gamma-1)}{1\over n}\eqno(5.14)$$
as $n\to\infty$.  The modes corresponding to $\lambda_n^+$ and $\lambda_n^-$ are called ${\rm p}_n$ and ${\rm g}_n$ respectively in stellar oscillations.  There is an additional mode which corresponds to the trivial solution $\Delta_0=0$ of equation (5.8).  This has
$$\xi_{r0}\propto\xi_{z0}\propto{\rm e}^{-x_1}\eqno(5.15)$$
and $\lambda=\lambda_0=1$.  It is a surface gravity mode and corresponds to the f mode of stellar oscillations.

Somewhat surprisingly, the p modes and g modes are equally compressive, sharing the same function $\Delta_0$, and only the f mode has $\nabla\!\cdot\!\bxi=0$ at leading order.  Nevertheless, the g modes do survive the incompressible limit.  When $\gamma\to\infty$,
$$\lambda_n^+\to\infty\qquad\hbox{while}\qquad\lambda_n^-\to{1\over2(\Gamma-1)n+1}.\eqno(5.16)$$
Conversely, when $\gamma\to\Gamma$,
$$\lambda_n^+\to2(\Gamma-1)n+1\qquad\hbox{while}\qquad\lambda_n^-\to0.\eqno(5.17)$$

The notation `f', `p', `g', etc., for stellar oscillations can be readily applied to thin accretion discs provided that an additional distinction is made between even and odd modes.  This distinction is lost in the limit $k\zeta_{\rm s}\to\infty$, as described above.  For general values of $k$, one should refer to the modes as ${\rm f}^{\rm e}$, ${\rm f}^{\rm o}$, ${\rm p}_n^{\rm e}$, ${\rm p}_n^{\rm o}$, ${\rm g}_n^{\rm e}$ and ${\rm g}_n^{\rm o}$, where $n\in\bbbn$ is a vertical mode number and the superfix denotes the parity of the mode.\note{$^6$}{An alternative classification scheme is possible in which ${\rm p}_n^{\rm e}$ and ${\rm p}_n^{\rm o}$ are renamed ${\rm p}_{2n}$ and ${\rm p}_{2n-1}$, respectively, and similarly for other classes of modes.  This is closer to the notation of KP, and has the advantage that the superfix is not required.  However, the frequency eigenvalues do not then fall in sequence for all classes of modes.  Also, to refer to ${\rm f}_1$ and ${\rm f}_2$ rather than ${\rm f}^{\rm o}$ and ${\rm f}^{\rm e}$ is perhaps less helpful since it obscures the uniqueness of the f mode.}

The r modes behave quite differently in the limit $k\zeta_{\rm s}\to\infty$.  In a convectively stable disc, the r modes become trapped in a layer of extent $O\left((k\zeta_{\rm s})^{-1/2}\right)$ centred on the equatorial plane, and have $\omega^2=O\left((k\zeta_{\rm s})^{-1}\right)$.  The asymptotic analysis proceeds in a similar way, with a stretched coordinate
$$x_2=(k\zeta_{\rm s})^{1/2}(\zeta/\zeta_{\rm s})\eqno(5.18)$$
and scalings
$$\xi_r(r,\zeta)\sim(k\zeta_{\rm s})^{-1/2}\xi_{r0}(r,x_2),\eqno(5.19)$$
$$\xi_\phi(r,\zeta)\sim\xi_{\phi0}(r,x_2),\eqno(5.20)$$
$$\xi_z(r,\zeta)\sim\xi_{z0}(r,x_2)\eqno(5.21)$$
and
$$\omega^2\sim(k\zeta_{\rm s})^{-1}\Omega^2\lambda.\eqno(5.22)$$
The leading-order equations are
$${\rm i}\xi_{r0}+{\partial\xi_{z0}\over\partial x_2}=0,\eqno(5.23)$$
$$\xi_{\phi0}=-2{\rm i}\lambda^{-1/2}\xi_{r0}\eqno(5.24)$$
and
$${\partial^2\xi_{z0}\over\partial x_2^2}+\left[\lambda-{2(\gamma-\Gamma)\over\gamma(\Gamma-1)}x_2^2\right]\xi_{z0}=0.\eqno(5.25)$$
A bounded solution exists if and only if
$$\lambda=(2N+1)\sqrt{2(\gamma-\Gamma)\over\gamma(\Gamma-1)},\eqno(5.26)$$
where $N$ is a non-negative integer; the solution is then
$$\xi_{z0}\propto\exp(-{\textstyle{1\over2}}\alpha^2x_2^2)H_N(\alpha x_2),\eqno(5.27)$$
where
$$\alpha=\left[{2(\gamma-\Gamma)\over\gamma(\Gamma-1)}\right]^{1/4},\eqno(5.28)$$
and $H_N$ is an Hermite polynomial (e.g. Erd\'elyi et~al. 1953b).  If $N$ is odd, the mode is even and may be called ${\rm r}_n^{\rm e}$, where $N=2n-1$.  If $N$ is even, the mode is odd and may be called ${\rm r}_n^{\rm o}$, where $N=2n-2$.  The r modes have $\nabla\!\cdot\!\bxi=0$ at leading order, and their limiting behaviour depends as much on buoyancy as on inertial forces.  In the marginally stable case $\gamma=\Gamma$, the r modes have $\omega^2=O\left((k\zeta_{\rm s})^{-2}\right)$ and the eigenfunctions are not localized.

These asymptotic results are all verified numerically, as shown in Fig.~2.  It should be emphasized that this classification scheme is based entirely on the behaviour of the modes for large values of $k\zeta_{\rm s}$ and may not accurately reflect the properties of the modes for smaller $k\zeta_{\rm s}$.

\beginfigure{2}
\centerline{\epsfysize=8cm\epsfbox{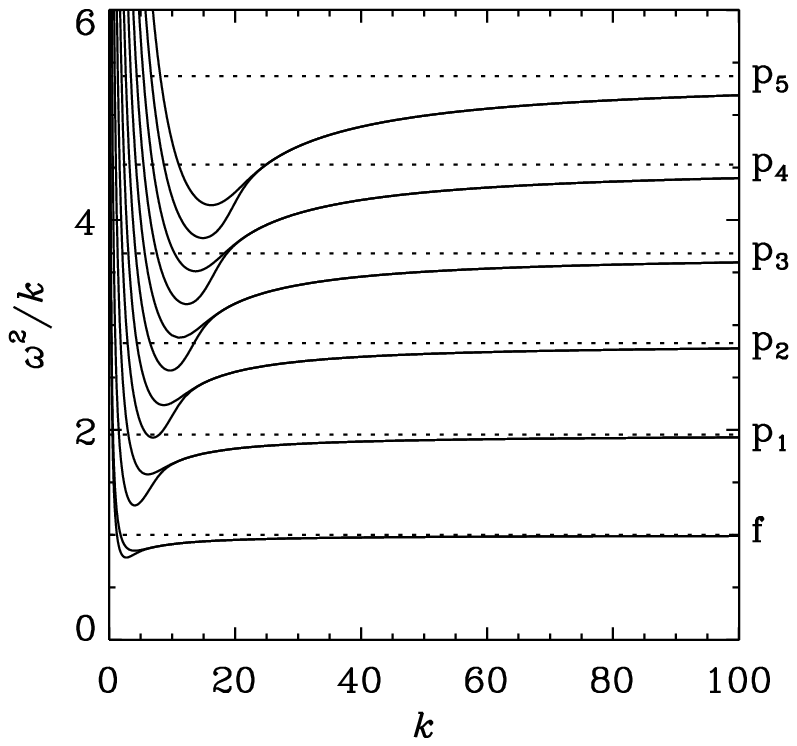}\qquad\qquad\epsfysize=8cm\epsfbox{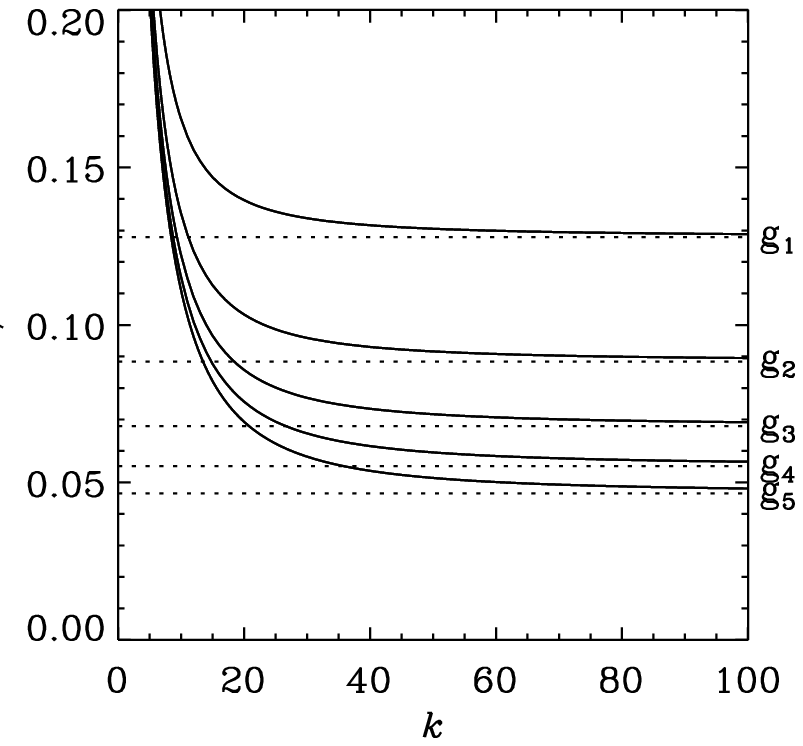}}
\centerline{\epsfysize=8cm\epsfbox{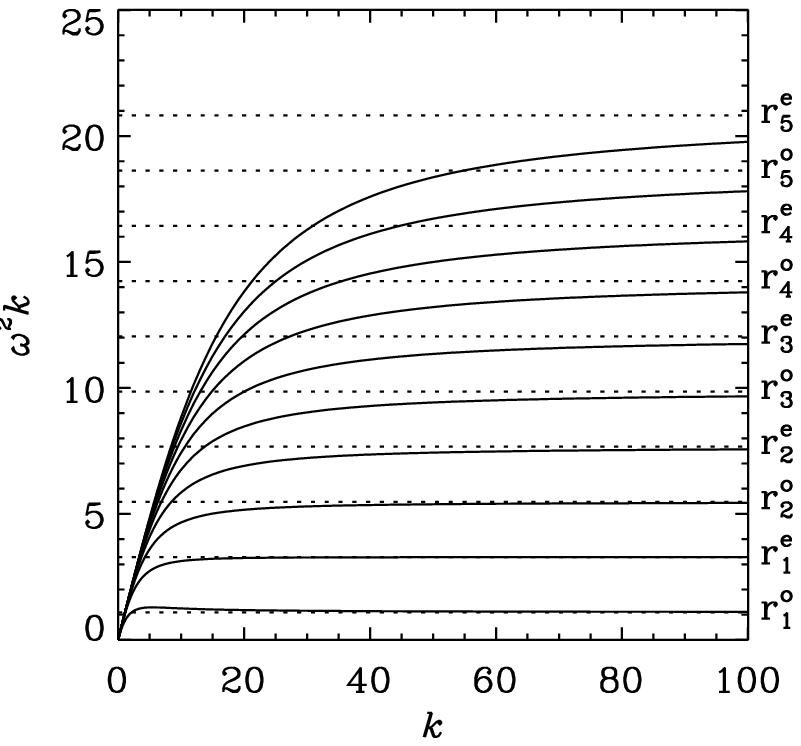}}
\caption{{\bf Figure~2.}  Asymptotic behaviour of the local dispersion relation for large radial wavenumber $k$, for the same disc as in Fig.~1.  The f modes and p modes (top left) and the g modes (top right) have $\omega^2=O(k)$ as $k\to\infty$, while the r modes (bottom) have $\omega^2=O(k^{-1})$.  In each case the dotted lines indicate the asymptotic limits derived in the text.  Note that the vertical axis is different in each plot.}
\endfigure

\subsection{Magnetized discs}

The presence of a poloidal magnetic field in the disc has profound implications for the spectrum of waves and instabilities.  As well as modifying the modes that occur in a hydrodynamic disc, the magnetic field gives rise to additional modes which are potentially unstable.  Furthermore, the branches are no longer separated in the dispersion diagram, but undergo avoided crossings.  This makes the classification of modes difficult, although there is one limit which can be investigated semi-analytically.  This is the case $k=0$ in a disc with a purely vertical magnetic field, as was considered by Gammie \& Balbus (1994).

When the magnetic field is purely vertical, the Lorentz force vanishes and equations (5.1)--(5.3) for the equilibrium are valid.  If also $k=0$, equations (3.13)--(3.15) simplify considerably because the horizontal and vertical components of $\bxi$ become decoupled.  There exist purely horizontal modes, with
$$\xi_r(r,\zeta)=a_r(r)u_N(r,\zeta)\eqno(5.29)$$
and
$$\xi_\phi(r,\zeta)=a_\phi(r)u_N(r,\zeta),\eqno(5.30)$$
where $u_N$ is one of the eigenfunctions of equation (4.1), and the components $a_r$ and $a_\phi$ satisfy the equation
$$\left[\matrix{\omega^2+3\Omega^2-\omega_N^2&-2{\rm i}\omega\Omega\cr 2{\rm i}\omega\Omega&\omega^2-\omega_N^2\cr}\right]\left[\matrix{a_r\cr a_\phi\cr}\right]=\left[\matrix{0\cr 0\cr}\right].\eqno(5.31)$$
The frequency eigenvalues are the roots of the equation
$$\omega^4-(2\omega_N^2+\Omega^2)\omega^2+\omega_N^2(\omega_N^2-3\Omega^2)=0,\eqno(5.32)$$
viz.
$$\omega^2=\omega_N^2+{\textstyle{1\over2}}\Omega^2\left[1\pm\left(1+16\omega_N^2/\Omega^2\right)^{1/2}\right],\eqno(5.33)$$
so that there is an unstable mode if and only if $0<\omega_N^2<3\Omega^2$.  Since the eigenvalues $\{\omega_N^2\}$ are ordered, the $N=1$ mode is unstable if any mode is unstable.  The maximum possible growth rate is the Oort parameter $A={\textstyle{3\over4}}\Omega$, which is achieved if $\omega_N^2={\textstyle{15\over16}}\Omega^2$.  In the limit $B_z\to0$, the eigenvalues $\omega_N^2\to0$ (although not uniformly with respect to $N$), and equation (5.33) reduces to
$$\omega^2=\Omega^2\quad\hbox{or}\quad0.\eqno(5.34)$$
Accordingly, the lesser solution of equation (5.33) may be called an m mode, since it is due entirely to the magnetic field.  Specifically, it is ${\rm m}_n^{\rm e}$ if $N=2n$, or ${\rm m}_n^{\rm o}$ if $N=2n-1$.  If $N=0$ the mode is trivial.  The greater solution of equation (5.33) may be called ${\rm r}_n^{\rm e}$ if $N=2n$, ${\rm r}_n^{\rm o}$ if $N=2n-1$, or ${\rm f}^{\rm e}$ if $N=0$.  However, the r and g modes really lose their identity when a magnetic field is introduced.

There are also purely vertical modes, which satisfy the equation
$$(\zeta_{\rm s}^2-\zeta^2){\partial^2\xi_z\over\partial\zeta^2}-(1+2q)\zeta{\partial\xi_z\over\partial\zeta}+N(N+2q)\xi_z=0,\eqno(5.35)$$
where
$$q={\Gamma+1\over2(\Gamma-1)}\eqno(5.36)$$
and
$$N(N+2q)={1\over\gamma}\left({2\Gamma\over\Gamma-1}\right)\left({\omega^2-\Omega^2\over\Omega^2}\right),\eqno(5.37)$$
Appropriate solutions, regular at $\zeta=\pm\zeta_{\rm s}$, are obtained when $N$ is a non-negative integer, in which case
$$\xi_z\propto C_N^q(\zeta/\zeta_{\rm s}),\eqno(5.38)$$
where $C_N^q$ is a Gegenbauer polynomial (e.g. Erd\'elyi et~al. 1953b).  Equation (5.37) then leads to the dispersion relation
$$\omega^2/\Omega^2=1+\gamma\left({\Gamma-1\over2\Gamma}\right)N\left[N+\left({\Gamma+1\over\Gamma-1}\right)\right].\eqno(5.39)$$
These modes may be identified as ${\rm p}_n^{\rm e}$ if $N=2n-1$, ${\rm p}_n^{\rm o}$ if $N=2n$, or ${\rm f}^{\rm o}$ if $N=0$.\note{$^7$}{In fact the ${\rm f}^{\rm o}$ and ${\rm r}_1^{\rm o}$ modes in a Keplerian disc without a magnetic field involve both horizontal and vertical displacements even when $k=0$.}

An example of the dispersion diagram for a disc containing a purely vertical magnetic field is shown in Fig.~3.  In this case the strength of the magnetic field is such that one unstable magnetorotational mode of each parity exists.  These modes are unstable only for sufficiently small $k$, and their growth rates are greatest for $k=0$.  The stable modes form a complicated dispersion diagram because of the avoided crossings that occur between different branches.  Broadly speaking, the pattern consists of branches which, like the f and p modes in a hydrodynamic disc, sweep upwards in the diagram, and other branches which are almost flat.  Where these would cross over each other, a closer examination reveals that each branch is in fact continuous; the two branches approach and then diverge without crossing.  The character of the eigenfunctions is exchanged smoothly in the neighbourhood of these avoided crossings.  This means that a classification of the modes based on continuity with $k=0$ would not be meaningful.  The reason for plotting even and odd modes separately is that avoided crossings occur only between modes of equal parity.  The fact that parity is a discrete property which cannot be exchanged smoothly implies that modes of opposite parity cannot interact in this way, and so branches of even and odd modes cross freely over one another.  Similar avoided crossings in the spectrum of a magnetized isothermal atmosphere (without the symmetry of reflection in a horizontal plane) have been analysed by Hasan \& Christensen-Dalsgaard (1992).

\beginfigure{3}
\centerline{\epsfysize=8cm\epsfbox{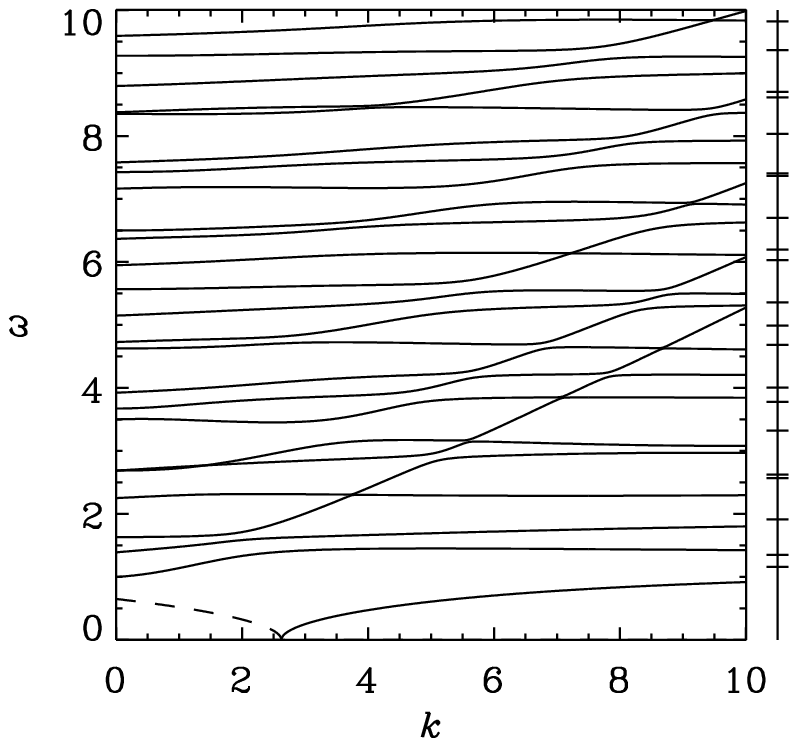}\qquad\qquad\epsfysize8cm\epsfbox{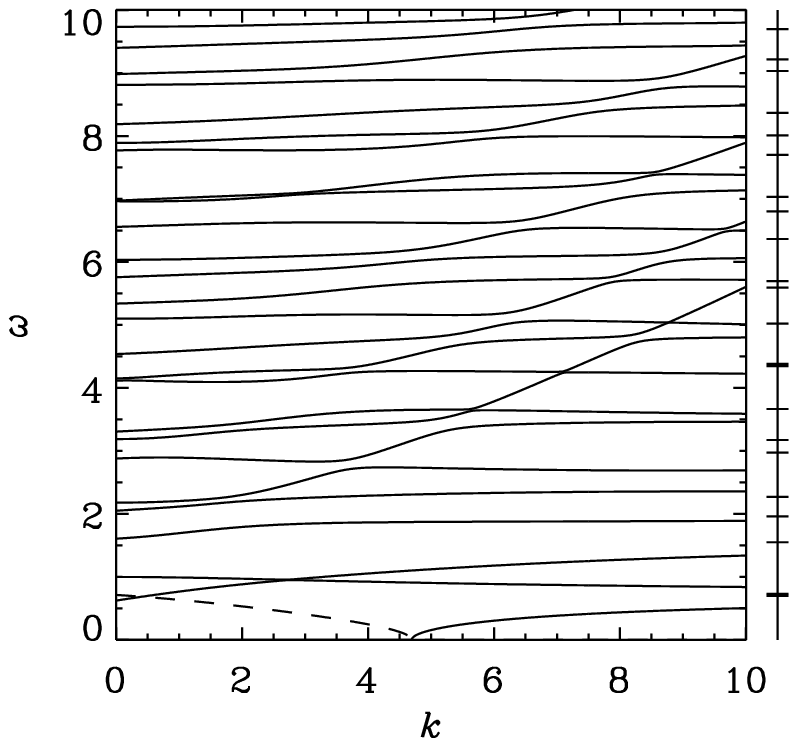}}
\caption{{\bf Figure~3.}  Part of the local dispersion relation for a thin disc containing a purely vertical magnetic field.  The parameters are $\Gamma=4/3$, $\gamma=5/3$ and $B_z=0.01$.  Modes with even and odd symmetry are shown in the left and right panels, respectively.  The frequency eigenvalues of various branches of modes, in units of the local angular velocity, are plotted against the radial wavenumber, in units of $H^{-1}$.  A dashed line indicates an unstable mode for which $\omega$ is imaginary.  In this case one unstable magnetorotational mode of each parity exists.  A maximum growth rate of $0.7151\Omega$ is achieved at $k=0$ by the ${\rm m}_1^{\rm o}$ mode.  The frequency eigenvalues of the continuous spectrum, which are all real, are marked on the scale at the right of each panel; these are the eventual limits of the branches as $k\to\infty$.}
\endfigure

When the magnetic field is weaker, more unstable modes exist but their branches fit neatly without mode interactions, as shown in Fig.~4.  It is well known that the addition of a weak magnetic field to a hydrodynamic disc constitutes a highly singular perturbation if ideal MHD is assumed.  This is reflected in the fact that the convergence of the frequency eigenvalues of ${\rm m}_n$ modes to zero as ${\bmath B}\to{\bf0}$ is not uniform with respect to $n$.  Modes of large $n$ clutter the dispersion diagram even when the magnetic field is exceedingly weak, and can be unstable with large growth rates.

\beginfigure{4}
\centerline{\epsfysize=8cm\epsfbox{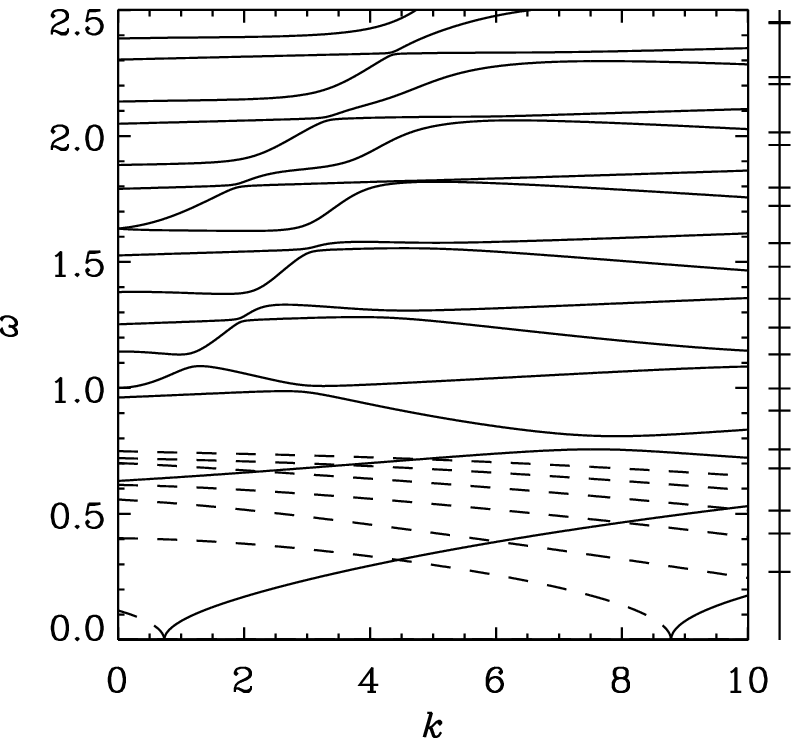}\qquad\qquad\epsfysize=8cm\epsfbox{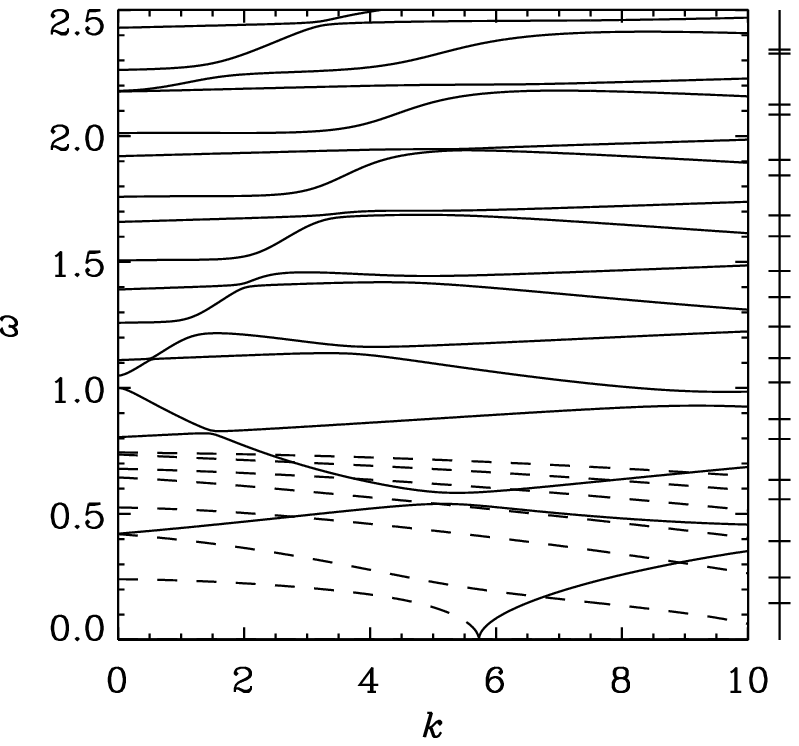}}
\caption{{\bf Figure~4.}  Part of the local dispersion relation for a thin disc containing a purely vertical magnetic field.  The parameters are $\Gamma=4/3$, $\gamma=5/3$ and $B_z=0.002$.  Modes with even and odd symmetry are shown in the left and right panels, respectively.  In this case seven unstable magnetorotational modes of each parity exist.  A maximum growth rate of $0.7495\Omega$ is achieved at $k=0$ by the ${\rm m}_4^{\rm e}$ mode.  The frequency eigenvalues of the continuous spectrum, which are all real, are marked on the scale at the right of each panel.}
\endfigure

Finally, an example of the dispersion diagram for a disc containing a bending poloidal magnetic field is shown in Fig.~5.  The avoided crossings are wider than for a purely vertical magnetic field, which implies that the coupling between different branches of modes is stronger when the field lines bend.

\beginfigure{5}
\centerline{\epsfysize=8cm\epsfbox{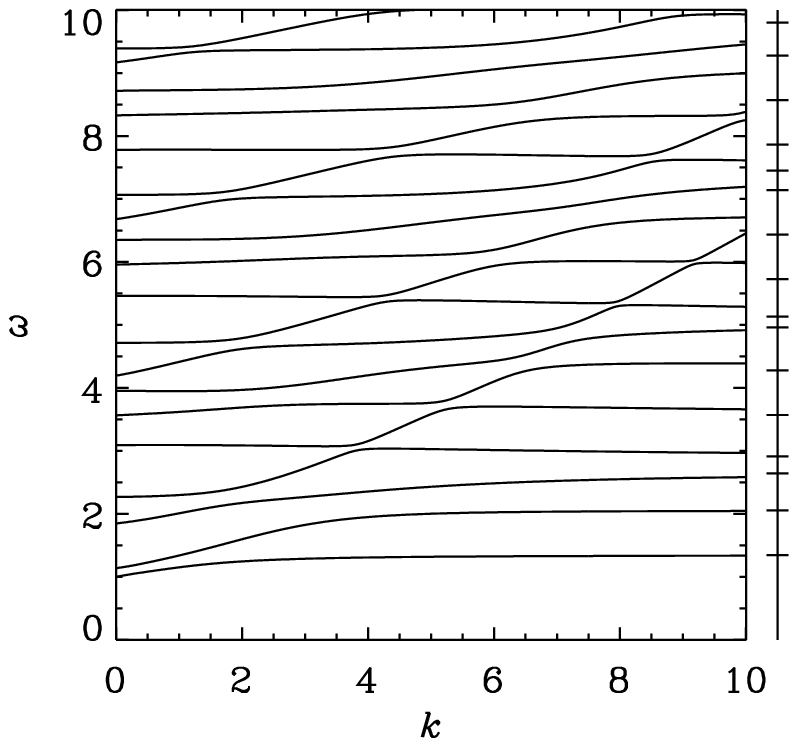}\qquad\qquad\epsfysize=8cm\epsfbox{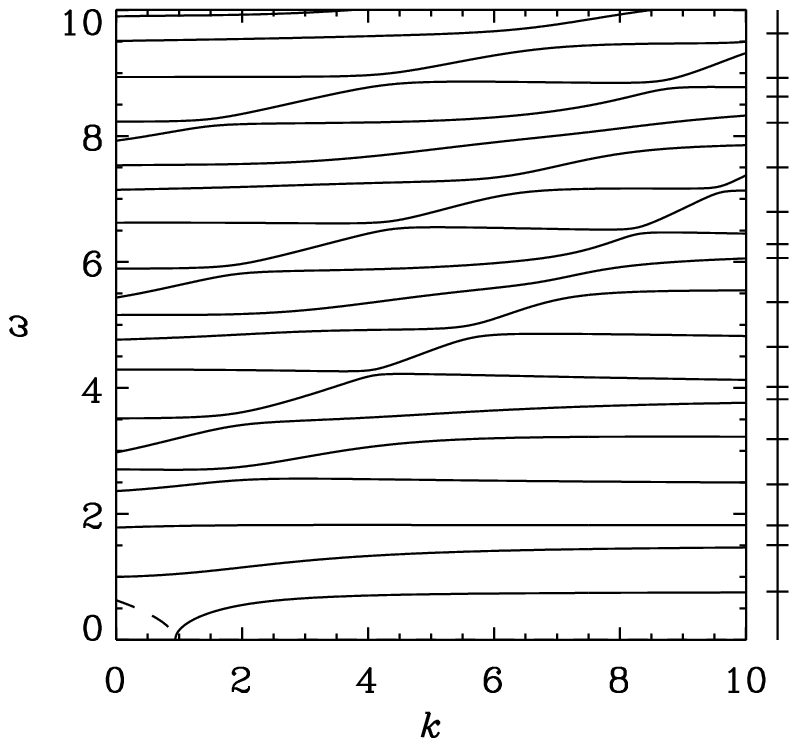}}
\caption{{\bf Figure~5.}  Part of the local dispersion relation for a thin disc containing a bending poloidal magnetic field.  The parameters are $\Gamma=4/3$, $\gamma=5/3$, $B_z=0.015$ and $B_{r{\rm s}}=0.005$, with $\Omega_{1{\rm s}}\approx0.3376$.  Modes with even and odd symmetry are shown in the left and right panels, respectively.  In this case only the ${\rm m}_1^{\rm o}$ mode is unstable, achieving a maximum growth rate of $0.6019\Omega$ at $k=0$.  The frequency eigenvalues of the continuous spectrum, which are all real, are marked on the scale at the right of each panel.}
\endfigure

The behaviour of branches in the limit $k\to\infty$ is quite different from the hydrodynamic case.  If any one branch is followed continuously, it undergoes a finite number of avoided crossings and then $\omega$ approaches a finite limit.  It can be shown that these limits are the frequency eigenvalues of the continuous spectrum discussed in Section~2.  In this limit the Lagrangian displacement is confined entirely within the magnetic surface and the magnetorotational instability, which is associated with the gradient of angular velocity perpendicular to the magnetic field, is lost.

\section{Stability criteria for magnetized discs}

The stability of a weakly magnetized thin disc to axisymmetric perturbations can be decided, in principle, by computing the local dispersion relation at each radius separately; the disc is unstable if and only if an imaginary frequency eigenvalue exists, for any real value of $k$, at any radius.  However, a more efficient algorithm is required in practice.  In the case of polytropic discs, the parameter space divides into stable and unstable regions separated by a curve of marginal stability.  The location of the marginal curve depends on a further parameter which is the adiabatic exponent $\gamma$.  Equilibria on the marginal curve possess a mode with $\omega=0$, but this is not sufficient to determine the curve, since all unstable equilibria also possess modes with $\omega=0$ (cf. Figs~3--5).  In principle, a marginal curve can be drawn for each m mode and for every value of $k$.

Before describing the marginal curves, it is appropriate to give a more detailed account of the solutions of the equilibrium equations for polytropic discs.  When $B_{r{\rm s}}$ and $B_z$ are specified, equations (3.1)--(3.5) constitute a non-linear eigenvalue problem for $\Omega_{1{\rm s}}$.  The solutions of these equations lie on a two-dimensional manifold in the three-dimensional parameter space of $(B_{r{\rm s}},B_z,\Omega_{1{\rm s}})$.  Some of these solutions are physically acceptable and may be called `regular' equilibria: the magnetic field lines bend only once when passing through the disc, and the density and pressure decrease monotonically from the equatorial plane to the surface.  In the remaining solutions, which may be called `irregular' equilibria, the field lines bend more than once; some of these solutions also have density inversions.  In Paper~I only the class of regular equilibria (described as the `principal branch') for the case $\Gamma=5/3$ was mentioned.  It is now to be shown that all irregular equilibria are unstable, so that the marginal curves need be drawn only on the principal branch.
\smallskip
\noindent{\sl Theorem 1}.\quad All irregular equilibria possess an unstable mode of odd symmetry at $k=0$.
\smallskip
\noindent{\sl Proof}.\quad When $k=0$, the form of the functional $R[\xi_r,\xi_z;k]$ for purely horizontal trial displacements of odd symmetry is, from equation (4.13),
$$R[\xi_r,0;0]=\int_{-\zeta_{\rm s}}^{\zeta_{\rm s}}\left(B_z^2\left|{\partial\xi_r\over\partial\zeta}\right|^2-3\rho\Omega^2|\xi_r|^2\right)\,{\rm d}\zeta,\eqno(6.1)$$
and the trial function must satisfy $\partial\xi_r/\partial\zeta=0$ at $\zeta=\pm\zeta_{\rm s}$.  Given that the equilibrium is irregular, $B_r$ must have zeros, other than at $\zeta=0$, at points $\zeta=\pm\zeta_n$, where $0<\zeta_1<\zeta_2<\cdots<\zeta_N<\zeta_{\rm s}$ and $N\geq1$.  Then consider the Sturm--Liouville equation
$$B_z^2{\partial^2y\over\partial\zeta^2}+\lambda\rho y=0\eqno(6.2)$$
on $-\zeta_1<\zeta<\zeta_1$, subject to the boundary conditions
$${\partial y\over\partial\zeta}=0\qquad\hbox{at}\quad\zeta=\pm\zeta_1.\eqno(6.3)$$
Since the equation is symmetrical about $\zeta=0$, the eigenfunctions $y_0$, $y_1$, $y_2$, \dots\ alternate between even and odd symmetry, and the eigenvalues $\lambda_0$, $\lambda_1$, $\lambda_2$, \dots\ form an ordered, increasing sequence of distinct, non-negative real numbers.  The lowest eigenvalue is $\lambda_0=0$, corresponding to $y_0=\hbox{constant}$.  Now it follows from equations (3.1)--(3.4) for the equilibrium that $\Omega_1$ satisfies the equation
$$B_z^2{\partial^2\Omega_1\over\partial\zeta^2}+3\rho\Omega^2\Omega_1=0,\eqno(6.4)$$
and also the boundary conditions (6.3).  Moreover, since $\zeta_1$ is the first zero of $B_r$ in $0<\zeta<\zeta_{\rm s}$, $B_r$ has exactly two extrema, and $\Omega_1$ has exactly two zeros, in the domain of the Sturm--Liouville equation.  It follows that $\Omega_1$ is the eigenfunction $y_2$, and has eigenvalue $\lambda_2=3\Omega^2$.  The properties of the Sturm--Liouville equation then ensure that the first odd eigenfunction $y_1$ has an eigenvalue satisfying the (strict) inequalities $0<\lambda_1<3\Omega^2$.  Now construct the trial function
$$f(\zeta)=\cases{y_1(\zeta),&$|\zeta|<\zeta_1$,\cr
{\rm sgn}(\zeta)y_1(\zeta_1),&$|\zeta|\geq\zeta_1$,\cr}\eqno(6.5)$$
which is continuous throughout $-\zeta_{\rm s}<\zeta<\zeta_{\rm s}$ and satisfies the boundary conditions for the variational principle.  After an integration by parts one obtains
$$R[f,0;0]=-2\int_0^{\zeta_1}(3\rho\Omega^2-\lambda_1)|f|^2\,{\rm d}\zeta-2\int_{\zeta_1}^{\zeta_{\rm s}}3\rho\Omega^2|f|^2\,{\rm d}\zeta,\eqno(6.6)$$
and, since this is clearly negative, the theorem is proved.

In Fig.~6, the different branches of equilibria, now for the case $\Gamma=4/3$, are shown in a number of suitably chosen sections through the parameter space.  It is found that, as $B_z$ is reduced, an increasingly large number of branches of irregular equilibria appear as the solution manifold folds and divides into many leaves.  There is a critical value of $B_z$, approximately 0.01283, below which the principal branch becomes separated from the `vertical' solution (which has $B_r=0$ and $\Omega_{1{\rm s}}=0$) by a stretch of irregular equilibria.  (The vertical solution, however, is technically always regular.)  As $B_z$ is reduced further, a bifurcation disconnects the principal branch from the vertical solution.  An infinite number of further bifurcations occur as $B_z\to0$.

\beginfigure{6}
\centerline{\epsfysize=5cm\epsfbox{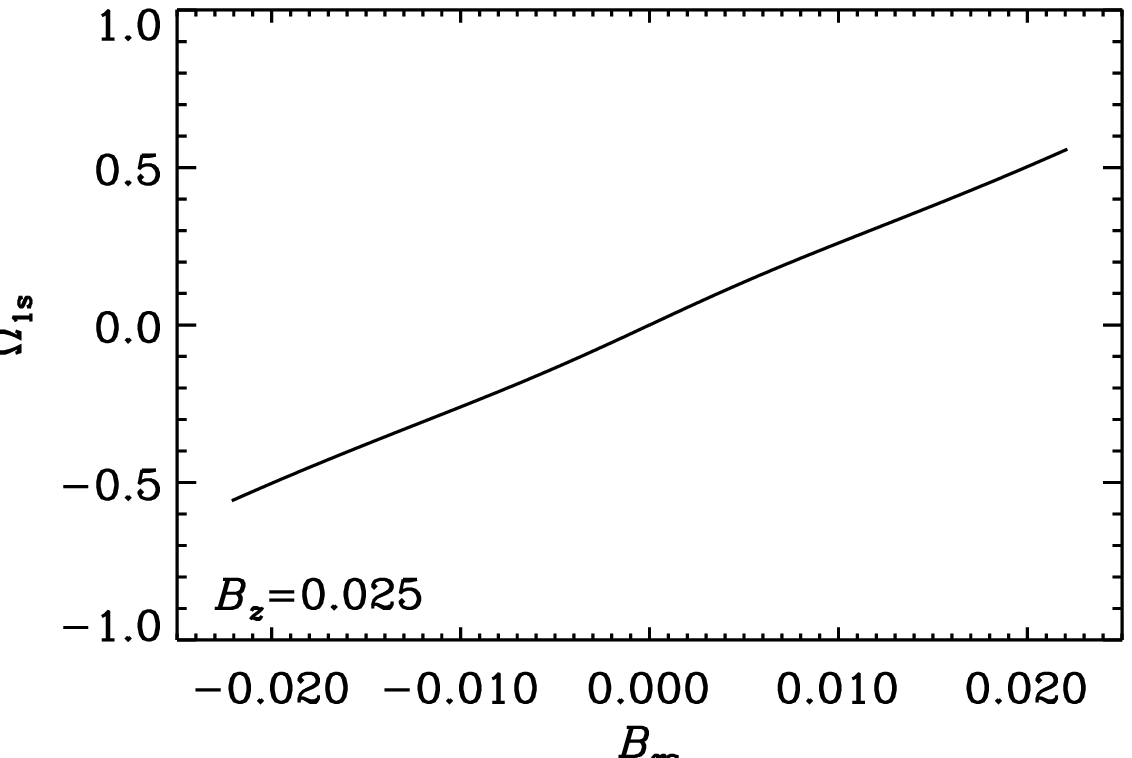}\qquad\qquad\epsfysize=5cm\epsfbox{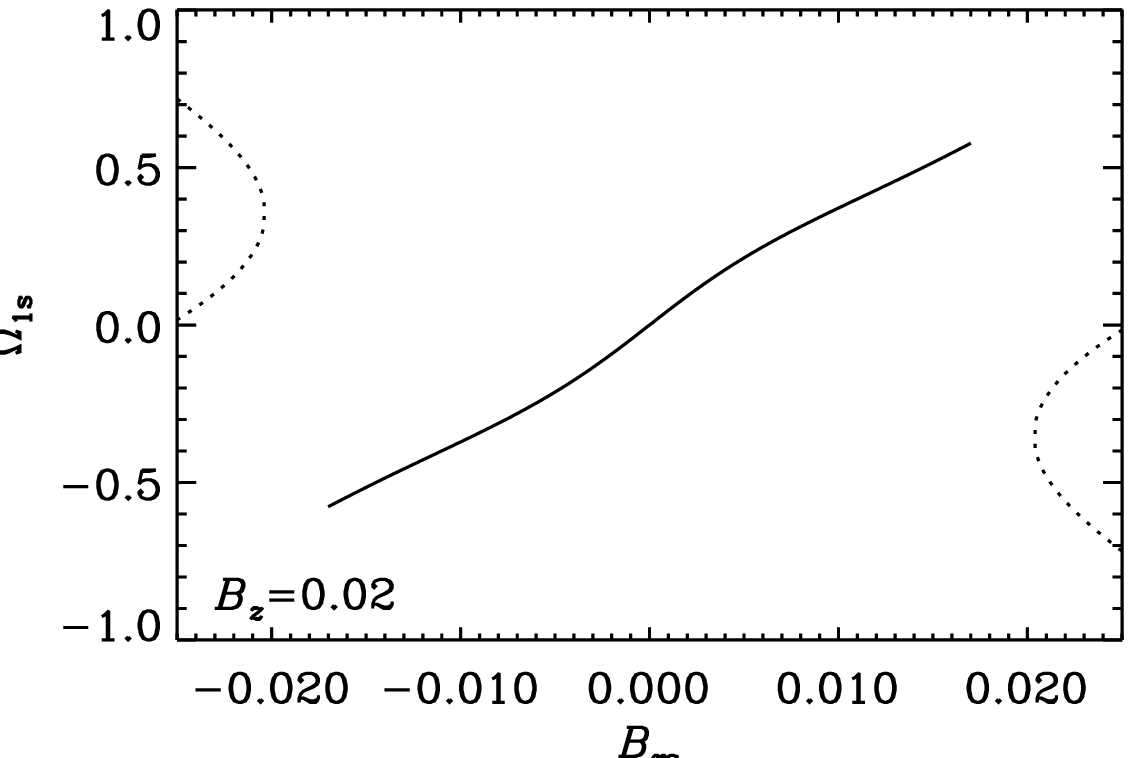}}
\centerline{\epsfysize=5cm\epsfbox{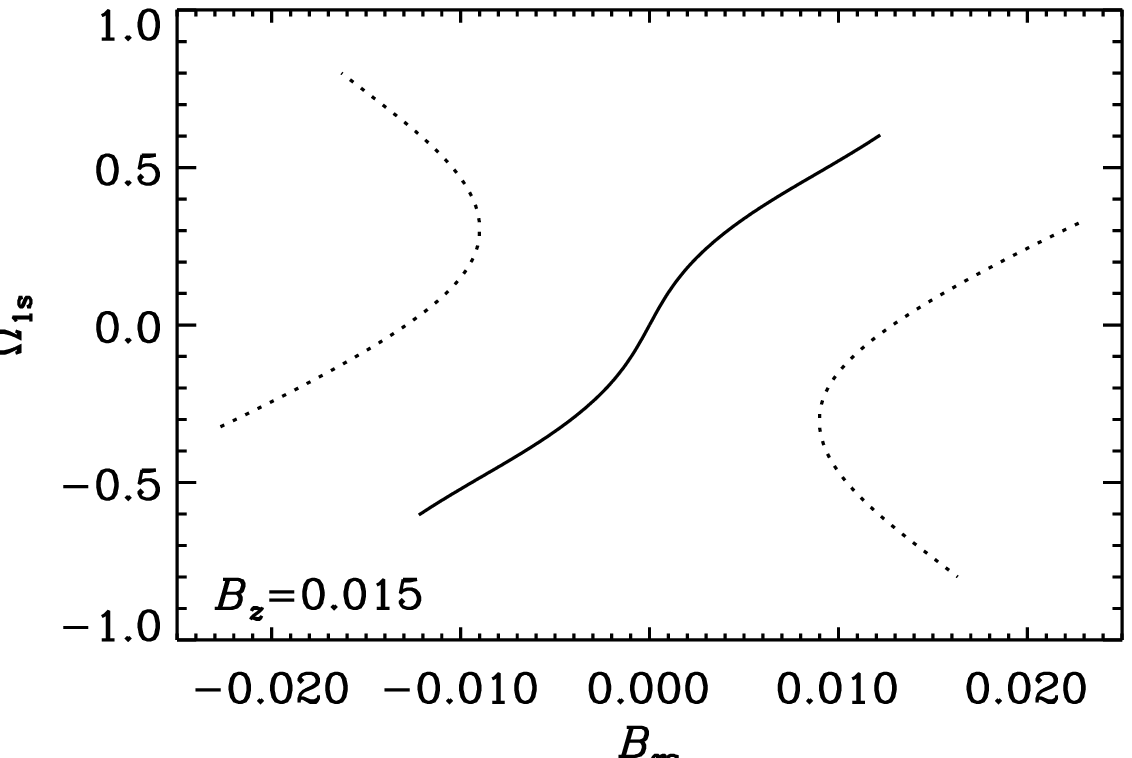}\qquad\qquad\epsfysize=5cm\epsfbox{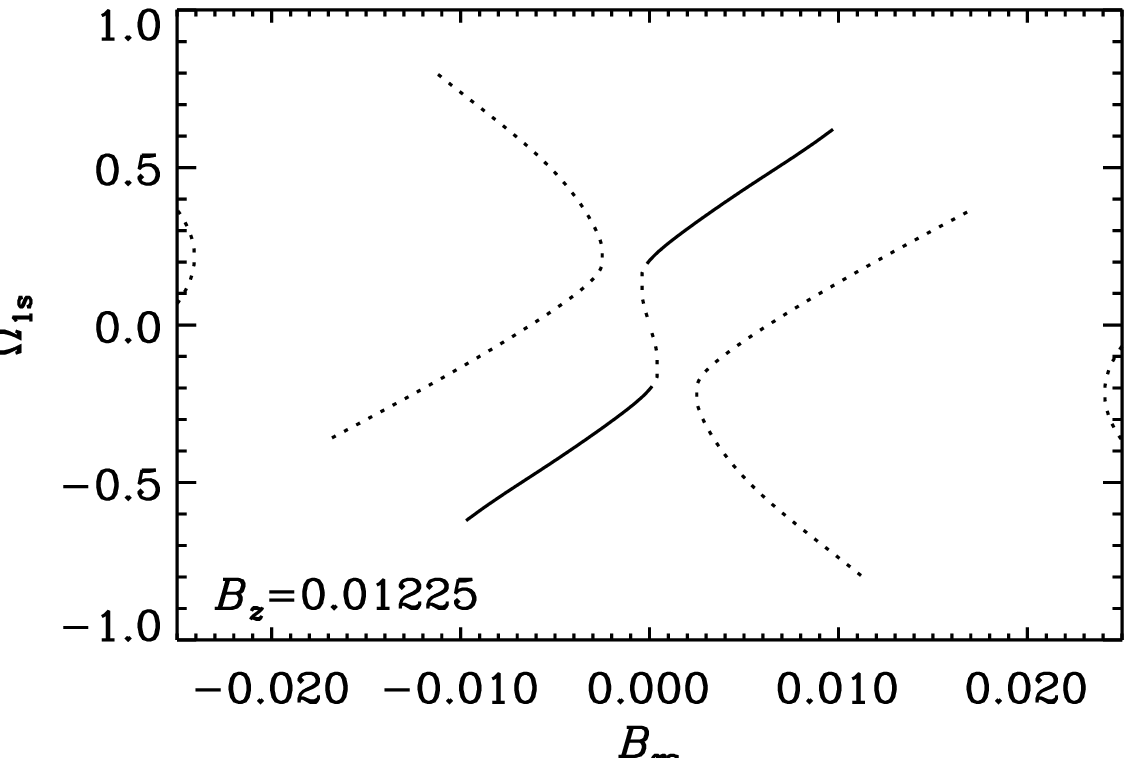}}
\centerline{\epsfysize=5cm\epsfbox{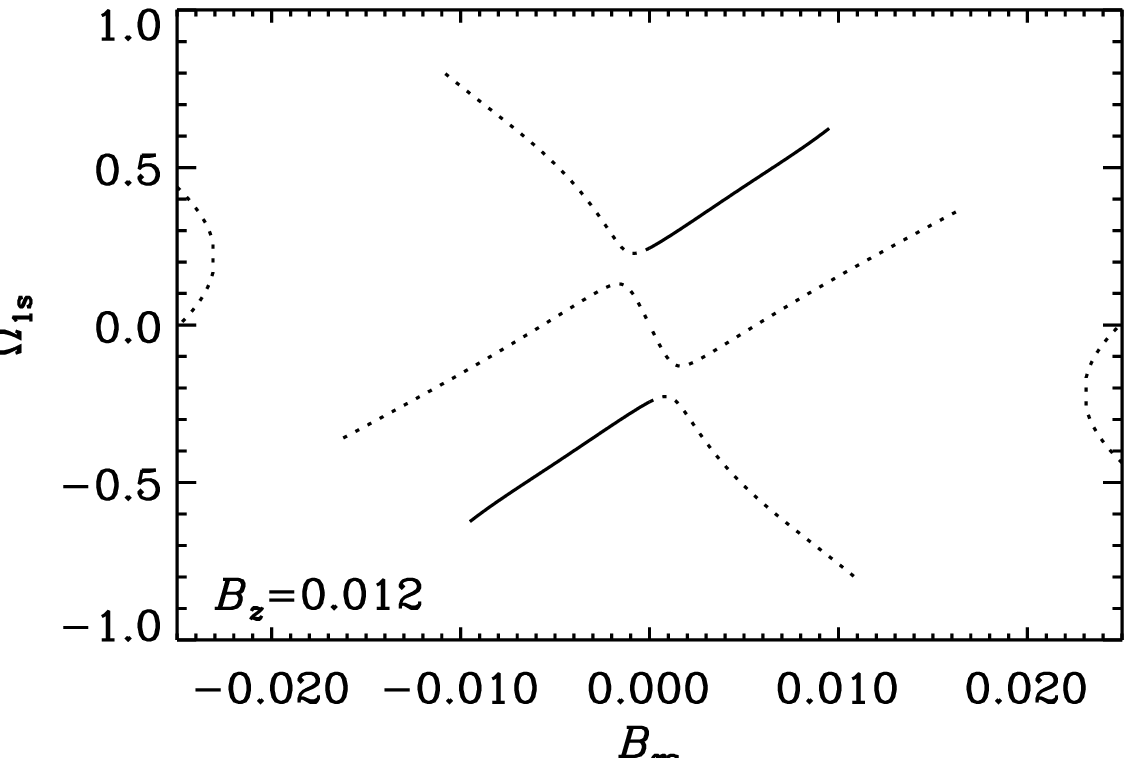}\qquad\qquad\epsfysize=5cm\epsfbox{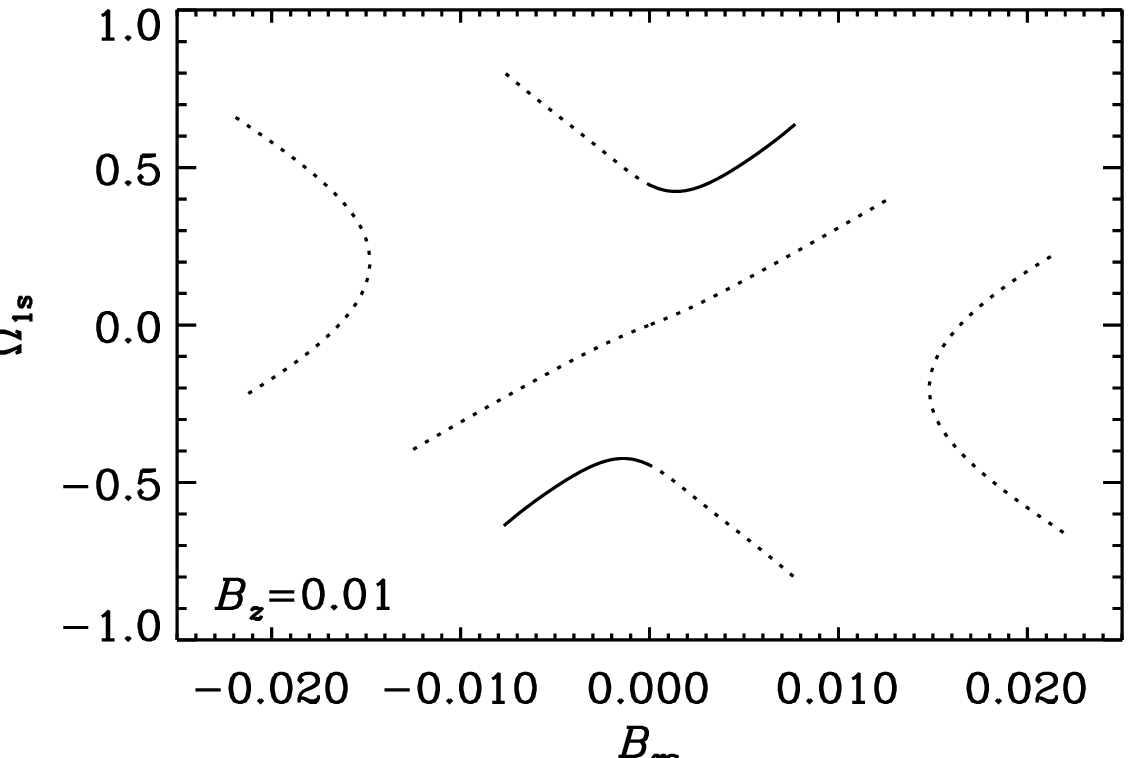}}
\centerline{\epsfysize=5cm\epsfbox{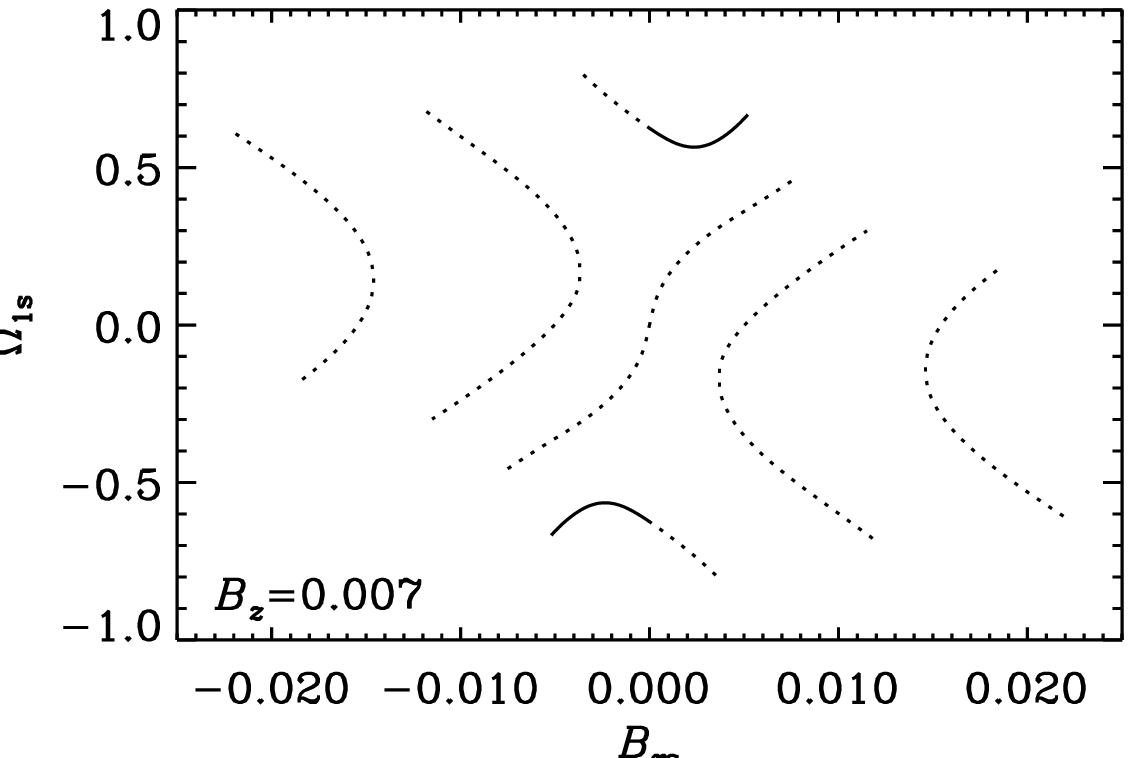}\qquad\qquad\epsfysize=5cm\epsfbox{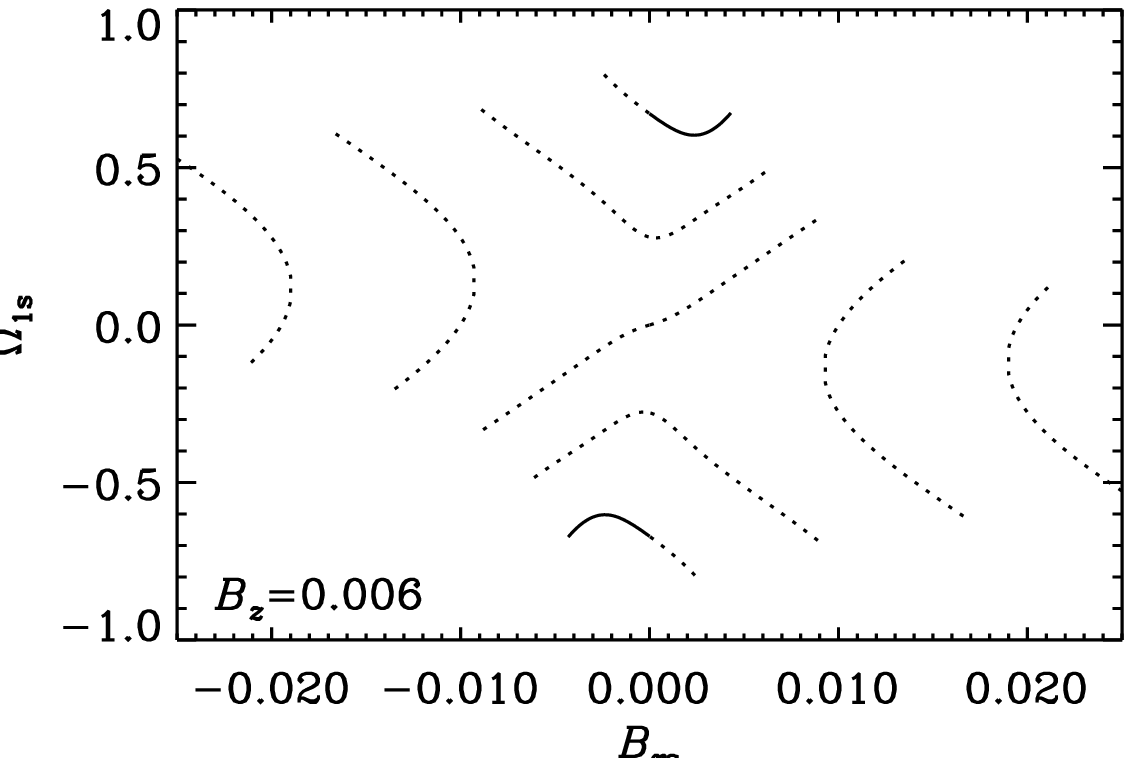}}
\caption{{\bf Figure~6.}  Solution curves for weakly magnetized, polytropic thin discs with $\Gamma=4/3$.  The `eigenvalue' $\Omega_{1{\rm s}}$ is plotted against $B_{r{\rm s}}$ for various values of $B_z$ (as indicated in each panel).  All solutions in the range $-0.025<B_{r{\rm s}}<0.025$ are shown.  Regular and irregular equilibria are indicated by solid and dotted lines, respectively.  As $B_z$ is decreased, the solution manifold folds and divides into many leaves, and the branch of regular equilibria becomes separated from the solution at the origin which has a purely vertical magnetic field.}
\endfigure

It is observed in the numerical calculations that, for regular equilibria, the last mode to be stabilized is the ${\rm m}_1^{\rm o}$ mode, and it is stabilized last at $k=0$.  This supports the following conjecture.
\smallskip
\noindent{\sl Conjecture}.\quad If a regular equilibrium possesses an unstable mode at any value of $k$, then it possesses an unstable mode at $k=0$, assuming that $\gamma\geq\Gamma$.\note{$^8$}{When $\gamma<\Gamma$ there may be a (magneto-) convective instability and the result cannot be expected to hold.  In fact, a condition differing slightly from $\gamma\geq\Gamma$ may be required in order to prove the conjecture.}
\smallskip\noindent While this conjecture has not yet been proved in the most general circumstances envisaged, the following two theorems support it strongly.
\smallskip
\noindent{\sl Theorem 2}.\quad The conjecture holds in the case of an incompressible fluid, provided there is no density inversion.
\smallskip
\noindent{\sl Proof}.\quad An appropriate version of the functional $R[\xi_r,\xi_z;k]$ for an incompressible fluid is
$$R[\xi_r,\xi_z;k]=4\Omega^2|a_0|^2+\int_{-\infty}^\infty\left(|\GIOscriptD\xi_r|^2+|\GIOscriptD\xi_z|^2-3\rho\Omega^2|\xi_r|^2-\Omega^2\zeta{\partial\rho\over\partial\zeta}|\xi_z|^2\right)\,{\rm d}\zeta,\eqno(6.7)$$
where the trial displacement $\bxi$ is defined on $-\infty<\zeta<\infty$ and is subject to the constraint $\Delta=0$.  Given that there exists an unstable eigenfunction $\hat\bxi(\zeta)$ at some wavenumber $\hat k\neq0$, it follows that
$$R[\hat\xi_r,\hat\xi_z;\hat k]<0.\eqno(6.8)$$
When $k=0$, the constraint $\Delta=0$ can be satisfied by using a purely horizontal trial displacement $\xi_r=f$, where the function $f(\zeta)$ is defined by
$$\hat\xi_r(\zeta)=f(\zeta)\exp\left[-(2\Omega_1/3\Omega){\rm i}\hat kr\right],\eqno(6.9)$$
so that
$$\GIOscriptD_{(\hat k)}\hat\xi_r={\rm i}\hat kB_r\hat\xi_r+B_z{\partial\hat\xi_r\over\partial\zeta}=B_z\left({\partial f\over\partial\zeta}\right)\exp\left[-(2\Omega_1/3\Omega){\rm i}\hat kr\right]=\left(\GIOscriptD_{(0)}f\right)\exp\left[-(2\Omega_1/3\Omega){\rm i}\hat kr\right],\eqno(6.10)$$
where the suffix on the operator $\GIOscriptD$ indicates the value of $k$ used.  Then
$$|\hat\xi_r|^2=|f|^2\qquad\hbox{and}\qquad|\GIOscriptD_{(\hat k)}\hat\xi_r|^2=|\GIOscriptD_{(0)}f|^2.\eqno(6.11)$$
Also, the coefficients $\{a_n\}$ in the eigenfunction expansion (4.8) are the same when $\xi_r=f$ and $k=0$ as they are when $\xi_r=\hat\xi_r$ and $k=\hat k$.  Thus the difference
$$R[\hat\xi_r,\hat\xi_z;\hat k]-R[f,0;0]=\int_{-\infty}^\infty\left(|\GIOscriptD_{(\hat k)}\hat\xi_z|^2-\Omega^2\zeta{\partial\rho\over\partial\zeta}|\hat\xi_z|^2\right)\,{\rm d}\zeta\eqno(6.12)$$
is non-negative, and so
$$R[f,0;0]\leq R[\hat\xi_r,\hat\xi_z;\hat k]<0,\eqno(6.13)$$
which proves that an unstable mode must exist with $k=0$.
\smallskip
\noindent{\sl Theorem 3}.\quad The conjecture holds for a compressible fluid when the magnetic field is purely vertical.
\smallskip
\noindent{\sl Proof}.\quad In this case one may take
$$\eqalignno{R[\xi_r,\xi_z;k]=&4\Omega^2|a_0|^2&\cr
&+\int_{-\infty}^\infty\left\{{|\delta\Pi|^2\over\gamma p+B^2}+B^2\left|{\partial\xi_r\over\partial\zeta}\right|^2+\left({\gamma p\over\gamma p+B^2}\right)B^2\left|{\partial\xi_z\over\partial\zeta}-\left({\rho\Omega^2\zeta\over\gamma p}\right)\xi_z\right|^2\right.&\cr
&\left.\qquad\qquad\qquad-3\rho\Omega^2|\xi_r|^2-\left[\Omega^2\zeta{\partial\rho\over\partial\zeta}+{(\rho\Omega^2\zeta)^2\over\gamma p}\right]|\xi_z|^2\right\}\,{\rm d}\zeta,&(6.14)\cr}$$
where
$$\delta\Pi=-(\gamma p+B^2)\Delta+\rho\Omega^2\zeta\xi_z+B^2{\partial\xi_z\over\partial\zeta}.\eqno(6.15)$$
Given that there exists an unstable eigenfunction $\hat\bxi(\zeta)$ at some wavenumber $\hat k\neq0$, it follows that
$$R[\hat\xi_r,\hat\xi_z;\hat k]<0.\eqno(6.16)$$
When $k=0$, the purely horizontal displacement $\xi_r=\hat\xi_r$ should be used, and $\delta\Pi$ then vanishes.  The coefficient of $|\xi_z|^2$ in the integrand of equation (6.14),
$$-\left[\Omega^2\zeta{\partial\rho\over\partial\zeta}+{(\rho\Omega^2\zeta)^2\over\gamma p}\right]=\left({\gamma-\Gamma\over\Gamma}\right){(\rho\Omega^2\zeta)^2\over\gamma p},\eqno(6.17)$$
is non-negative, since it is assumed that $\gamma\geq\Gamma$.  Thus
$$R[\hat\xi_r,0;0]\leq R[\hat\xi_r,\hat\xi_z;\hat k]<0,$$
which proves that an unstable mode must exist with $k=0$.

If the conjecture is accepted on the basis of these lemmata and the numerical evidence, it implies that the overall stability boundary in the parameter space is the marginal curve for a mode (in fact, the ${\rm m}_1^{\rm o}$ mode) at $k=0$, drawn on the principal branch.  This curve has been computed directly by solving the equations for an equilibrium (with a given value of $B_z$) and a mode (with $\omega=0$ and $k=0$) as an eigenvalue problem, yielding a value for $B_{r{\rm s}}$.  The stability boundaries for equilibria with $\Gamma=4/3$ and $\Gamma=5/3$ are shown in Fig.~7, using $\gamma=5/3$ for the adiabatic exponent.  In comparing the two panels it is important to note that the unit of magnetic field strength depends on $\Gamma$, according to equation (4.22).  It is more convenient, therefore, to compare a dimensionless quantity such as the plasma beta $\beta=2p/B^2$.  The plasma beta of the marginal equilibria, evaluated on the equatorial plane, is plotted against $B_z$ in Fig.~8.  It is seen that all marginal equilibria have values of $\beta$ reasonably close to unity.  However, $\beta$ is not generally a reliable guide to stability, especially for equilibria in which the angle of inclination of the magnetic field exceeds $\pi/6$.

The emergence and disappearance of equilibrium solutions as the parameters are varied is not necessarily governed by conventional bifurcation theory.  Typically, an equilibrium ceases to exist when the parameters are changed in such a way that would make the enthalpy become negative at some point.  This would correspond to a branch point of the polytropic relation (3.5), and non-analytic behaviour is to be expected.  In fact, the equilibrium may continue to exist in a complex-valued sense, even though this has no physical meaning.  This means that there is little hope of using the bifurcations of the solution manifold to infer the stability of the solutions.  Indeed, two further obstacles to such a method present themselves: first, only the stability to modes of even parity could be considered, since the equilibria are all symmetric; and secondly, the effect of the adiabatic exponent $\gamma$ on stability could not be taken into account.

The magnetorotational modes are not affected greatly by buoyancy or compressibility.  In Fig.~9 the stability boundaries for equilibria with $\Gamma=4/3$ and $\Gamma=5/3$ are plotted for different values of the adiabatic exponent.  The unstable region reduces slightly in size as $\gamma$ is increased from $\Gamma$ to $\infty$, demonstrating the mildly stabilizing effect of a sub-adiabatic stratification.  The effect is most pronounced for equilibria in which the magnetic field bends significantly, and is zero for those with a purely vertical magnetic field.

\beginfigure{7}
\centerline{\epsfysize=8cm\epsfbox{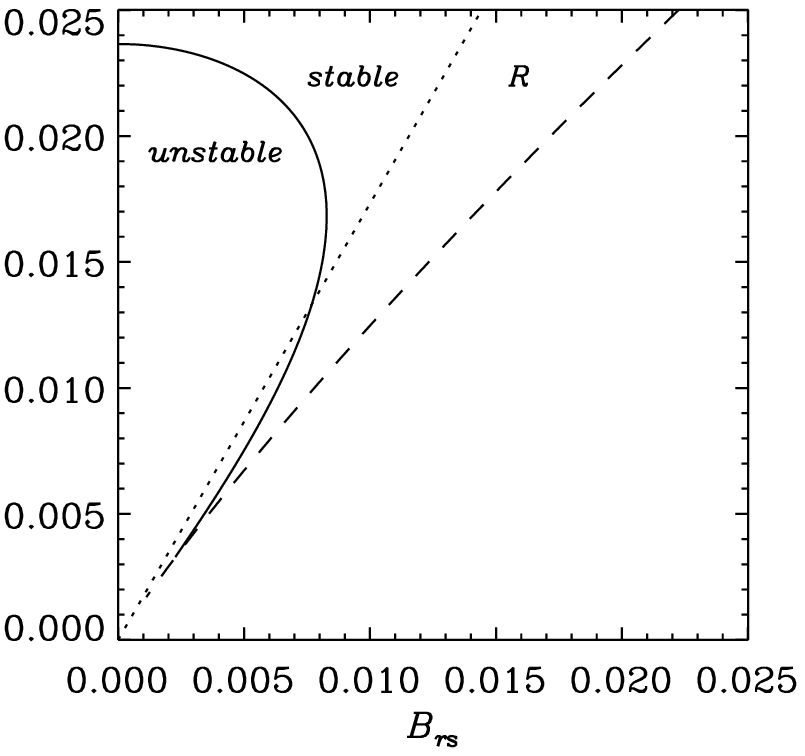}\qquad\qquad\epsfysize=8cm\epsfbox{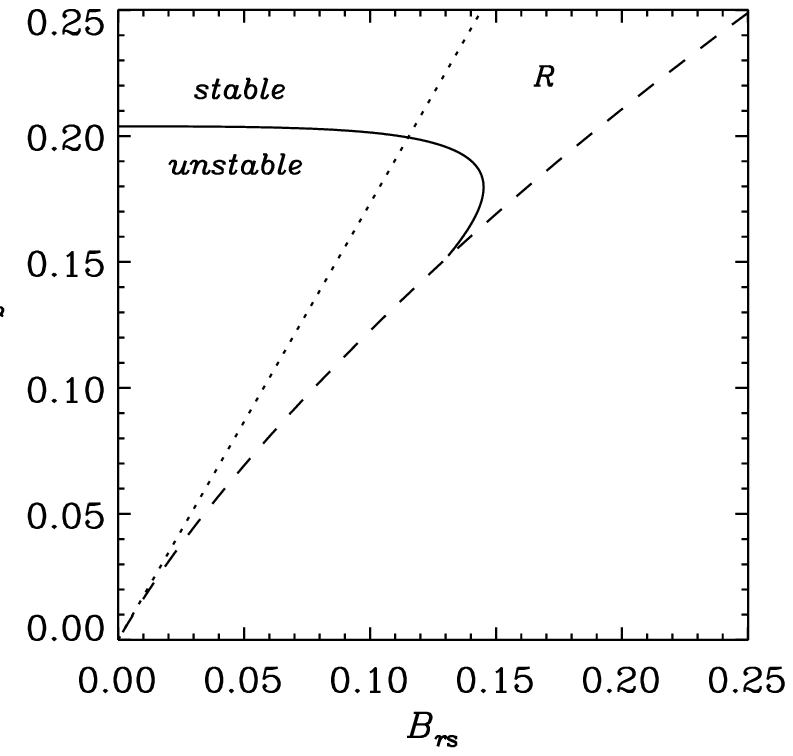}}
\caption{{\bf Figure~7.}  Stability boundaries in the parameter spaces for weakly magnetized, polytropic thin discs with $\Gamma=4/3$ (left) and $\Gamma=5/3$ (right).  In each case, the dotted line indicates an angle of inclination $i=\pi/6$.  The dashed line is the critical curve for the existence of the principal solution branch.  The solid line is the marginal curve, at $k=0$, for the ${\rm m}_1^{\rm o}$ mode, when $\gamma=5/3$.  This is the overall stability boundary of the equilibria to axisymmetric perturbations.  In the regions {\it R\/}, which extend indefinitely beyond the upper right of each figure, there exist equilibria which are capable of driving a wind and are also stable to the magnetorotational instability.  Note the different scales used in each plot.}
\endfigure

\beginfigure{8}
\centerline{\epsfysize=8cm\epsfbox{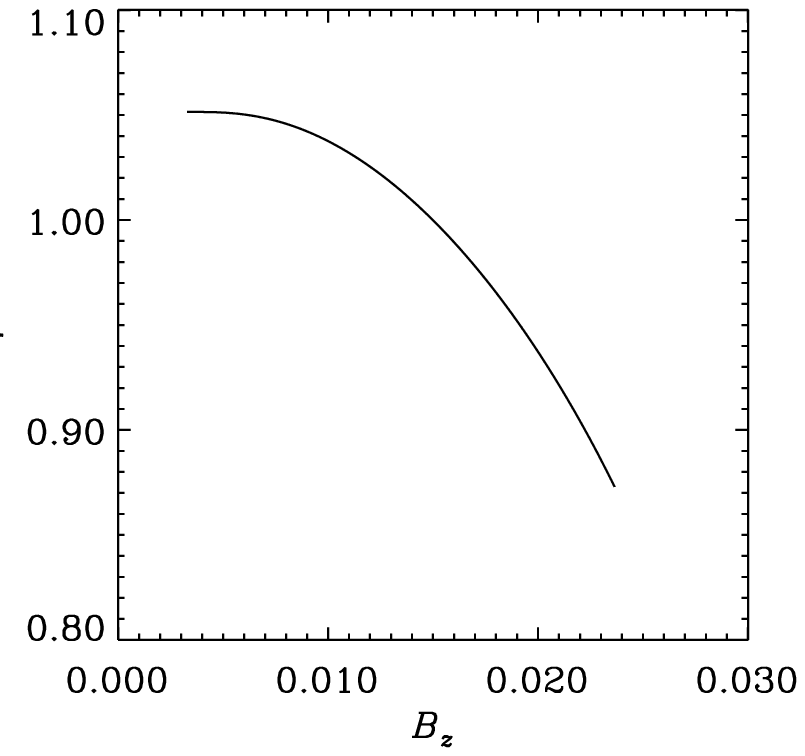}\qquad\qquad\epsfysize=8cm\epsfbox{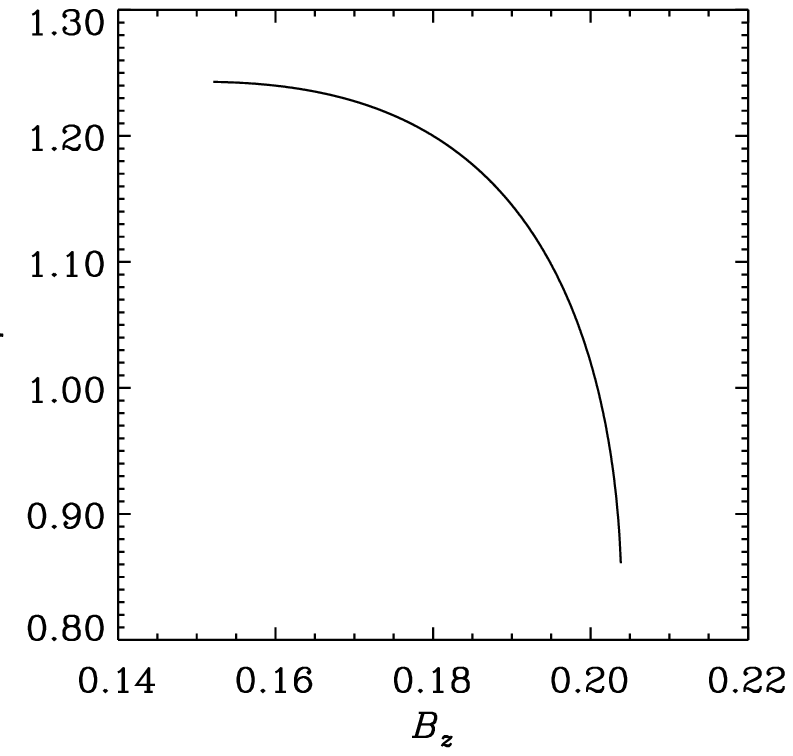}}
\caption{{\bf Figure~8.}  The plasma beta of the marginal equilibria, evaluated on the equatorial plane, plotted against the vertical magnetic field strength in the interval in which a marginal equilibrium exists.  The two panels correspond to the two panels of Fig.~7, i.e. equilibria with $\Gamma=4/3$ (left) and with $\Gamma=5/3$ (right), where $\gamma=5/3$ in each case.  Marginal equilibria typically have values of $\beta$ that are reasonably close to unity.}
\endfigure

\beginfigure{9}
\centerline{\epsfysize=8cm\epsfbox{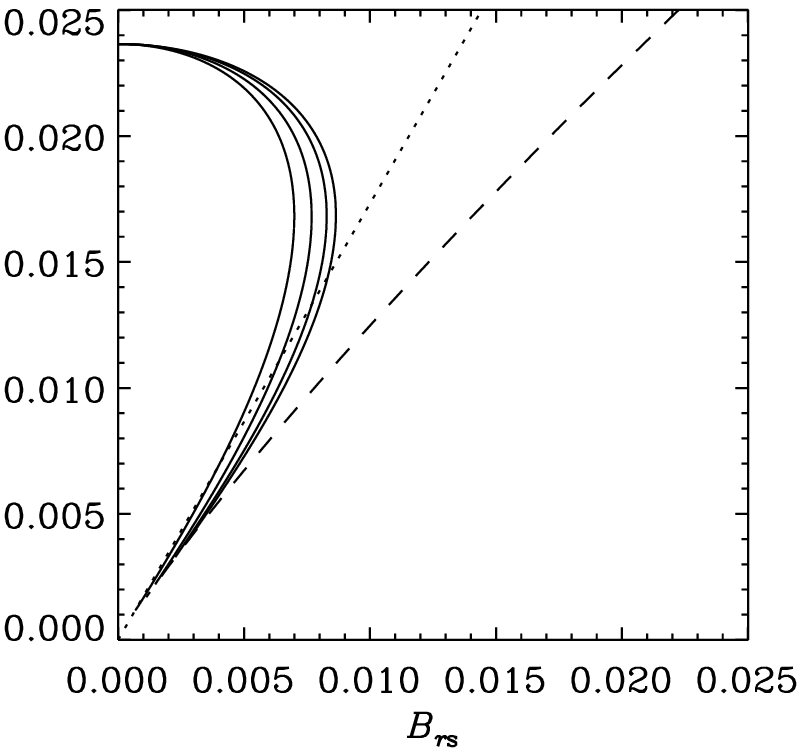}\qquad\qquad\epsfysize=8cm\epsfbox{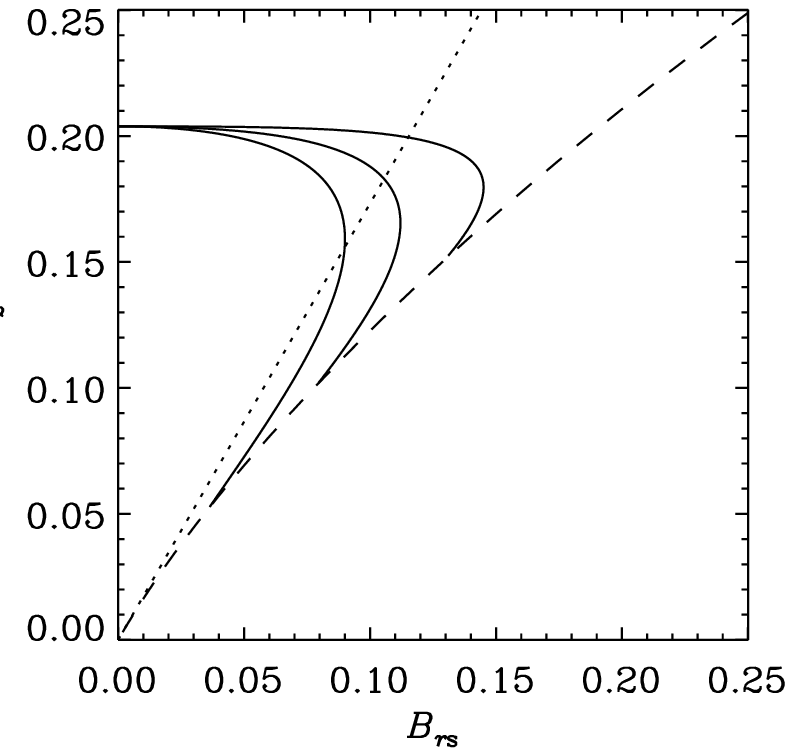}}
\caption{{\bf Figure~9.}  The effect of sub-adiabatic stratification on the magnetorotational instability.  Left: stability boundaries for weakly magnetized, polytropic thin discs with $\Gamma=4/3$ are plotted for four different values of the adiabatic exponent $\gamma$.  The four solid lines are the stability boundaries for $\gamma=4/3$, $5/3$, $3$ and $\infty$ (from right to left).  Right: similar results for equilibria with $\Gamma=5/3$.  The three solid lines are the stability boundaries for $\gamma=5/3$, $3$ and $\infty$ (from right to left).}
\endfigure

\section{Non-ax\-isym\-met\-ric modes}

The analysis of Section~3 may be extended to include non-ax\-isym\-met\-ric waves and instabilities of azimuthal wavenumber $m$.  A distinction must be made between modes with $m=O(\epsilon^{-1})$ and those with $m=O(1)$.  In the former case, the azimuthal wavenumber is comparable to the radial and vertical wavenumbers, and the equations of Section~3 are not valid in any sense.  These modes are expected to resemble the global non-ax\-isym\-met\-ric instabilities of thick tori studied by Papaloizou \& Pringle (1984).  The strong differential Doppler shift means that these modes are probably localized in a region of small radial extent about their corotation radius, and require a boundary of the disc to be present in this small region if they are to have dynamical [$O(1)$] growth rates.

For modes with $m=O(1)$, however, the equations of Section~3 require very few changes.  The eigenfunctions have the form
$$\bxi({\bmath r},t)\sim{\rm Re}\left\{\bxi_0(r,\zeta)\exp\left[-{\rm i}\omega t+{\rm i}m\phi+{\rm i}\epsilon^{-1}\int k(r)\,{\rm d}r\right]\right\},\eqno(7.1)$$
and the frequency eigenvalue $\omega$ must be replaced with the intrinsic frequency $\hat\omega=\omega-m\Omega$.  In this way LP were able to discuss the propagation of non-ax\-isym\-met\-ric waves in an isothermal accretion disc.  Provided that the dispersion relation yields real values of $\hat\omega$ for real values of $k$, there is no difficulty because $\omega$, $\hat\omega$ and $k$ are all real within the wave region of a mode.

There remains the possibility that the dispersion relation yields imaginary values of $\omega$ for real values of $k$, as a result of the magnetorotational instability or a convective instability.  In that case one cannot simply replace $\omega$ with $\hat\omega$, for the following reason.  Since $\omega$ must be constant for a normal mode, it follows that the real part of $\hat\omega$ can vanish at (at most) one radius, the corotation radius of the mode.  Therefore $\hat\omega$ is imaginary at the corotation radius, but anywhere else it cannot be a solution of the dispersion relation for any real value of $k$.  The conclusion is that, for a non-ax\-isym\-met\-ric unstable mode, $k$ is real only at the corotation radius.

When $k$ is not real, the variational principle of Section~4.2 is not valid.  Otherwise, the only change that need be made in Section~3 is to replace $|k|$ with the function
$$p(k)=\cases{+k,&${\rm Re}(k)>0$,\cr -k,&${\rm Re}(k)<0$,\cr}\eqno(7.2)$$
which is an analytic function apart from a branch cut along the imaginary axis.  It then follows that each branch of the dispersion relation is an analytic function, except on ${\rm Re}(k)=0$.  The derivative $\partial\omega/\partial k$ is known for real values of $k$ from the real dispersion relation, and is imaginary for an unstable branch.  Let $k=k_{\rm c}$ be the real value of the radial wavenumber at the corotation radius $r=r_{\rm c}$ of such a mode.  Then, at a neighbouring radius $r$, the wavenumber is (to first order)
$$k\approx k_{\rm c}+\left[{3m\Omega_{\rm c}\over2r_{\rm c}}\bigg/\!\left({\partial\omega\over\partial k}\right)_{\!\!\rm c}\,\right](r-r_{\rm c}),\eqno(7.3)$$
assuming that the derivative $(\partial\omega/\partial k)_{\rm c}$ does not vanish.  The quantity in square brackets is purely imaginary; depending on the sign of the various terms, ${\rm Im}(k)$ is either an increasing function or a decreasing function of $(r-r_{\rm c})$.  If it is an increasing function, then the mode will take the form of a localized Gaussian wave packet around $r=r_{\rm c}$, according to equation (7.1).  If it is a decreasing function, the function (7.1) would increase greatly in magnitude away from $r=r_{\rm c}$ and cannot satisfy the radial boundary conditions.

In conclusion, non-ax\-isym\-met\-ric unstable modes are localized around the corotation radius, and obey the real dispersion relation only at that point.  The modes have either positive or negative values of $m$ depending on the sign of the imaginary group velocity.  At any given point on an unstable branch of the real dispersion relation, non-ax\-isym\-met\-ric modes of this kind can be found.

\section{Discussion}

This paper has addressed a number of issues relating to the spectrum of waves and instabilities in an accretion disc containing a poloidal magnetic field.  The general analysis of the continuous spectrum in Section~2 demonstrates that the only type of possible instability that is truly localized on a single magnetic surface is an essentially axisymmetric magnetoconvective instability associated with the component of gravity parallel to the magnetic field.  Unlike the case of uniform (or zero) rotation, the interchange instability, if present, is not manifest in the continuous spectrum.  Neither is the magnetorotational instability or any other instability associated with differential rotation.  These must instead be sought in the normal modes of the system.

The extension of the asymptotic methods of Paper~I leads to a WKB description of axisymmetric waves and instabilities in a weakly magnetized thin disc in terms of a local dispersion relation which generalizes earlier work by LP and KP.  The waves propagate radially as in a slowly varying waveguide.  The modes in a hydrodynamic disc can be classified as f, p, g and r modes according to their behaviour in the limit of large radial wavenumber.  When a magnetic field is introduced, the dispersion diagram is complicated by a large number of avoided crossings which make a systematic classification difficult.  If the magnetic field is relatively weak, unstable magnetorotational modes also occur for sufficiently small values of the radial wavenumber.  The overall stability boundary in the parameter space of polytropic equilibria is obtained by computing the marginal curve for the first magnetorotational mode of odd symmetry.  For a given value of the vertical magnetic field, the addition of a radial component (which makes the field lines bend) has a stabilizing influence.  A sub-adiabatic stratification also has a mild stabilizing influence if the magnetic field is not purely vertical.

This analysis shows for the first time that it is possible to construct stable equilibria which are capable of driving a wind.  Indeed, increasing the magnetic field strength not only tends to stabilize the equilibria but also makes it easier to construct equilibria in which the magnetic field lines at the surface of the disc are inclined to the vertical at angles significantly greater than $\pi/6$.  It should be emphasized, however, that the stability analysis applies only to the disc and not to any wind solution that may be superimposed on it.

It is also important to note that the analysis is valid only for axisymmetric modes and non-ax\-isym\-met\-ric modes of small azimuthal wavenumber [$m=O(1)$], and that it can only detect instabilities with dynamical [$O(1)$] growth rates.  It is expected that all models are subject to global non-ax\-isym\-met\-ric instabilities which depend strongly on the radial boundaries of the disc, resembling the Papaloizou--Pringle instability, but these probably have very small growth rates in a thin disc.  The radial interchange instability (Spruit et al. 1995), if it is not entirely stabilized by the differential rotation, would also have a sub-dynamical growth rate in a weakly magnetized thin disc.  The equilibria described here as `stable' may therefore not exist as truly laminar flows but could be subject to weak turbulence or at least fluctuations.  However, that would be quite different from equilibria that are locally unstable to the magnetorotational instability, which are bound to degenerate into strong MHD turbulence.

There are many ways in which this analysis could be improved and extended.  In order to study the radial propagation of waves, an explicit global equilibrium model of a magnetized disc should be constructed by choosing the functional forms of $\zeta_{\rm s}(r)$, $\psi_0(r)$ and $K_0(r)$, say, and solving for the equilibrium at each radius.  The variation of the radial wavenumber with radius could then be determined for any mode by following the dispersion relation.  The dependence of the amplitude of the mode on radius could be found either by solving the linearized equations at the next order in $\epsilon$, or, more simply, by appealing to a wave-action conservation relation (cf. LP).  Other areas to be explored are the possible stabilization of a super-adiabatically stratified disc by a sufficiently strong magnetic field, the propagation of the $m=1$ `tilt' mode in a magnetized disc, and the effects of self-grav\-i\-ta\-tion on the equilibria and spectra of both hydrodynamic and magnetized discs.

\section*{Acknowledgments}

I would like to thank Jim Pringle, Douglas Gough and Ulf Torkelsson for helpful discussions.  A research studentship from the Particle Physics and Astronomy Research Council is acknowledged.

\bye